\documentclass[pre,twocolumn,10pt,aps,longbibliography,nofootinbib]{revtex4-1}

 \usepackage{verbatim}
 \usepackage{amsmath}
 \usepackage{amssymb}
 \usepackage{dsfont}
 \usepackage{latexsym}
 \usepackage{amsfonts}
 \usepackage{epsfig}
 \usepackage{epstopdf}
 \usepackage{color}
 \definecolor{darkblue}{rgb}{0,0,.5}
 \usepackage[linktocpage, colorlinks=true, linkcolor=darkblue, citecolor=darkblue]{hyperref}
 \usepackage[all]{hypcap}

\newcommand{\C}[1]{{\cal{#1}}}
\newcommand{\bb}[1]{\textbf{#1}}
\newcommand{\ket}[1]{\left.\left|#1\right.\right>}

\newcommand{\rl}[0]{{\rangle\langle}}

\begin{document}

\title{Quantum and Information Thermodynamics: \\
A Unifying Framework based on Repeated Interactions}

\author{Philipp Strasberg}
\email{phist@physik.tu-berlin.de}
\author{Gernot Schaller}
\author{Tobias Brandes}
\affiliation{Institut f\"ur Theoretische Physik, Technische Universit\"at Berlin, Hardenbergstr. 36, D-10623 Berlin, Germany}

\author{Massimiliano Esposito}
\affiliation{Complex Systems and Statistical Mechanics, Physics and Materials Science, University of Luxembourg, L-1511 Luxembourg, Luxembourg}

\date{\today}

\begin{abstract}
 We expand the standard thermodynamic framework of a system coupled to a thermal reservoir by considering a stream 
 of independently prepared units repeatedly put into contact with the system. These units can be in any nonequilibrium 
 state and interact with the system with an arbitrary strength and duration. We show that this stream constitutes an 
 effective resource of nonequilibrium free energy and identify the conditions under which it behaves as a heat, work 
 or information reservoir. We also show that this setup provides a natural framework to analyze information erasure 
 (``Landauer's principle'') and feedback controlled systems (``Maxwell's demon''). In the limit of a short system-unit 
 interaction time, we further demonstrate that this setup can be used to provide a thermodynamically sound 
 interpretation to many effective master equations. We discuss how non-autonomously driven systems, micromasers,
 lasing without inversion, and the electronic Maxwell demon, can be thermodynamically analyzed within our framework. 
 While the present framework accounts for quantum features (e.g. squeezing, entanglement, coherence), 
 we also show that quantum resources do not offer any advantage compared to classical ones in terms of the maximum 
 extractable work. 
\end{abstract}

\maketitle

\section{Introduction}

Thermodynamics was traditionally designed to understand the laws which govern the behavior of macroscopic systems at 
equilibrium in terms of few macroscopic variables (e.g.  temperature, pressure, chemical potential, energy, volume, 
particle number, etc.) containing very limited information about the microscopic state of the system. 
Remarkable progress has been done over the last decades to understand under which conditions the laws of 
thermodynamics emerge for small-scale systems where quantum and stochastic effects dominate and which 
are usually far away from thermal equilibrium. 
This includes a consistent thermodynamic framework for driven systems weakly coupled to large and fast thermal reservoirs, 
which are described by a microscopically derived (quantum) master equation (ME)~\cite{KosloffEntropy2013, 
SchallerBook2014, GelbwaserKlimovskyNiedenzuKurizkiAdv2015, BulnesCuetaraEspositoSchallerEntropy2016}. 
Such MEs can be also used as a basis to establish universal fluctuation relations which replace the traditional second 
law formulated as an inequality by an exact symmetry that fluctuations must satisfy arbitrarily
far from equilibrium~\cite{EspositoHarbolaMukamelRMP2009, CampisiHaenggiTalknerRMP2011, 
JarzynskiAnnuRevCondMat2011, SeifertRPP2012, VandenBroeckEspositoPhysA2015}.
This theory has been very successful to study small systems in a large variety of fields ranging from biophysics to 
electronics and many of its predictions have been verified experimentally~\cite{BustamanteLiphardtRitortPhysTod2005, 
GeQianQianPR2012, CilibertoGomezSolanoPetrosyanAnnuRevCondMat2013, PekolaNatPhys2015, ThierschmannEtAlCompRend2016, 
SerraGarciaEtAlPRL2016, RossnagelEtAlScience2016}.

Yet, many situations encountered nowadays force us to go beyond the setup of driven systems weakly coupled to 
thermal reservoirs. 
Notable examples include the thermodynamic description of computation and information processing using 
feedback controls (``Maxwell demons") where different members of the statistical ensemble undergo different 
drivings~\cite{BennettIJTP1982, LeffRexBook2003, MaruyamaNoriVedralRMP2009, SagawaJSM2014, ParrondoHorowitzSagawaNatPhys2015},  
systems interacting with reservoirs prepared in nonequilibrium states~\cite{ScullyEtAlScience2003, DillenschneiderLutzEPL2009, 
RossnagelEtAlPRL2014, LiEtAlPRE2014, AlickiGelbwaserNJP2015, ManzanoEtAlPRE2016, NiedenzuEtAlNJP2016} 
or non-Gibbsian equilibrium states~\cite{GardasDeffnerPRE2015}, 
and systems described by non-Hermitian dynamics \cite{GardasDeffnerSaxenaSciRep2016}. 

In this paper, we extend the traditional framework of thermodynamics by considering a system which, in addition 
of being in contact with a thermal reservoir, interacts with a stream of external systems which we call ``units''. 
Each independently prepared unit interacts for a certain time with the system before being replaced by another 
one and no additional assumption about the state of the units nor the system-unit interaction is required. 
In the most general picture, this stream of units will be shown to constitute a \emph{resource of nonequilibrium 
free energy} modifying the traditional energetic and entropic balances. We will study the limits in which the 
stream of units effectively reproduces the effect of a heat, work or information reservoir. We will also explore 
limits giving rise to an effective closed dynamics for the system which still allows for a consistent thermodynamic 
description. We will focus on the ensemble averaged description and not on fluctuations. 

The benefit of our generalized thermodynamic framework is that it provides a unified perspective and 
encompasses many previously considered setups.  
In modern physics, such setups have probably first been used in quantum optics, theoretically as well as experimentally, 
to model a maser in which a stream of atoms is injected into a cavity in order to macroscopically populate 
it~\cite{MeschedeWaltherMuellerPRL1985, FilipowiczJavanainenMeystrePRA1986, ScullyZubairyBook1997}. 
Such setups have also been used to stabilize photon number states via measurement based quantum feedback control~\cite{SayrinEtAlNature2011}. 
\begin{figure*}
 \centering\includegraphics[width=0.97\textwidth,clip=true]{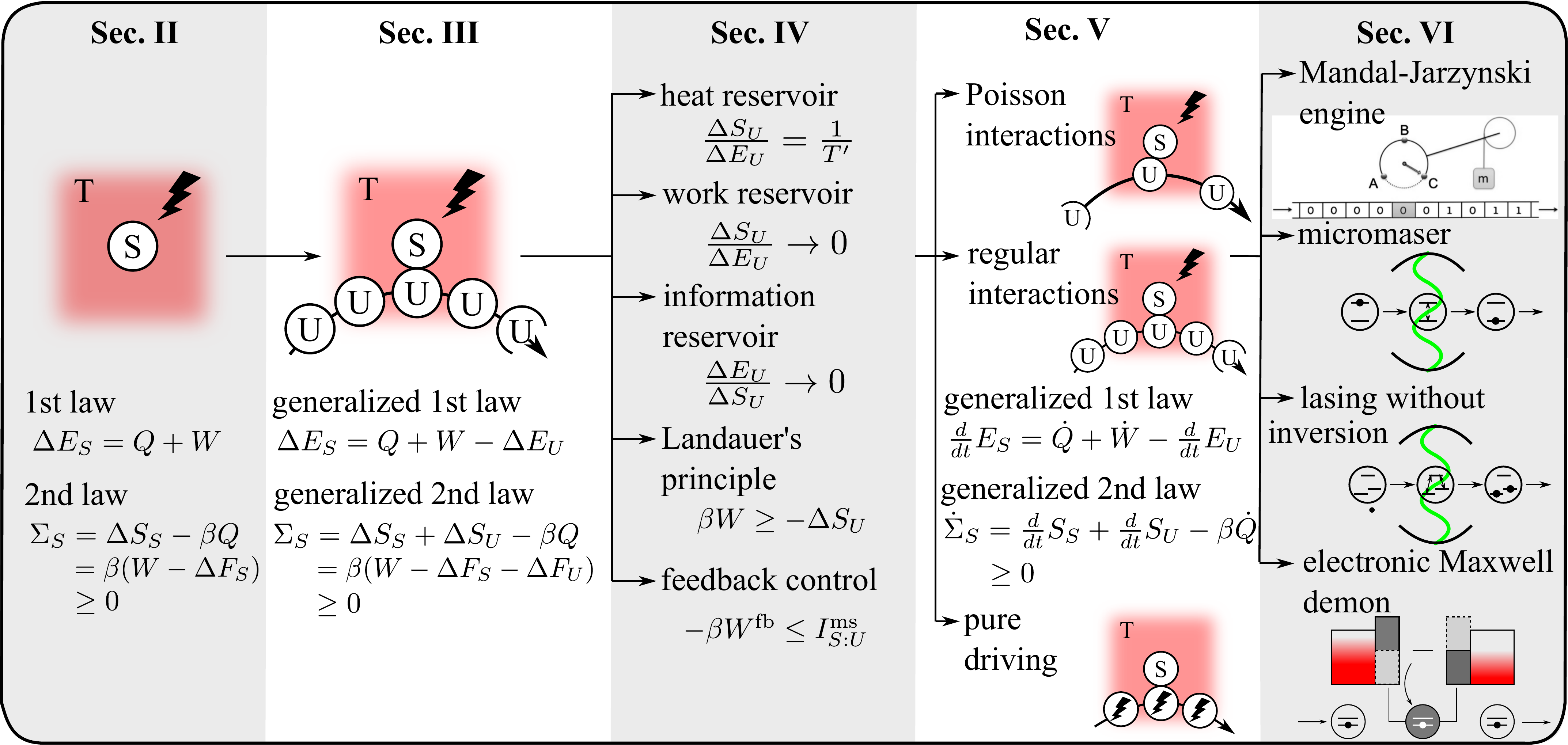}
 \label{fig outline} 
 \caption{(Color Online) Overview of the structure of the article covering its main results and applications. 
 Here and in the following, $E_{S,U}$, $S_{S,U}$ and $F_{S,U}$ denote the energy, entropy and non-equilibrium free 
 energy of the system ($S$) or unit ($U$). $Q$ is the heat flowing from the reservoir at inverse temperature 
 $\beta = T^{-1}$ ($k_B \equiv 1$) and $W$ is the work done on the system. Furthermore, $\Sigma_S$ denotes the 
 entropy production and $I_{S:U}$ the mutual information between system and unit. The picture of the 
 Mandal-Jarzynski engine was taken from Ref.~\cite{MandalJarzynskiPNAS2012}. }
\end{figure*}
In theoretical works, an ``information reservoir'' made of a stream of bits was proposed to extract work from a 
single heat reservoir~\cite{MandalJarzynskiPNAS2012, MandalQuanJarzynskiPRL2013}, a picture that also closely 
resembles a Turing machine~\cite{BennettIJTP1982, StrasbergEtAlPRE2015}. 
The setup is also close to resource theoretic formulations of thermodynamics, in which one investigates which system 
transformations are possible given a set of freely available states and resources (all other 
states)~\cite{GourMuellerEtAlPR2015, GooldEtAlJPA2016}. 
Further analogies can be drawn with biomolecular motors or enzymes~\cite{GeQianQianPR2012} which manipulate, e.g.  
nucleic acids (units) on a DNA strand, or with scattering theory where incoming and outgoing wave packets 
(units) interact for a short time with the scatterer (the system)~\cite{UzdinKosloffNJP2014, WhitneyPRL2014}. 

\subsection*{Outline}

The structure of the paper and some of its main results are summarized in Fig.~\ref{fig outline}. 
We start very generally in Sec.~\ref{sec standard thermodynamic description} by considering two interacting systems and review 
under what conditions the laws of thermodynamics can be established if one of the systems is initially in a thermal state
and plays the role of the reservoir while the other is the driven system of interest. Besides establishing notation, 
this section also sets the basis for Sec.~\ref{sec generalized thermodynamic framework}, were the system now also 
interacts with an external stream of units. Generalized laws of thermodynamics are established which show that the 
stream of units effectively constitutes a resource of nonequilibrium free energy. 
We then consider various limiting cases in Sec.~\ref{sec applications part I} where the stream of units respectively 
behaves as a heat, work and information reservoir. Furthermore, Landauer's principle is derived as well as a quantum 
version of the second law of thermodynamics under feedback control. 
We go one step further in Sec.~\ref{sec effective master equation} by considering scenarios leading to a closed reduced 
dynamics for the system when tracing out the units, but where one still retains a consistent thermodynamic description. 
More specifically, we consider the limit of infinitesimally short interactions which are either Poissonian or regularly 
distributed and which lead to effective MEs. 
We also analyze the limit where the units effectively generate a time-dependent Hamiltonian for the system. 
Specific models, apparently unrelated in the literature, are then analyzed within our unifying framework in the subsequent 
section (Sec.~\ref{sec applications part II}). These include the Mandal-Jarzynski engine, the micromaser, lasing without 
inversion where work is extracted from quantum coherence, and the electronic Maxwell demon. 
We finally close the paper with Sec.~\ref{sec final remarks}, where we first show that our generalized second law of 
thermodynamics does not conflict with the traditional Kelvin-Planck statement of the second law, we then prove that 
quantum thermodynamics offers \emph{a priori} no advantages compared to classical thermodynamics within our framework, 
and we finally give a short summary and outlook. 

\tableofcontents

\section{Energy and entropy balance of an open quantum system}
\label{sec standard thermodynamic description}

\subsection{Two interacting systems}
\label{sec two interacting systems}

To introduce important concepts and notation, we start by considering two interacting systems $X$ and $Y$, 
which are initially (at time $t=0$) decorrelated, i.e. 
\begin{equation}\label{EQ:initcond}
 \rho_{XY}(0) = \rho_X(0)\otimes\rho_Y(0) \equiv \rho_X(0)\rho_Y(0).
\end{equation}
Here, $\rho_{XY}$ denotes the density operator of the compound system $X$ and $Y$ whereas $\rho_X$ ($\rho_Y$) 
describes the reduced state of $X$ ($Y$). In order to make a statement about the first law of thermodynamics, 
we associate a Hamiltonian to the setup which we decompose as 
\begin{equation}
 \begin{split}
  H_\text{tot}(t)	&=	H_X(t)\otimes \mathbf{1}_Y + \mathbf{1}_X\otimes H_Y(t) + H_{XY}(t)	\\
			&\equiv H_X(t) + H_Y(t) + H_{XY}(t).
 \end{split}
\end{equation}
Here, $\mathbf{1}_X$ ($\mathbf{1}_Y$) denotes the identity on the Hilbert space of system $X$ ($Y$), which we 
usually suppress to simplify notation, and $H_{XY}(t)$ describes the interaction between system $X$ and $Y$. 
Furthermore, all terms can, in principle, be explicitly time-dependent. 

The time evolution of the compound system is governed by the Liouville-von Neumann equation 
$d_t\rho_{XY}(t) = -i[H_\text{tot}(t),\rho_{XY}(t)]$ ($\hbar\equiv1$ throughout the text). 
Introducing the unitary time evolution operator $U_t \equiv \C T_+ \exp\left[-i\int_0^t ds H_\text{tot}(s)\right]$ 
(where $\C T_+$ denotes the time ordering operator), the state of the compound system at time $\tau$ is given by 
\begin{equation}
 \rho_{XY}(\tau) = U_\tau \rho_X(0)\rho_Y(0) U_\tau^\dagger,
\end{equation}
which is in general correlated.

To obtain the reduced state of system $X$, we must trace over system $Y$. Using the spectral decomposition of the 
initial density matrix of system $Y$, $\rho_Y(0) = \sum_l p_l |l\rangle_Y\langle l|$, we arrive at 
\begin{equation}\label{eq Kraus map}
 \begin{split}
  \rho_X(\tau)	&=	\mbox{tr}_Y\{\rho_{XY}(\tau)\}	\\
		&=	\sum_{kl} T_{kl}\rho_X(0)T_{kl}^\dagger \equiv \Phi_X(\tau)\rho_X(0),
 \end{split}
\end{equation}
where $T_{kl} \equiv \sqrt{p_l}\langle k|U_\tau|l\rangle_Y$. The $T_{kl}$ are still operators in the Hilbert space 
of system $X$ and fulfill the completeness relation 
\begin{equation}
 \sum_{kl} T_{kl}^\dagger T_{kl} = \mathbf{1}_X.
\end{equation}
The map $\Phi_X(\tau)$ in Eq.~(\ref{eq Kraus map}) is known as a Kraus map or quantum operation and it is the most 
general map (preserving the density matrix properties) for a quantum system which was initially 
uncorrelated~\cite{KrausBook1983, NielsenChuangBook2000,VedralRMP2002, WisemanMilburnBook2010}. 
We note that the representation of $\Phi_X$ in terms of the Kraus operators 
$T_{kl}$ is \emph{not} unique. 

The energy of the compound system at any point in time is $E_{XY}(t) \equiv \mbox{tr}_{XY}\{H_\text{tot}(t)\rho_{XY}(t)\}$. 
Because the compound system is isolated, the dynamics is unitary and the energy change is solely due to the time dependence 
in the Hamiltonian and can thus be identified as work 
\begin{equation}\label{eq 1st law isolated}
 d_t E_{XY}(t) = \mbox{tr}_{XY}\left\{\rho_{XY}(t)d_tH_\text{tot}(t)\right\} \equiv \dot W(t).
\end{equation}
The rate of work injection $\dot W$ is positive if it increases the energy of the compound system. 
Eq.~(\ref{eq 1st law isolated}) is the first law of thermodynamics for an \emph{isolated} system. 

We now consider the entropy balance by defining the von Neumann entropy of a system $X$, which we 
interpret as a measure for our lack of knowledge about the state $\rho_{X}$, as usual by 
\begin{equation}
 S_X \equiv -\mbox{tr}_X\{\rho_X\ln\rho_X\}.
\end{equation}
Because the joint von Neumann entropy $S_{XY}(t) \equiv -\mbox{tr}_{XY}\{\rho_{XY}(t)\ln\rho_{XY}(t)\}$ 
does not change under unitary time evolution, we have $S_{XY}(\tau) = S_{XY}(0)$. We further introduce the non-negative 
(quantum) mutual information 
\begin{equation}
 I_{X:Y}(t) \equiv S_X(t) + S_Y(t) - S_{XY}(t) \ge 0\,,
\end{equation}
which measures the amount of correlations shared between $X$ and $Y$~\cite{NielsenChuangBook2000}. 
For the initially decorrelated state~(\ref{EQ:initcond}) we have $S_{XY}(0) = S_X(0) + S_Y(0)$. 
Hence, in terms of the mutual information we can write that
\begin{equation}\label{eq entropy balance X Y}
 I_{X:Y}(\tau) = \Delta S_X(\tau) + \Delta S_Y(\tau) \ge 0.
\end{equation}
Thus, the mutual information tells us how the sum of the marginal entropies $\Delta S_X(\tau) \equiv S_X(\tau) - S_X(0)$ 
and $\Delta S_Y(\tau) \equiv S_Y(\tau) - S_Y(0)$ can change. 

Introducing the relative entropy between two density matrices $\rho$ and $\sigma$~\cite{NielsenChuangBook2000}
\begin{equation}\label{eq Kleins inequality}
 D(\rho||\sigma) \equiv \mbox{tr}\{\rho(\ln\rho-\ln\sigma)\} \ge 0
\end{equation}
which is non-negative by Klein's inequality, the mutual information can also be written as 
\begin{equation}\label{eq def mutual information}
 I_{X:Y}(t) = D[\rho_{XY}(t)||\rho_X(t)\rho_Y(t)] \geq 0.
\end{equation}
By measuring the local entropy changes in $X$ and $Y$, the mutual information is therefore 
also a measure of the information lost when disregarding the correlation established over time $t$ between 
$X$ and $Y$ while keeping full knowledge of $X$ and of $Y$ separately in the description.
Note that relative entropy is not symmetric, i.e. $D(\rho||\sigma) \neq D(\sigma||\rho)$ in general, but mutual 
information fulfills $I_{X:Y} = I_{Y:X}$. Furthermore, it is important to mention that the action of any Kraus map 
$\Phi$ can never increase the relative entropy~\cite{LindbladCMP1975, VedralRMP2002}, i.e. 
\begin{equation}\label{eq contractivity Kraus map}
 D(\Phi\rho||\Phi\sigma) \le D(\rho||\sigma).
\end{equation}

\subsection{System coupled to a thermal reservoir}
\label{sec a general approach}

To make further contact with thermodynamics we now consider the case where the system $Y$ is supposed to play the role 
of a thermal reservoir. For this purpose we relabel $Y$ by $R$ and make the two assumptions that the 
Hamiltonian $H_R$ is time-independent and that the initial state of the reservoir is thermal: 
\begin{equation}\label{IniThermalRes}
 \rho_R(0) = \rho^R_\beta \equiv \frac{e^{-\beta H_R}}{Z_R}, ~~~ Z_R = \mbox{tr}(e^{-\beta H_R}).
\end{equation}
Similar treatments were presented, e.g.  in Refs.~\cite{EspositoLindenbergVandenBroeckNJP2010, 
DeffnerJarzynskiPRX2013, ReebWolfNJP2014, GooldPaternostroModiPRL2015}. 

Following the rational of Eq.~(\ref{eq 1st law isolated}), the energy change in the total system 
is identified as the work done by the external time dependent driving on the system  
\begin{equation}
 d_t E_{XY}(t) = \mbox{tr}_{XR} \left\{\rho_{XR}(t) d_t [ H_\text{X}(t) + H_{XR}(t) ] \right\} \equiv \dot W(t).
\end{equation}
The energy flowing out of the reservoir is in turn identified as the heat flow into the system $X$ 
at time $t$ (positive if it increases the system energy)
\begin{equation}
 \dot Q(t) \equiv - \mbox{tr}_R \left\{H_R d_t\rho_R(t)\right\}.
\end{equation}
Consequently, the internal energy of the system $X$ is identified as
\begin{equation}
 E_X(t) \equiv \mbox{tr}_{XR}\{[H_X(t)+H_{XR}(t)]\rho_{XR}(t)\}
\end{equation}
so that
\begin{equation}
 d_t E_X(t) = \dot W(t) + \dot Q(t).
\end{equation}
This constitutes the first law of thermodynamics for a closed system. We use the conventional terminology 
of thermodynamics where a system exchanging only energy (but not matter) with a reservoir is called 
``closed'' though one would rather call it ``open'' from the perspective of open quantum system theory. An ``open'' 
system in the thermodynamic sense (exchanging also matter with its environment) can be considered by introducing a 
chemical potential for reservoir $R$, which is then described by an initial grand canonical equilibrium state. 
Integrating the first law over an interval $[0,\tau]$ gives 
\begin{equation}\label{eq 1st law general}
 \Delta E_X(\tau) = E_X(\tau)-E_X(0) = W(\tau) + Q(\tau)
\end{equation}
with 
\begin{equation}
 \begin{split}\label{eq def Q general evolution}
 & W(\tau) \equiv \int_0^\tau dt \dot W(t) = E_{XR}(\tau) - E_{XR}(0), \\
 & Q(\tau) \equiv \int_0^\tau dt \dot Q(t) = -\mbox{tr}_R\{H_R[\rho_R(\tau)-\rho_R(0)]\}. 
 \end{split}
\end{equation}

By analogy with the second law of phenomenological nonequilibrium thermodynamics~\cite{KondepudiPrigogineBook2007} 
which states that the non-negative entropy production characterizing the irreversibility of a process is given 
by the sum of the entropy change in the system and in the (macroscopic ideal and always equilibrated) reservoir, 
we follow Ref.~\cite{EspositoLindenbergVandenBroeckNJP2010} and define entropy production as 
\begin{equation}\label{eq 2nd law general}
 \Sigma(\tau) \equiv \Delta S_X(\tau) - \beta Q(\tau).
\end{equation}
Since the initial reservoir state~(\ref{IniThermalRes}) is thermal, 
the non-negativity of $\Sigma$ can be shown by noting the identities~\cite{EspositoLindenbergVandenBroeckNJP2010} 
\begin{equation}
 \begin{split}\label{eq 2nd law relative entropy}
  \Sigma(\tau)	&=	D[\rho_{XR}(\tau)||\rho_X(\tau)\otimes\rho^R_\beta]	\\
		&=	D[\rho_R(\tau)||\rho^R_\beta] + I_{X:R}(\tau) \ge 0.
 \end{split}
\end{equation}
It relies on the assumption that the initial total system state is decorrelated. 
We emphasize that $\Sigma(\tau)\ge0$ holds for any reservoir size and can thus not be considered 
as strictly equivalent to the phenomenological second law of nonequilibrium thermodynamics. 
We also note that expression (\ref{eq 2nd law relative entropy}) provides interesting insight 
into the difference between the way in which we treated the reservoir $R$ in this section 
compared to the way in which we treated system $Y$ in the previous section: 
the entropy production does not only measure the information lost in the correlations between the system 
and the reservoir via the mutual information $I_{X:R}$, it also measures the information lost in not 
knowing the state of the reservoir after the interaction via $D[\rho_R(\tau)||\rho^R_\beta]$.
This translates the obvious fact in thermodynamics that one has no access to the state 
of the reservoir and that one only knows the energy that has flown in it as heat.

We define an \emph{ideal heat reservoir} as a reservoir which remains close to thermal equilibrium during its interaction with the 
system, i.e. $\rho_R(\tau) = \rho_{\beta}^R + \epsilon \sigma_R$, where $\epsilon$ is a small parameter and $\mbox{tr}_R(\sigma_R) = 0$. 
Using the exact identity 
\begin{equation}\label{FreeEnergRes}
 T D[\rho_R(\tau)||\rho_{\beta}^R] = - Q(\tau) - T \Delta S_R(\tau)  \geq 0\,,
\end{equation}
this means that the information lost by not knowing the reservoir state becomes 
negligible because $D[\rho_R(\tau)||\rho^R_\beta] = \C O(\epsilon^2)$.\footnote{\label{footnote 1}Proving 
$D[\rho_R(\tau)||\rho_{\beta}^R] = \C O(\epsilon^2)$ can be done by expanding in a power series 
$D[\rho_R(\tau)||\rho_{\beta}^R] = D_0 + \epsilon D_1 + \C O(\epsilon^2)$. Clearly, 
$D_0 = D[\rho_{\beta}^R||\rho_{\beta}^R] = 0$ and $D_1 = 0$ due to Klein's inequality~(\ref{eq Kleins inequality}) 
because $\epsilon$ can be positive as well as negative.}
Consequently, the entropy change in the reservoir can be solely expressed in terms of the heat flowing 
in it via Clausius equality, i.e. $\Delta S_R(\tau) = -\beta Q(\tau)$ where these two quantities are generically
of first order in $\epsilon$ and only differ from each other to second order in $\epsilon$. 
Using (\ref{eq 2nd law relative entropy}), it also means that the entropy production due to an ideal heat reservoir 
coincides (to second order in $\epsilon$) with the lost mutual information between the system and the reservoir, 
i.e. $\Sigma(\tau) = I_{X:R}(\tau)$.

Finally, it will turn out to be useful to introduce the concept of a \emph{nonequilibrium free energy}. 
Following Refs.~\cite{GaveauSchulmanPLA1997, CrooksPRE2007, EspositoLindenbergVandenBroeckNJP2010, 
EspositoVandenBroeckEPL2011}, we define 
\begin{equation}\label{eq def non eq free energy}
 F_X(t) \equiv E_X(t) - T S_X(t),
\end{equation}
where $T$ is the temperature of the initial reservoir attached to the system. Since it is fixed, 
$dF_X(t)$ is still an exact differential. Using this quantity, we can write the second law as 
\begin{equation}\label{eq 2nd law general 2}
  \Sigma(\tau) = \beta[W(\tau) - \Delta F_X(\tau)] \ge 0.
\end{equation}
Explicit processes where $\Sigma(\tau)$ can be made arbitrarily close 
to zero have been considered, e.g. in Refs.~\cite{JacobsPRA2009, EspositoVandenBroeckEPL2011, HorowitzParrondoNJP2011, 
ReebWolfNJP2014, SkrzypczykShortPopescuNatComm2014}.

\subsection{The weak coupling limit and master equations}
\label{sec thermodynamics weak coupling limit}

The results above provide a general way to formally derive the laws of thermodynamics as exact identities.
Remarkably, these relations hold even if the reservoir $R$ is arbitrarily small and strongly influenced by the presence of 
the system. However, while the entropy production $\Sigma(\tau)$ is proven to be nonnegative, its rate can be negative. 
Indeed, as finite-size quantum systems evolve quasi-periodically, the exact rate of entropy 
production must become negative at some time to restore the initial state. Furthermore, these 
identities are also of limited practical use because computing any particular expression 
requires to solve the full Liouville-von Neumann dynamics for the joint system and reservoir. 

The weak coupling limit between the system and the reservoir circumvents these limitations and is of great practical relevance. 
It has been used since a long time to study quantum thermodynamics~\cite{SpohnLebowitzAdvChemPhys1979, AlickiJPA1979}
(see also Refs.~\cite{KosloffEntropy2013, GelbwaserKlimovskyNiedenzuKurizkiAdv2015} for recent reviews). 
Within this limit the system does not perturb the reservoir over any relevant time-scale and it is further 
assumed that the reservoir behaves memory-less (i.e. Markovian). This allows to close the equation of motion 
for the system density matrix $\rho_X(t)$. The resulting dynamics is called a (quantum) master equation (ME). 
In this limit, the general results of Sec.~\ref{sec a general approach} reduce to the well-known 
ME formulation of quantum thermodynamics~\cite{EspositoLindenbergVandenBroeckNJP2010}, where all 
thermodynamic quantities can be expressed solely in terms of system operators.

More specifically, after applying the Born-Markov-secular approximations~\cite{SpohnLebowitzAdvChemPhys1979, 
BreuerPetruccioneBook2002, EspositoHarbolaMukamelRMP2009, KosloffEntropy2013, SchallerBook2014} which is 
usually justified in the weak coupling limit, the ME can be put into the form\footnote{As a technical sideremark 
one should note that the system Hamiltonian $H_X(t)$ usually gets renormalized due to Lamb shift 
terms~\cite{BreuerPetruccioneBook2002}. In the following we will not explicitly mention them tacitly assuming that they 
were already conveniently absorbed in the definition of $H_X(t)$. } 
\begin{equation}\label{eq generic ME for X}
 \begin{split}
  d_t\rho_X(t)	&=	-i[H_X(t),\rho_X(t)] + \C L_\beta(t)\rho_X(t)	\\
			&\equiv	\C L_X(t)\rho_X(t),
 \end{split}
\end{equation}
where $\C L_\beta(t)$ and $\C L_X(t)$ denote superoperators which act linearly on the space of system operators.
In order to derive Eq.~(\ref{eq generic ME for X}) one also has to assume that the driving of $H_X(t)$ 
is slow compared to the relaxation time of the reservoir, though this does not imply that the driving 
must be adiabatic~\cite{HauptEtAlPSSB2013, YamaguchiYugeOgawaPRE2017}.\footnote{Other MEs can be derived for 
(strong) periodic driving using Floquet theory~\cite{BreuerPetruccioneBook2002, 
KosloffEntropy2013, GelbwaserKlimovskyNiedenzuKurizkiAdv2015} and also give rise to a consistent nonequilibrium 
thermodynamics~\cite{SzczygielskiGelbwaserKlimovskyAlickiPRE2013, KosloffEntropy2013, BulnesCuetaraEngelEspositoNJP2015, 
GelbwaserKlimovskyNiedenzuKurizkiAdv2015}. If the driving is neither slow nor periodic, no method is currently known to 
derive a ME although the thermodynamic quantities defined here remain meaningful, see also 
Ref.~\cite{DeffnerLutzPRL2011}.}
For a system-reservoir coupling of the form $H_{XR} = \sum_k A_k\otimes B_k$ with hermitian system and reservoir 
operators $A_k$ and $B_k$, the superoperator $\C L_\beta(t)$ reads (see, e.g. Sec.~3.3 in 
Ref.~\cite{BreuerPetruccioneBook2002} for a microscopic derivation)
\begin{eqnarray}\label{eq thermal dissipator explicit}
 &&\C L_\beta(t)\rho(t) = \sum_\omega \sum_{k,\ell} \gamma_{k\ell}(\omega) \\
 &&\hspace{1.4cm} \times\left( A_\ell(\omega)\rho(t)A^\dagger_k(\omega) 
- \frac{1}{2}\{A_k^\dagger(\omega)A_\ell(\omega),\rho(t)\}\right). \nonumber
\end{eqnarray}
Here, $\{\omega = \epsilon - \epsilon'\}$ denotes the set of transition frequencies of the Hamiltonian 
$H_S(t) = \sum_\epsilon \epsilon(t) \Pi_{\epsilon(t)}$ with instantaneous eigenenergies $\epsilon(t)$ 
and corresponding projection operators $\Pi_{\epsilon(t)}$. Omitting the explicit time-dependence in 
the notation, $A_k(\omega)$ is defined as 
\begin{equation}
 A_k(\omega) \equiv \sum_{\epsilon-\epsilon'=\omega}\Pi_\epsilon A_k \Pi_{\epsilon'}.
\end{equation}
Furthermore, the rate $\gamma_{k\ell}(\omega)$ is the Fourier transformed reservoir correlation function 
\begin{equation}
 \gamma_{k\ell}(\omega) = \int_{-\infty}^\infty dt e^{i\omega t}\mbox{tr}_R\{B_k(t)B_\ell(0)\rho_\beta^R\},
\end{equation}
which can be shown to be non-negative~\cite{BreuerPetruccioneBook2002}.
The thermal generator~(\ref{eq thermal dissipator explicit}) fulfills two important properties. 
First, it is of so-called Lindblad form~\cite{LindbladCMP1976, GoriniEtAlJMP1976}. 
This means that the time evolution of Eq.~(\ref{eq generic ME for X}),
\begin{equation}\label{eq time evo ME general}
 \rho_X(\tau) = \C T_+\exp\left[\int_0^\tau ds \C L_X(s)\right]\rho_X(0)
\end{equation}
(or simply $\rho_X(\tau) = e^{\C L_X \tau}\rho_X(0)$ if $\C L_X$ is time-independent), can be written as a Kraus 
map~(\ref{eq Kraus map}). Whereas each Lindblad ME defines a particular Kraus map, the inverse is not necessarily true, 
i.e. for a given Kraus map we cannot associate a unique ME in Lindblad form.
Second, and very important from the thermodynamic point of view, the rates satisfy the property 
of \emph{local detailed balance}
\begin{equation}\label{eq local detailed balance 2}
 \gamma_{k\ell}(-\omega) = e^{-\beta\omega}\gamma_{\ell k}(\omega),
\end{equation}
which follows from the Kubo-Martin-Schwinger (KMS) relation of the reservoir correlation 
functions~\cite{EspositoHarbolaMukamelRMP2009, SpohnLebowitzAdvChemPhys1979, KosloffEntropy2013, SchallerBook2014, 
BreuerPetruccioneBook2002}. It also plays a crucial role to establish a consistent nonequilibrium 
thermodynamics for classical stochastic processes~\cite{SeifertRPP2012, VandenBroeckEspositoPhysA2015}. 
More importantly for us here, it implies that the steady state of the system corresponds to the canonical 
equilibrium (or Gibbs) state~\cite{BreuerPetruccioneBook2002}: 
\begin{equation}\label{eq local detailed balance}
 \C L_\beta(t) \rho_\beta^X(t) = 0, ~~~ \rho_\beta^X(t) = \frac{e^{-\beta H_X(t)}}{Z_X(t)}.
\end{equation}

Since the system-reservoir interaction is now negligible, the system energy reads 
$E_X(t) \equiv \mbox{tr}\{H_X(t)\rho_X(t)\}$ and the first law takes the usual form, 
\begin{equation}\label{eq 1st law general ME}
 d_tE_X(t) = \dot W(t) + \dot Q(t),
\end{equation}
where the work rate on the system is given by 
\begin{equation}\label{eq def W general ME}
  \dot W(t)	=	\mbox{tr}_X\left\{\rho_X(t) d_t H_X(t)\right\}
\end{equation}
and the heat flow into the system is caused by the dissipative part of the evolution 
\begin{equation}\label{eq def Q general ME}
 \begin{split}
  \dot Q(t)	&=	\mbox{tr}_X\left\{H_X(t)d_t\rho_X(t)\right\}	\\
		&=	\mbox{tr}_X\{H_X(t)\C L_X(t)\rho_X(t)\}.
 \end{split}
\end{equation}
The second law is now more stringent since it not only ensures the nonnegativity of the entropy 
production but also of its \emph{rate} 
\begin{equation}\label{eq 2nd law general ME}
 \dot \Sigma(t) = d_t S_X(t) - \beta \dot Q(t) \ge 0.
\end{equation} 
Mathematically, this result follows from Spohn's inequality~\cite{SpohnJMP1978} 
\begin{equation}\label{eq Spohns inequality}
 -\mbox{tr}\{[\C L_X(t)\rho_X(t)][\ln\rho_X(t)-\ln\bar\rho_X(t)]\} \ge 0.
\end{equation}
This is true for any superoperator $\C L_X(t)$ of Lindblad form with steady state $\bar\rho_X(t)$, i.e. 
$\C L_X(t)\bar\rho_X(t) = 0$, and corresponds to the differential version of Eq.~(\ref{eq contractivity Kraus map}). 
For the case considered here, we have $\bar\rho_X(t) = \rho_\beta^X(t)$ [see Eq.~(\ref{eq local detailed balance})] 
and Spohn's inequality gives after a straightforward manipulation~(\ref{eq 2nd law general ME}).

\section{Thermodynamics of repeated interactions}
\label{sec generalized thermodynamic framework}

\subsection{Idea}

Traditional thermodynamic setups consist of a system in contact with thermal reservoirs, as reviewed in the previous section. 
However, as motivated in the introduction, many experimental (and theoretical) setups today do not fit into this picture 
and make use of much more structured resources. 
For instance, a micromaser (as reviewed in Sec.~\ref{sec micromaser}) makes use of a stream of flying atoms prepared in 
non-equilibrium states interacting sequentially with a cavity. Another example is the measurement and subsequent 
manipulation of a device by an external observer (``Maxwell's demon''), see Secs.~\ref{sec feedback control} 
and~\ref{sec electronic Maxwell demon}. 

Although both examples are quite different, we will see that we can treat them within a unified framework 
by introducing a new form of ``reservoir'' which can be prepared in arbitrary non-equilbrium states. 
The basic setup consists of a stream of additional systems which we will generically call ``units''.
These units are identically and independently prepared and interact one after another with the device 
(i.e. the system of interest) without ever coming back. This framework of repeated interactions provides sufficient 
structure to formulate a meaningful and tractable dynamics and thermodynamics, as we will now demonstrate. 

\subsection{Formal setup}
\label{sec formal setup}

\begin{figure}
 \centering\includegraphics[width=0.40\textwidth,clip=true]{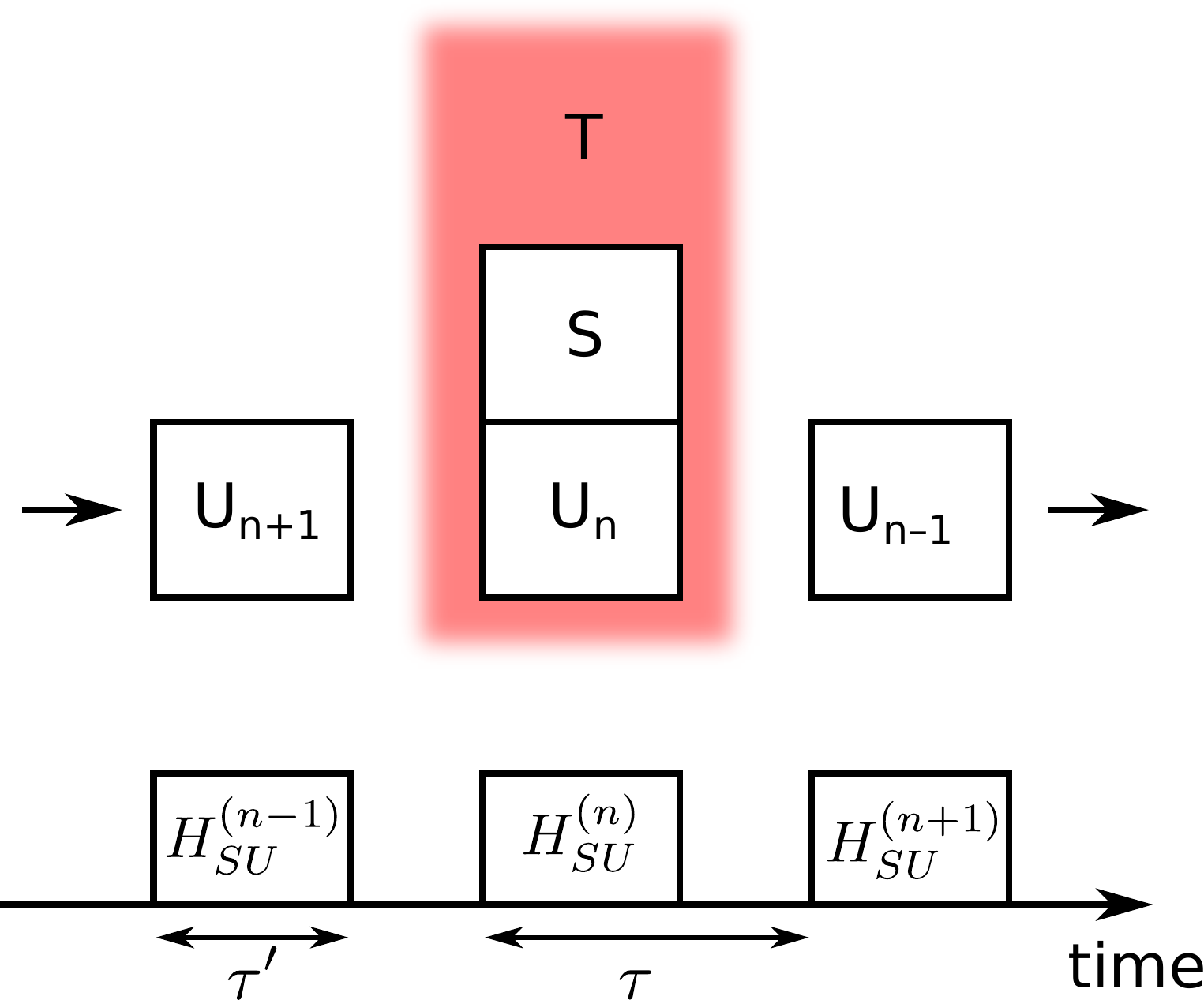}
 \label{fig interaction time series} 
 \caption{(Color Online) Sketch of a stream of units interacting with a system in contact with a heat reservoir at inverse temperature 
 $\beta$. The lower panel shows the switching on and off of the system-unit interaction as a function of time. Note that 
 $\tau$ denotes the full interaction period whereas the system and unit are only physically coupled during a time 
 $\tau'$.}
\end{figure}

The system $X$ that we considered until now is split in two as $X = S\oplus U$, where $S$ denotes 
the system of interest whereas $U$ is an auxiliary system called the unit. 
We assume that this unit interacts with the system $S$ for a time $\tau'$ and is afterwards decoupled 
from it and replaced at time $\tau > \tau'$ by a new unit. This new unit interacts with $S$ from $\tau$ to 
$\tau+\tau'$ before it is replaced by yet another new unit at $2\tau$. The process is then repeated over and over.
This means that the system is interacting sequentially with a stream of units $\{U_n\}$, where the index $n$ 
refers to the $n$'th interaction taking place in the interval $[n\tau,n\tau + \tau)$ as sketched in 
Fig.~\ref{fig interaction time series}. 

The Hamiltonian $H_X(t)$ of the joint system and \emph{all} units can be formally written as
\begin{equation}\label{eq system unit Hamiltonian general}
 H_X(t) = H_S(t) + H_U + H_{SU}(t),
\end{equation}
{where the system Hamiltonian $H_S(t)$ may or may not be time dependent.} 
The Hamiltonian of the stream of (non-interacting) units can be written as a 
sum of (time-independent) single unit Hamiltonians $H_U^{(n)}$
\begin{equation}
 H_U = \sum_n H_U^{(n)}
\end{equation}
and the system-unit interaction as 
\begin{equation}
 H_{SU}(t) = \sum_n \Theta(t-n\tau)\Theta(n\tau + \tau' - t) V_{SU}^{(n)}(t), 
\end{equation}
where $V_{SU}^{(n)}(t)$ is an arbitrary interaction between the system and the $n$'th unit. 
The Heaviside step function explicitly reads
\begin{equation}
 \Theta(t) \equiv \left\{\begin{array}{cc}
                          1	&	\text{if } t\ge0,	\\
                          0	&	\text{if } t<0.
                         \end{array}\right.
\end{equation}
This means that the interaction between the system and the $n$'th unit is switched on 
at time $t = n\tau$ and switched off at $t = n\tau + \tau'$ with $0<\tau'<\tau$ as explained above. 
During this interaction time the system and unit are both coupled to the reservoir $R$. 
Their dynamics can be described exactly or by a ME of the form~(\ref{eq generic ME for X}) in the weak coupling limit.
However, before or after the interaction with the system, when the unit is \emph{not in contact} with the system, 
it will evolve freely (i.e. unitarily) with $H_U^{(n)}$ and its energy 
$E^{(n)}_U(t) = \mbox{tr}_{U_n}\{H^{(n)}_U\rho^{(n)}_U(t)\}$ and entropy 
$S^{(n)}_U(t) = -\mbox{tr}_{U_n}\{\rho^{(n)}_U(t)\ln\rho^{(n)}_U(t)\}$ will remain constant. 

Our setup is meant to model situations where an experimentalist can prepare independent units in any desired state.
Therefore, a crucial but reasonable assumption that we use is that the incoming units are decorrelated 
(i.e. independently prepared) 
and that their statistical description is stationary in time, i.e. the density matrix of the incoming units 
fulfills $\rho_U^{(n)}(n\tau) = \rho_U^{(m)}(m\tau)$ for any 
$n, m$.\footnote{In Sec.~\ref{sec driven systems} we will discuss for a particular application which changes have to be 
made if this assumption is not fulfilled.} 
We further assume that the interaction Hamiltonian $V_{SU}^{(n)}(t)$ has always the same form (of course it
acts on different unit Hilbert spaces but for simplicity we will always denote it by $V_{SU}(t)$). 

Our goal will now be to formulate thermodynamic laws for the system where one regards the stream of units as a 
nonequilibrium reservoir, and to understand to what extend this latter modifies the traditional thermodynamic laws. 
{In the next section, we will focus on one fixed interval where the system interacts with a single 
unit only. For simplicity we choose the interval $[0,\tau)$ and drop the index $n=0$. Sec.~\ref{sec steady state regime} 
then discusses what happens if the system is repeatedly put into contact 
with subsequent units and whether one can expect the system to reach a stroboscopic steady state.} 

\subsection{Modified energy and entropy balance}
\label{sec modified energy and entropy balance}

To obtain the first law of thermodynamics, we can either take Eq.~(\ref{eq 1st law general}) or integrate 
Eq.~(\ref{eq 1st law general ME}), where care has to be taken with the definition of the time interval to 
correctly capture boundary effects. Thus, we define the global change of system and unit energy as 
\begin{equation}
 \Delta E_{SU} \equiv \lim_{\epsilon\searrow0} \int_{-\epsilon}^{\tau-\epsilon} dt \frac{dE_X(t)}{dt} = \Delta E_S + \Delta E_U
\end{equation}
such that the interaction term does not contribute. Here, 
$\Delta E_{S} = \mbox{tr}_{S}\{H_{S}(\tau)\rho_{S}(\tau)\} - \mbox{tr}_{S}\{H_{S}(0)\rho_{S}(0)\}$ 
and analogously for $\Delta E_U$. Integrating the rate of work yields two terms 
\begin{equation}
 W \equiv \lim_{\epsilon\searrow0} \int_{-\epsilon}^{\tau-\epsilon} dt \dot W(t) = W_X + W_\text{sw}.
\end{equation}
The first term is standard and results from the smooth time-dependence of $H_S(t)$ during the full interval and of 
$V_{SU}(t)$ during the interaction, i.e. 
\begin{equation}\label{eq def work X}
 \begin{split}
  W_X	=&~	\int_0^\tau dt \mbox{tr}_X\left\{\rho_X(t)d_tH_S(t)\right\}	\\
	&+	\lim_{\epsilon\searrow0}\int_\epsilon^{\tau'-\epsilon} dt' \mbox{tr}_X\left\{\rho_X(t)d_tV_{SU}(t)\right\}.
 \end{split}
\end{equation}
The second term, to which we will refer as the switching work, 
is a boundary term resulting from the sudden switching on and off of the interaction and reads 
\begin{equation}\label{eq def work switch}
 W_\text{sw} = \mbox{tr}_X\{V_{SU}(0)\rho_X(0) - V_{SU}(\tau')\rho_X(\tau')\}.
\end{equation}
Mathematically, it follows from $d_t\Theta(t) = \delta(t)$ where $\delta(t)$ is a Dirac delta distribution. 
Physically, we can interpret $W_\text{sw}$ as the \emph{work needed to pull the stream of units along the 
system} (we assume that the other units which do not interact with the system move in a frictionless way). 
Finally, by integrating the heat flow [Eqs.~(\ref{eq def Q general evolution}) or~(\ref{eq def Q general ME})] we get
\begin{equation}
 Q \equiv \lim_{\epsilon\searrow0} \int_{-\epsilon}^{\tau-\epsilon} dt \dot Q(t),
\end{equation}
and the first law of thermodynamics takes the form 
\begin{equation}\label{eq 1st law modified}
  \Delta E_S = W + Q - \Delta E_U.
\end{equation}

Using the second law of thermodynamics, Eq.~(\ref{eq 2nd law general}) or Eq.~(\ref{eq 2nd law general ME}) as well
as the factorization of the initial condition $\rho_X(0) = \rho_S(0)\rho_U(0)$, the entropy production of the system 
\emph{and} unit during each interaction period reads 
\begin{equation}\label{eq system dissipation}
 \Sigma  = \Delta S_S + \Delta S_U - I_{S:U}(\tau) - \beta Q \ge 0.
\end{equation}
where the unit entropy change $\Delta S_U$ and the final system-unit correlations $I_{S:U}(\tau)$ modify what 
would otherwise be the traditional second law for the system in contact with its reservoir. 
In view of our interpretation of Eq.~(\ref{eq 2nd law general}), Eq.~(\ref{eq system dissipation})
describes the dissipation of the joint system $SU$ to the reservoir. The units and the system are thus 
treated on the same footing. However, since each unit only interacts once with the system, the 
mutual information that they established by interacting with the system is never used or 
recovered.\footnote{In fact, if the $m$'th unit would be allowed to come back and interact again with the system in the 
$n$'th interval ($n>m$), then a simple description in terms of the system and single unit (as carried out above) would 
not be possible anymore.} 
A more meaningful definition of entropy production for our setup which accounts for these losses is thus
\begin{equation}\label{eq system dissipation 2}
 \Sigma_S \equiv \Delta S_S + \Delta S_U - \beta Q \ge I_{S:U}(\tau) \ge 0.
\end{equation}
This entropy production not only measures the lost information as system-reservoir mutual 
information and as relative entropy between the nonequilibrium and equilibrium reservoir state after the interaction 
(compare with Eq.~(\ref{eq 2nd law relative entropy})), but it also accounts for the information lost as mutual 
information by the units which never come back. 
Obviously, $\Sigma_S \ge \Sigma$. In the special case where the coupling to the reservoir is switched off, $Q=0$ and 
$\Sigma = 0$, the entropy production is solely given by the mutual information $\Sigma_S=I_{S:U}$ lost in our setup.

Using the first law~(\ref{eq 1st law modified}) together with the definition of the nonequilibrium 
free energy~(\ref{eq def non eq free energy}) with respect to the reservoir temperature $\beta$, 
(\ref{eq system dissipation 2}) can be rewritten as
\begin{equation}\label{eq 2nd law modified}
 \Sigma_S = \beta(W - \Delta F_S - \Delta F_U) \ge I_{S:U}(\tau) \ge 0.
\end{equation}
This form of the modified second law allows us to draw the important conclusion that the stream of 
units in its most general sense is nothing else than a resource of nonequilibrium free energy.

\subsection{Steady state regime}
\label{sec steady state regime}

We derived our modified laws of thermodynamics for an arbitrary initial system state over a single 
interaction interval. To treat many interaction intervals, we have to link the (thermo)dynamics between 
successive interaction intervals, i.e. the final system state $\rho_S(n\tau)$ of the $n$'th interaction interval 
has to be taken as the initial condition for the $(n+1)$'th interval. Because the incoming units are statistically 
independent and identically prepared, we can treat each interaction interval as above.

A particularly important case is the limit of multiple interactions where it is reasonable to assume 
that the system will eventually reach a stroboscopic steady state, given that the time-dependence (if any) of the system 
Hamiltonian $H_S(t)$ and of the interaction Hamiltonian $H_{XR}(t)$ with the heat reservoir is also $\tau$-periodic. 
We will often resort to this steady state assumption for the applications considered in this article, which reads
\begin{equation}\label{eq steady state system}
 \rho_S(0) = \rho_S(\tau)
\end{equation}
and implies $\Delta E_S = 0$ and $\Delta S_S = 0$. Then, the laws of thermodynamics simplify to 
\begin{align}
 0		&=	W + Q - \Delta E_U,	\\
 \Sigma_S	&=	\beta(W-\Delta F_U) \ge I_{S:U}(\tau) \ge 0.
\end{align}

To justify this steady state assumption, we assume that the reservoir is always in the same initial state 
at the beginning of every interaction interval $n$ so that the system and unit evolve according to the same Kraus map 
$\Phi_{SU}$ over each interaction interval. Physically this means that the reservoir remains virtually unaffected by 
the interactions with the system and unit. This assumption is for instance implicit if the system-unit dynamics is described 
by a ME. Without this reservoir resetting assumption, justifying existence of the steady state regime is a much harder task. 
Proceeding with this assumption, it is easy to show that there also exists a Kraus map $\Phi_S$ for the system alone
\begin{equation}
 \rho_S(n\tau+\tau) = \Phi_S\rho_S(n\tau) \equiv \mbox{tr}_U\{\Phi_{SU}\rho_S(n\tau)\rho_U(n\tau)\}.
\end{equation}
Importantly, $\Phi_S$ does not depend on the interaction interval $n$ because the initial state of the unit 
$\rho_U(n\tau)$ is always the same and because all relevant Hamiltonians are assumed $\tau$-periodic. 
Therefore, if a unique steady state exits, it must be $\tau$-periodic. The existence of a steady state is guaranteed by 
Eqs.~(\ref{eq Kleins inequality}) and~(\ref{eq contractivity Kraus map}) and its uniqueness can be proven if we have 
a strict inequality\footnote{{The proof is easily carried out by, first, noting that any 
system state must asymptotically (i.e. for $n\rightarrow\infty$) reach some steady state because the sequence 
$f_n \equiv D[\Phi_S^n(\rho_S)||\sigma_S]$ for some steady state $\sigma_S$ is monotonically decreasing and bounded from 
below, and second, by realizing that the existence of two different steady states $\sigma_S$ and $\sigma_S^*$ contradicts 
Eq.~(\ref{eq strict contractivity Kraus map}). }} 
\begin{equation}\label{eq strict contractivity Kraus map}
 D(\Phi\rho||\Phi\sigma) < D(\rho||\sigma) ~~~ (\text{for } \rho\neq\sigma).
\end{equation}
A precise mathematical condition for this strict inequality was worked out in Ref.~\cite{PetzRevMathPhys2003} but is hard 
to translate physically. For instance, if additional symmetries are present or for pure dephasing interactions (commuting 
with the system Hamiltonian), it is well known that there is no unique steady state~\cite{BreuerPetruccioneBook2002}.
Nevertheless, for most relevant scenarios the strict inequality~(\ref{eq strict contractivity Kraus map}) will be satisfied. 
In some examples to be considered below, the existence of a unique steady state is also a well-established experimental 
fact (e.g.  for the micromaser treated in Sec.~\ref{sec micromaser}).

Finally, let us stress that even when the steady state regime is guaranteed, solving the combined system-unit 
dynamics exactly is often out of reach, especially when the system and unit are complicated systems by themselves. 
This is why in Sec.~\ref{sec effective master equation} we go one step further and describe various limiting 
regimes (corresponding to special types of interaction $H_{SU}(t)$) where an effective ME can be derived 
for the system alone, with its corresponding thermodynamic interpretation.

\subsection{Discussion}
\label{sec discussion}

Let us summarize what we have achieved. 
By allowing the system to interact with units, we showed that a new term arises in the system 
energy and entropy balance, (\ref{eq 1st law modified}) and (\ref{eq system dissipation 2}) respectively. 
It describes the unit energy and entropy changes, $\Delta E_U$ and $\Delta S_U$ respectively, 
in addition to the traditional terms describing the energy and entropy changes in the reservoir in terms of heat.  
Consequently, the entropy production~$\Sigma_S$, which measures the irreversible losses in the system 
dynamics, now displays a new term which is given by the free energy change of the unit $\Delta F_U$. 
This term enables new transformations that would have been impossible without the units. 
From an operational perspective, evaluating this free energy requires preparing the units in 
a known state before the interaction and measuring their state after the interaction has ended. 
We will examine in the next section whether $\Delta E_U$ and $\Delta S_U$ can be linked 
to traditional thermodynamic notions of work or heat.

Since our generalized second law $\Sigma_S \ge 0$ provides a bound on the possibility to extract work or to convert 
different states into each other, it is worth mentioning that a number of different bounds have been established recently 
within the framework of \emph{resource theories}~\cite{HorodeckiOppenheimNatComm2013, BrandaoEtAlPNAS2014, 
LostaglioJenningsRudolphNatComm2015, BruschiEtAlPRE2015, WoodsNgWehnerArXiv2015, WilmingGallegoEisertPRE2016, 
KorzekwaEtAlNJP2016}. These studies also explicitly show ways to saturate these bounds. While the setups 
they consider share some similarities with ours, the bounds obtained are in general different from our 
second law and are derived under \emph{additional} restrictions imposed on the setup. 
For instance, within the ``resource theory of thermal operations'' (see Ref.~\cite{GooldEtAlJPA2016} for an 
overview about different resource theories), it is assumed that the global time evolution commutes respectively 
with the bare Hamiltonian of the system, of the unit, and of the reservoir. 
This assumption is not needed within the approach presented in this section. 
Given that our sole restriction is to consider initially decorrelated system-unit states, the problem of 
finding specific protocols which saturate the bound in Eq.~(\ref{eq 2nd law modified}) is in principle 
equivalent to finding protocols saturating the bound~(\ref{eq 2nd law general 2}) because the former 
is a consequence of the latter (see, e.g.  Refs.~\cite{JacobsPRA2009, EspositoVandenBroeckEPL2011, 
HorowitzParrondoNJP2011, ReebWolfNJP2014, SkrzypczykShortPopescuNatComm2014} for such optimal protocols). 
These optimal protocols might however correspond to highly idealized, if not unrealistic, situations. 
Instead of following a resource theory strategy by imposing restrictions from the start, 
we kept a general level of discussion in this section and will consider specific physical 
setups of greater experimental relevance in Sec.~\ref{sec applications part II}.

\section{Implications}
\label{sec applications part I}

\subsection{Thermal units and ideal heat reservoir}
\label{sec heat reservoir}

We consider thermal units initially prepared in an equilibrium state at an inverse 
temperature $\beta'$, i.e $\rho_U(0) = \rho_{\beta'}^U = e^{-\beta' H_U}/Z_U$. 

We say that these units behave as an ideal heat reservoir when 
\begin{equation}\label{eq ideal heat reservoir}
  \frac{\Delta S_U}{\Delta E_U} = \frac{1}{T'} ~~~ \text{(ideal heat reservoir)}.
\end{equation}
More insight is obtained using an argument similar to the one used in section \ref{sec a general approach} 
to define an ideal heat reservoir. Using the identity
\begin{equation}\label{eq help 15}
T' D[\rho_U(\tau)||\rho_{\beta'}^U] = \Delta E_U - T'\Delta S_U  \geq 0\,,
\end{equation}
we see that Eq.~(\ref{eq ideal heat reservoir}) is fulfilled when the state of the unit remains close 
to thermal equilibrium after the interaction, i.e. when $\rho_U(\tau) = \rho_{\beta'}^U + \epsilon\sigma_U$ 
where $\epsilon$ is a small parameter and $\mbox{tr}_U(\sigma_U) = 0$. 
Indeed, we then get that $D[\rho_U(\tau)||\rho_{\beta'}^U] = \C O(\epsilon^2)$ whereas $\Delta E_U$ 
and $\Delta S_U$ are in general non-zero and equal to first order in $\epsilon$.
Since the unit energy change can be interpreted as heat, $\Delta E_U = -Q_U$, the second 
law~(\ref{eq system dissipation 2}) becomes $\Sigma_S \equiv \Delta S_S -\beta'Q_U - \beta Q \ge 0$. 
We remark that saturating Eq.~(\ref{eq ideal heat reservoir}) away from the weak-coupling limit is in 
general non-trivial, but see Sec.~\ref{sec alternative setup ideal heat} for another class of ideal heat reservoirs.

As a simple application, we operate our setup in the steady state regime with hot thermal units $T'>T$. 
Using (\ref{eq help 15}) in (\ref{eq 2nd law modified}), the entropy production bound implies 
\begin{equation}\label{boundTapeEngine}
W - (1-T/T') \Delta E_U \geq T D[\rho_U(\tau)||\rho_{\beta'}^U] + T I_{S:U}(\tau).
\end{equation}
This shows that to operate as a heat engine, where work is extracted ($W < 0$), 
energy must be extracted from the units ($\Delta E_U < 0$). For thermal units constituting 
an ideal heat reservoirs, the thermodynamic efficiency of the engine is defined as $\eta = -W/\Delta E_U$ 
and is upper bounded by the Carnot efficiency $\eta \leq 1-T/T'$ due to (\ref{boundTapeEngine}), 
which reduces to $W - (1-T/T') \Delta E_U \geq 0$ in that case. 
However, non-ideal thermal units decreases the efficiency bound as
\begin{equation}
 \eta \le 1 - \frac{T}{T'} - T\frac{D[\rho_U(\tau)||\rho_{\beta'}^U] + I_{S:U}(\tau)}{-\Delta E_U} \le 1 - \frac{T}{T'}.
\end{equation}

While realizing that a thermal stream of unit can behave as an ideal reservoir is interesting,
the importance of our setup is that it allows to treat units initially prepared out-of-equilibrium.
One way to do it here is to consider an initial state $\rho_{\beta'}^U$ with negative $\beta'$. 
The efficiency of the heat engine can then formally \emph{exceed} $1$ \emph{without violating the second 
law of thermodynamics} because the entropy production~(\ref{eq system dissipation}) is still non-negative. 
However, in this case -- as the work output has to be compared with the total energy put into the system -- 
the first law of thermodynamics tells us that a correctly defined efficiency would still be bound by one.

\subsection{Work reservoir}
\label{sec work reservoir}

To make the units act as a work source, we should engineer their state and interaction with the system 
in such a way that only energy is exchanged but not entropy. Thus, we define 
\begin{equation}\label{eq ideal work reservoir}
 \frac{\Delta S_U}{\Delta E_U} \rightarrow 0 ~~~ \text{(ideal work reservoir).}
\end{equation}
For finite $\Delta E_U$ this implies that the change in nonequilibrium free energy is given by 
$\Delta F_U = \Delta E_U$ and~(\ref{eq 2nd law modified}) becomes at steady state ($\Delta F_S=0$)
\begin{equation}\label{eq help 1}
 \Sigma_S = \beta(W - \Delta E_U) \ge I_{S:U}(\tau) \ge 0.
\end{equation}
This means that we can extract work from the energy initially stored in the units, but not by 
extracting heat from the reservoir since Eq.~(\ref{eq help 1}) implies due to the first law that $Q < 0$. 
Definition~(\ref{eq ideal work reservoir}) can be fulfilled for many different situations depending on the precise 
nature of the interaction and the state and Hamiltonian of the units (we will treat particular examples in 
Secs.~\ref{sec Poisson heat work info dominated},~\ref{sec driven systems} and~\ref{sec micromaser}). 

Let us note that for the special case of an ideal heat reservoir which is thermal before and after the interaction 
(as discussed above), the form of Eqs.~(\ref{eq ideal work reservoir}) and ~(\ref{eq help 1}) can be obtained by 
choosing $T'\rightarrow\infty$ in Eqs.~(\ref{eq ideal heat reservoir}) or~(\ref{eq help 15}) while keeping 
$\Delta F_U$ finite. This confirms the 
colloquial saying that a work reservoir corresponds to an infinite temperature heat 
reservoir.\footnote{To quote Sommerfeld (p. 36 in~\cite{SommerfeldBook1923}): 
``\emph{Thermodynamics investigates the conditions that govern the transformation of heat into work. It teaches us to 
recognize temperature as the measure of the work-value of heat. Heat of higher temperature is richer, is capable of doing 
more work. Work may be regarded as heat of an infinitely high temperature, as unconditionally available heat.}'' } 

Finally, we note that the notion of work for small systems is subtle and has been 
debated~\cite{AllahverdyanNieuwenhuizenPRE2005,
TalknerLutzHaenggiPRE2007, WeimerEtAlEPL2008, SchroederMahlerPRE2010, CampisiHaenggiTalknerRMP2011, SalmilehtoSolinasMoettoenenPRE2014, 
HosseinNejadOReillyOlayaCastroNJP2015, GallegoEisertWilmingNJP2016, JarzynskiQuanRahavPRX2015}. 
This originates from the desire to explain work microscopically in contrast to the standard approach 
where it is usually incorporated ``by hand'' as a time-dependent part of the system Hamiltonian. 
We see that the repeated interaction framework brings an interesting perspective on this issue 
by defining work as the part of the energy exchange that does not induce any entropy change in 
the units. This approach agrees with the point of view advertised in Refs.~\cite{WeimerEtAlEPL2008, 
SchroederMahlerPRE2010, HosseinNejadOReillyOlayaCastroNJP2015}.
In Sec.~\ref{sec driven systems}, we will provide an explicit model where the units 
effectively mimic a time-dependent system Hamiltonian and fulfill Eq.~(\ref{eq ideal work reservoir}).

\subsection{Information reservoir}
\label{sec info reservoir}

We now consider the regime where the units operate as a pure information source by demanding that the exchange of 
energy $\Delta E_U$ vanishes whereas the exchange of entropy $\Delta S_U$ remains finite. Thus, in contrast 
to Eq.~(\ref{eq ideal work reservoir}), we demand that 
\begin{equation}\label{eq ideal info reservoir}
 \left(\frac{\Delta S_U}{\Delta E_U}\right)^{-1} \rightarrow 0 ~~~ \text{(ideal information reservoir).}
\end{equation}
Then, $\Delta F_U = -T\Delta S_U$ and the second law of thermodynamics~(\ref{eq 2nd law modified}) becomes at steady state ($\Delta F_S=0$)
\begin{equation}\label{eq 2nd law info reservoir}
 \Sigma_S = \beta W + \Delta S_U \ge I_{S:U}(\tau) \ge 0.
\end{equation}
This shows that it is possible to extract work ($W < 0$) while writing information to the units 
($\Delta S_U > 0$) in an energy neutral fashion ($\Delta E_U = 0$). Note that for this interpretation we have 
tacitly equated the entropy of a system with its \emph{information content} in spirit of Shannon's fundamental 
work~\cite{ShannonBSTJ1948}. Engines which are able to extract work only at 
the expense of information are also called \emph{information-driven engines}~\cite{FeynmanBook1985}. 

The idea that an information reservoir represents the opposite of a work reservoir also becomes manifest by 
considering the case of an ideal heat reservoir in the limit $T'\rightarrow0$. Rearranging Eq.~(\ref{eq help 15}) yields 
\begin{equation}
 \frac{\Delta E_U}{T'} = \Delta S_U + D[\rho_U(\tau)||\rho_{\beta'}^U].
\end{equation}
In the limit of an ideal reservoir the second term on the right-hand side becomes negligibly small. However, in order 
to keep $\Delta S_U$ finite while $T'\rightarrow0$ we automatically see that we obtain the requirement 
$\Delta E_U \rightarrow 0$. This can be achieved, for instance, by scaling the Hamiltonian of the units as 
$H_U = T'\tilde H_U$ (note that $\tilde H_U$ is dimensionless now). The same conclusion was also reached in 
Ref.~\cite{HoppenauEngelEPL2014}. 

The notion of an ``information reservoir'' was introduced in a classical context by Deffner and Jarzynski 
in Ref.~\cite{DeffnerJarzynskiPRX2013}, where each single informational state corresponds to a set of 
microscopic states which are assumed to internally rapidly equilibrate. If the free energy barriers 
between the different informational states are large, this enables a stable encoding of the information. 
Here instead, we equate each microstate of the unit with an informational state. 
In this respect we do not impose any stability condition on our information, but also take 
all changes at the microscopic level into account. A correspondence between the two approaches 
can be established using a coarse graining procedure similar to Ref.~\cite{EspositoPRE2012}. 
Furthermore, the thermodynamics of information reservoirs has attracted a lot of attention recently 
as number of model studies shows~\cite{ScullyPRL2001, MandalJarzynskiPNAS2012, BaratoSeifertEPL2013, 
MandalQuanJarzynskiPRL2013, DeffnerPRE2013, HoppenauEngelEPL2014, LuMandalJarzynskiPT2014, 
BaratoSeifertPRE2014, StrasbergEtAlPRE2014, CaoGongQuanPRE2015}. 
In Sec.~\ref{sec appendix MJ engine} we will propose a microscopic model for the Mandal-Jarzynski 
engine~\cite{MandalJarzynskiPNAS2012} where the extracted work is shown to correspond to 
$W_\text{sw}$~(\ref{eq def work switch}). 

\begin{figure}
 \centering\includegraphics[width=0.40\textwidth,clip=true]{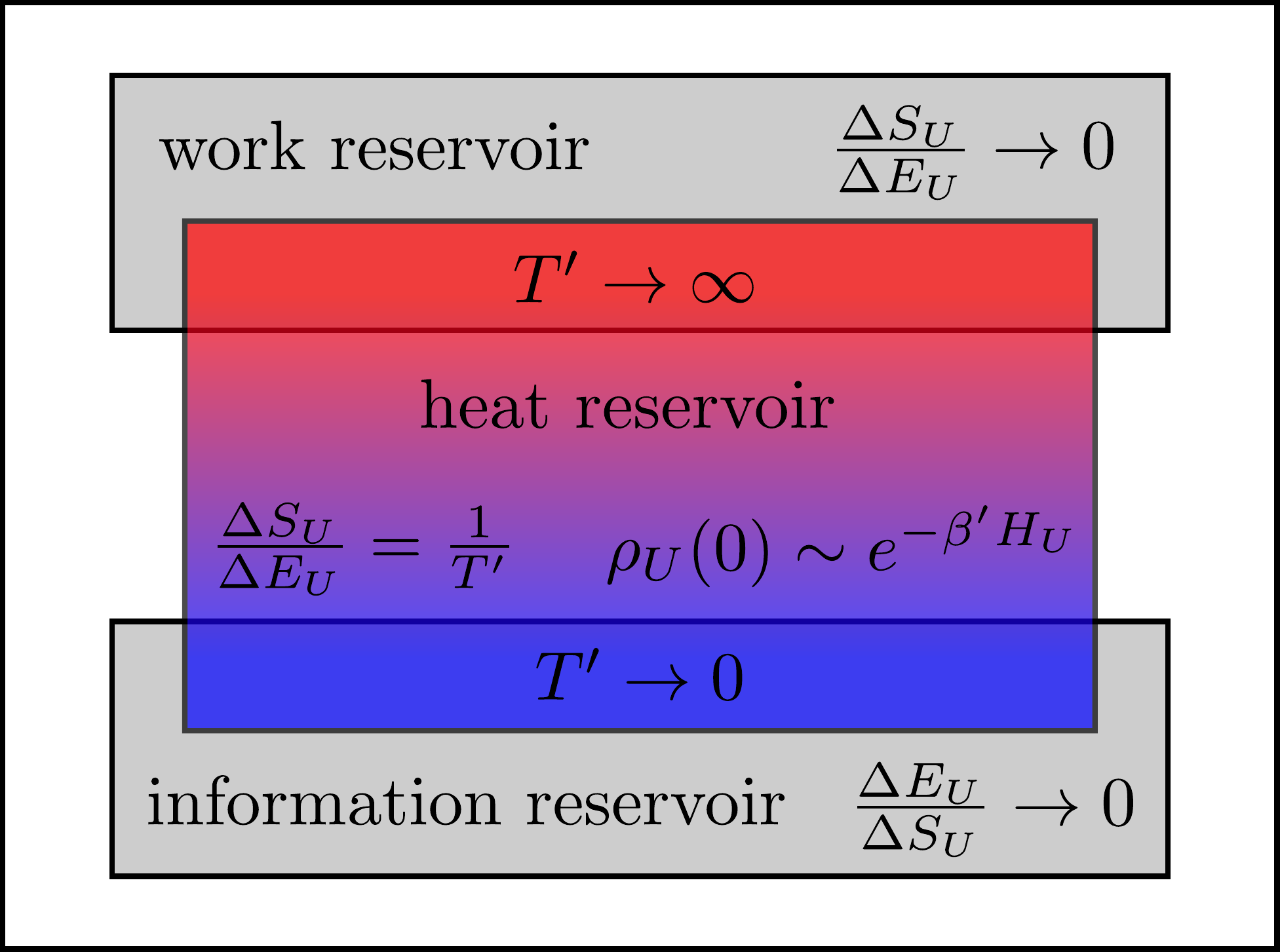}
 \label{fig reservoir diagram} 
 \caption{(Color Online) Venn diagram of the thermodynamic role of the stream of units. In general, the interaction 
 can be arbitrary, but if the initial state of the units is thermal, they can mimic an ideal heat reservoir when 
 they fulfill the Clausius equality~(\ref{eq ideal heat reservoir}). In the limiting case where $T'\rightarrow\infty$ we 
 obtain a work reservoir (Sec.~\ref{sec work reservoir}). The converse is not true, i.e. not every work reservoir 
 can be obtained as a limiting case of a heat reservoir. Similarly, we can obtain an information 
 reservoir (Sec.~\ref{sec info reservoir}) out of a heat reservoir for $T'\rightarrow0$, but again the converse is 
 not true. We note that some setups do not fit in either of the three categories. }
\end{figure}

An overview of the the last three sections is represented in Fig.~\ref{fig reservoir diagram}. 

\subsection{Landauer's principle}
\label{sec Landauers principle}

Landauer's principle colloquially states that logically irreversible computation, 
more specifically the erasure of a bit, has a fundamental work cost. 
This result was first derived for a particular model by Landauer in 1961~\cite{LandauerIBM1961}. 
Since then many groups made this statement more precise by deriving it from a more general 
context, partly also within a repeated interaction framework, and by extending it to finite-time 
operations~\cite{KeyesLandauerIBM1970, LikharevIJTP1982, BennettIJTP1982, ShizumePRE1995, PiechocinskaPRA2000, 
PlenioVitelliCP2001, SagawaUedaPRL2009, EspositoVandenBroeckEPL2011, AurellEtAlJSP2012, DianaBagciEspositoPRE2013, 
ReebWolfNJP2014, SagawaJSM2014, GooldPaternostroModiPRL2015, LorenzoEtAlPRL2015, HansonEtAlCMP2016, 
PezzuttoPaternostroOmarNJP2016}. 
The modern understanding is that this principle immediately follows from 
the nonequilibrium version of the second law as we will show below.
There is also growing experimental evidence in favor of it~\cite{OrlovEtAlJJAP2012, BerutEtAlNature2012, 
JunGavrilovBechhoeferPRL2014, BerutPetrosyanCilibertoJSM2015, SilvaEtAlPRSA2016, HongEtAlSciAdv2016, 
GavrilovBechhoeferPRL2016}. It nevertheless remains debated~\cite{NortonEntropy2013}. 

Within our framework, Landauer's principle can be formulated as follows: 
\emph{Changing the information content (that is to say the Shannon or von Neumann entropy) 
of a unit by $-\Delta S_U > 0$ requires a work expenditure of at least:} 
\begin{equation}\label{eq Landauers principle}
\beta(W - \Delta E_U) \ge -\Delta S_U > 0.
\end{equation}
This statement immediately follows from our generalized second law~(\ref{eq 2nd law modified}), 
where $\Delta E_S=0$ because we focus on the steady state regime. Note that, since 
$-\Delta S_U = S_U(0) - S_U(\tau) > 0$, we are indeed \emph{erasing} information, 
i.e. we lower the Shannon or von Neumann entropy of the unit. 
Furthermore, we see that we recover the standard statement $\beta W \ge -\Delta S_U$ for $\Delta E_U = 0$, 
which is automatically fulfilled if the states of the units are energetically degenerate as it is usually 
assumed (treatments including energetic changes can be found in Refs.~\cite{SagawaUedaPRL2009, 
EspositoVandenBroeckEPL2011} and are in agreement with our result and were also confirmed 
experimentally in Ref.~\cite{GavrilovBechhoeferPRL2016}). 

We emphasize that the initial product state of the system and unit, $\rho_X(0) = \rho_S(0)\rho_U(0)$, 
is essential for deriving Landauer's bound. In fact, we regard the unit (functioning as a memory in this case) as  
an auxiliary system to which the experimenter has free access. If the memory was initially correlated with 
the system, it should be treated as part of the system instead~\cite{WisemanMilburnBook2010}. In presence of initial 
correlations, it is well-known that Landauer's bound does \emph{not} 
hold~\cite{DelRioEtAlNature2011, ReebWolfNJP2014}.\footnote{This statement is also true in the classical context and 
does not require any quantum entanglement.} 

We end with some remarks. As pointed out in Ref.~\cite{SagawaJSM2014}, erasing information is \emph{not} necessarily 
a thermodynamically irreversible process because when reaching the equality in Eq.~(\ref{eq Landauers principle}),
the process becomes thermodynamically \emph{reversible} (i.e. with no entropy production). 
The inverse operation of erasure corresponds to a \emph{randomization} of the 
memory back to its initial state while absorbing heat from the reservoir. 
This can be viewed as creating information in the sense of Shannon. 
However, it is not a computational process, i.e. a deterministic operation on the set of logical 
states which cannot increase the Shannon entropy of the state during 
computation.\footnote{If randomization was a computational process, one could build 
computers that are perfect random number generators, which is not the case.} 
It is only in this sense that a logically irreversible computer can be said to produce irretrievable losses of energy. 
In fact, the information-driven engines introduced in Sec.~\ref{sec info reservoir} can be seen as an implementation 
of the reverse process. The duality between work extraction and information erasure was also noticed in 
Refs.~\cite{MandalJarzynskiPNAS2012, BaratoSeifertEPL2013, MandalQuanJarzynskiPRL2013, DeffnerJarzynskiPRX2013, 
HoppenauEngelEPL2014, LuMandalJarzynskiPT2014, BaratoSeifertPRE2014, StrasbergEtAlPRE2014, CaoGongQuanPRE2015}.

\subsection{The second law of thermodynamics for discrete feedback control}
\label{sec feedback control}

Feedback control describes setups where one manipulates the dynamics of a system 
based on the information that one obtains by measuring it. Several groups 
have established that for a system undergoing feedback control in contact with a thermal 
reservoir at inverse temperature $\beta$, the amount of extractable work $W^\text{fb}$ is 
bounded by (details about the assumptions are stated below)~\cite{LloydPRA1989, 
TouchetteLlyodPRL2000, TouchetteLloydPhysA2004, SagawaUedaPRL2008, CaoFeitoPRE2009, 
JacobsPRA2009, SagawaUedaPRL2010, DeffnerLutzArXiv2012, PonmuruganPRE2010, 
HorowitzVaikuntanathanPRE2010, AbreuSeifertPRL2012, SagawaUedaPRL2012, SagawaUedaPRE2012}
\begin{equation}\label{eq 2nd law feedback control}
 -\beta W^\text{fb} \le I_{S:U}^\text{ms},
\end{equation}
where $I_{S:U}^\text{ms}$ is the classical mutual information (which can be obtained from 
Eq.~(\ref{eq def mutual information}) by replacing the von Neumann by the Shannon entropy) 
between the system and the memory in which the measurement result is stored \emph{after} the measurement. 
Eq.~(\ref{eq 2nd law feedback control}) is also called the second law of thermodynamics for discrete feedback control. 
It was confirmed experimentally in Refs.~\cite{ToyabeEtAlNatPhys2010, KoskiEtAlPNAS2014, KoskiEtAlPRL2014}. 

To be more specific, the inequality~(\ref{eq 2nd law feedback control}) holds under special conditions. 
For instance, the bound is known to be different for quantum systems~\cite{JacobsPRA2009}, and 
even classically, additional requirements are imposed on the measurement which are seldomly stated 
explicitly. Within our framework, we will show that we are able to provide a very transparent 
and clean proof of Eq.~(\ref{eq 2nd law feedback control}). 

The memory used to store the measurement of the system will be a unit in our setup. 
The assumption of an initially decorrelated system-memory state complies with 
the notion of a memory used in Sec.~\ref{sec Landauers principle}. 
We assume that the Hamiltonian $H_U$ of the memory is completely degenerate 
so that the change in energy of the memory is always zero, $\Delta E_U = 0$. 
The stream of memories can thus be viewed as the information reservoir introduced in Sec.~\ref{sec info reservoir}. 
Including changes in the energy of the memory poses no fundamental challenge, but would just lengthen the equations below. 

\begin{figure}
 \centering\includegraphics[width=0.48\textwidth,clip=true]{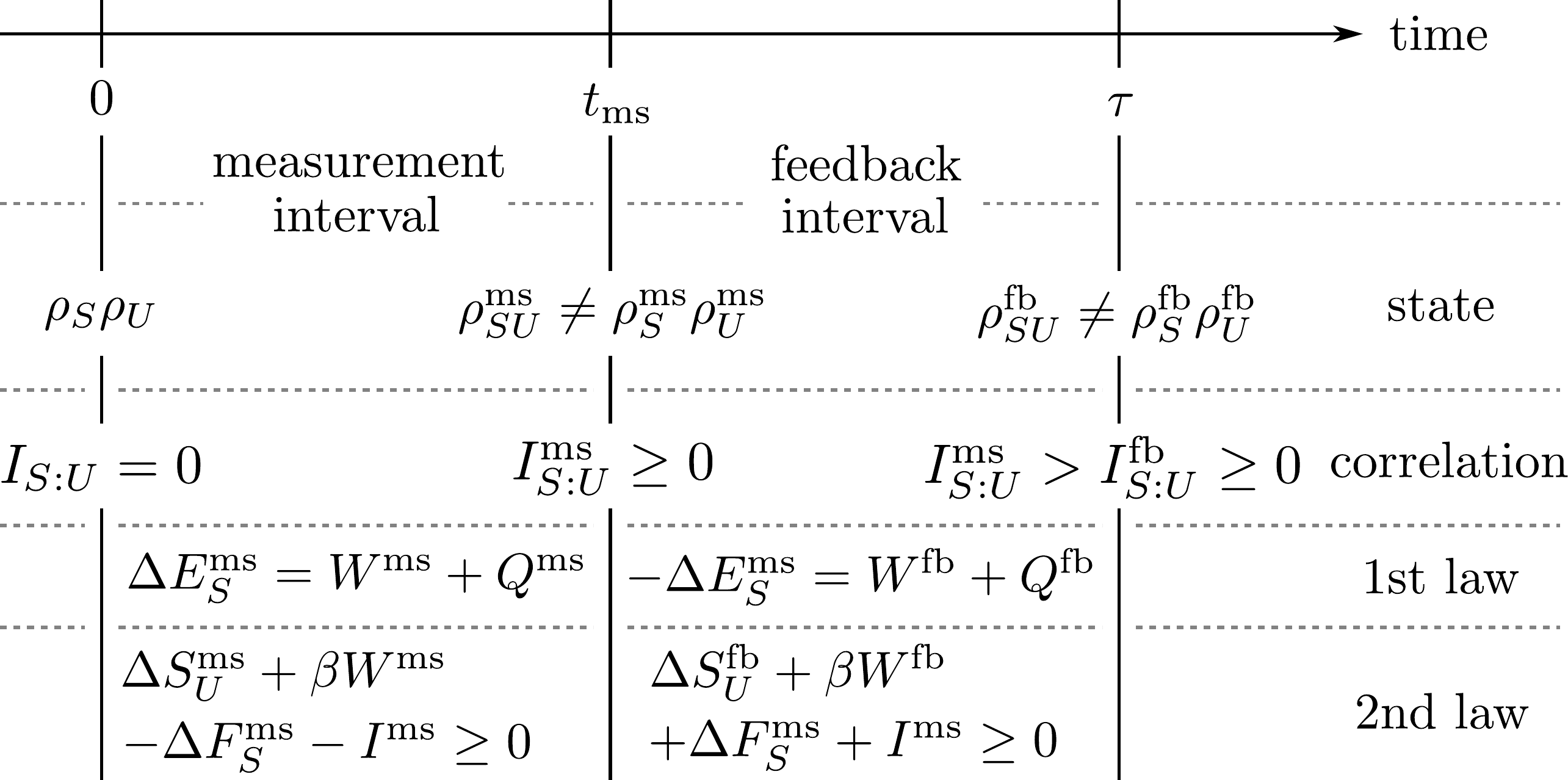}
 \label{fig meas and fb} 
 \caption{Overview of the various quantities involved during the 
          measurement and feedback step in the steady state regime.}
\end{figure}

We now divide the interaction interval in two parts $[0,\tau) = [0,t_\text{ms}) \cup [t_\text{ms},\tau)$ 
with $t_\text{ms}\in(0,\tau)$, as illustrated on Fig.~\ref{fig meas and fb}. 
The measurement is performed during $[0,t_\text{ms})$ whereas the feedback step is performed during $[t_\text{ms},\tau)$. 

One possibility is to treat an instantaneous measurement, $t_\text{ms}\rightarrow0$. 
In this case, the measurement consists of a delta-function time dependence of the interaction Hamiltonian, 
$H_{SU}(t) = \delta(t) V_{SU}^\text{ms}$, which generates a sudden unitary operation $U_\text{ms}$ acting 
on the joint system-memory space. The state of the system and memory after such a measurement reads 
\begin{equation}
 \rho^\text{ms}_{SU} = U_\text{ms}\rho_S(0)\rho_U U_\text{ms}^\dagger ,
\end{equation}
where $U_\text{ms} = \exp(-iV_{SU}^\text{ms})$ and $\rho^\text{ms}_{SU}$ will be in general correlated. 
During this short time window the system and memory are effectively decoupled from the reservoir 
and the measurement acts in an entropy-preserving fashion. 
As a result during $[0,t_\text{ms})$, the first and second law respectively read 
\begin{align}
 \Delta E_S^\text{ms} & = W^\text{ms}, \label{eq 1st law unitary meas} \\
 \Sigma^\text{ms}     &= \Delta S_{SU}^\text{ms} = \Delta S_S^\text{ms}
                       + \Delta S_U^\text{ms} - I_{S:U}^\text{ms} = 0. \label{eq 2nd law unitary meas}
\end{align}
The non-negative mutual information which has been created during the measurement step, 
$I_{S:U}^\text{ms}$, will constitute the resource during the feedback step. 

On the other hand, 
if the measurement time $t_\text{ms}$ remains finite, we are effectively implementing an ``environmentally-assisted 
measurement''. If $\C L_\text{ms}(t)$ denotes the superoperator governing the time evolution of the system and memory 
in weak contact with the reservoir during $[0,t_\text{ms})$, then their initial state will be mapped into the 
generically correlated state [see Eq.~(\ref{eq time evo ME general})]
\begin{equation}
 \rho^\text{ms}_{SU} = \left[\C T_+\exp\int_0^{t_\text{ms}} ds \C L_\text{ms}(s)\right] \rho_S(0)\rho_U .
\end{equation}
In this case heat exchanges with the reservoir $Q^\text{ms}$ will occur and the first and second law respectively read 
\begin{align}
 \Delta E_S^\text{ms}	&= W^\text{ms} + Q^\text{ms}, \label{eq 1st law ms interval}	\\
 \Sigma^\text{ms}	&= \Delta S_{SU}^\text{ms} - \beta Q^\text{ms} \nonumber \\	
			&= \Delta S_S^\text{ms} + \Delta S_U^\text{ms} - I_{S:U}^\text{ms} 
                          - \beta Q^\text{ms} \ge 0. \label{eq 2nd law ms interval}
\end{align}
By combining these two laws, we find that the measurement work is bounded by 
\begin{equation}\label{eq 2nd law ms interval final}
 -\beta W^\text{ms} \le - \beta \Delta F_S^\text{ms} + \Delta S_U^\text{ms} - I_{S:U}^\text{ms}.
\end{equation}
A reversible implementation of the measurement, $\Sigma^\text{ms} = 0$, is possible 
in the limit $t_\text{ms} \rightarrow \infty$ as examined explicitly by Bennett and 
others~\cite{KeyesLandauerIBM1970, BennettIJTP1982, LikharevIJTP1982, FeynmanBook1985}. 

We remark that from the system's perspective, the measurement simply changes its state from $\rho_S(0)$ to 
$\rho_S^\text{ms} = \mbox{tr}_U(\rho_{SU}^\text{ms}) \equiv \Phi_S^\text{ms}\rho_S(0)$, where 
$\Phi_S^\text{ms}$ denotes the Kraus map of the measurement. So far our approach is very general since it 
includes any kind of measurement scenario (including measurements on classical systems) compatible with 
an initially decorrelated system-memory state~\cite{NielsenChuangBook2000, WisemanMilburnBook2010}. 

We now turn to the feedback step. 
In a macroscopic setting, the observer would make a projective measurement 
of the memory in the \emph{computational basis} $|o \rangle_U$. 
After reading out the outcome $o$ of the measurement, she would subsequently perform a 
feedback step by accordingly changing the system Hamiltonian and/or the system-reservoir coupling. 
In the inclusive approach which we now follow, invoking a macroscopic agent is not necessary 
since the same resulting dynamics can be obtained by using a total Hamiltonian of the 
form\footnote{
The unitary evolution operator associated to $H_\text{fb}(t)$ is then of the form 
$U_\text{fb} = \sum_o \Pi_o U^{(o)}_{SR}$ where $U^{(o)}_{SR}$ acts exclusively in the system-reservoir space
and the reduced system-reservoir state is obtained by tracing over the unit Hilbert space.}
\begin{equation}\label{eq Hamiltonian fb}
 H_\text{fb}(t) = \sum_o \Pi_o \otimes \left[H_S^{(o)}(t) + H_{SR}^{(o)}\right] +  H_R.
\end{equation}
Here, $\Pi_o = |o \rangle_U \langle o|$ denotes the projector onto the unit subspace corresponding to outcome $o$. 
The unit plays the role of a minimal description of the external agent or feedback controller. 
We note that the idea to describe a measured and feedback controlled system in a larger space without 
having to rely on \emph{explicit} measurements is not new and is at the heart of coherent or autonomous feedback in 
quantum mechanics~\cite{WisemanPhD1994, WisemanMilburnPRA1994b, LloydPRA2000}. 
It also works classically~\cite{BechhoeferRMP2005}.
More details on such descriptions and on strategies to optimize 
work extraction can be found in Ref.~\cite{JacobsPRA2009}. 
We now proceed with the first and second law of thermodynamics during the feedback 
step which take the form {(independently of the measurement scheme)}\footnote{
We note that Eq.~(\ref{eq 2nd law fb interval}) does not rely on the assumption that the system-unit 
state is decorrelated at the beginning of the feedback interval (see Eqs.~(\ref{eq 2nd law general}) 
or~(\ref{eq 2nd law general ME})). This assumption is only used initially (before the measurement) and 
enables us to consider a regime of repeated interactions. } 
\begin{align}
 \Delta E_S^\text{fb}	&=	W^\text{fb} + Q^\text{fb},	\label{eq 1st law fb interval}	\\
 \Sigma_S^\text{fb}	&=	\Delta S_{S}^\text{fb} + \Delta S_U^\text{fb} - \beta Q^\text{fb} 
                         + I_{S:U}^\text{ms} \geq I_{S:U}^\text{fb} \ge 0 , \label{eq 2nd law fb interval}
\end{align}
where $I_{S:U}^\text{ms}$ is the system-memory mutual information at the end of the measurement interval 
while $I_{S:U}^\text{fb} \equiv I_{S:U}(\tau)$ is the remaining mutual information at the end of the 
entire interval. We note that this latter is a left out resource which will always diminish 
the amount of extractable work.
By combining Eq.~(\ref{eq 2nd law fb interval}) with Eq.~(\ref{eq 1st law fb interval}), we get 
\begin{equation}\label{eq help 11}
 -\beta W^\text{fb} \le -\beta \Delta F_S^\text{fb} + \Delta S_U^\text{fb} + I_{S:U}^\text{ms}.
\end{equation}
Assuming now that we operate in the steady state regime where 
$\Delta F_S = \Delta F_S^\text{ms} + \Delta F_S^\text{fb} = 0$, Eq.~(\ref{eq help 11}) becomes
\begin{equation}\label{eq 2nd law fb interval final}
 -\beta W^\text{fb} \le \beta\Delta F_S^\text{ms} + \Delta S_U^\text{fb} + I_{S:U}^\text{ms}.
\end{equation}
This result can be regarded as the \emph{generalized second law for discrete feedback control}. 

To recover (\ref{eq 2nd law feedback control}) from (\ref{eq 2nd law fb interval final}), 
one needs to consider a non-disturbing classical measurement in which the state of the system 
before and after the measurement is the same~\cite{WisemanMilburnBook2010} and the information 
stored in the memory is classical. 
These assumptions have been used implicitly or explicitly in the classical context~\cite{LloydPRA1989, 
TouchetteLlyodPRL2000, TouchetteLloydPhysA2004, CaoFeitoPRE2009, SagawaUedaPRL2010, DeffnerLutzArXiv2012, 
PonmuruganPRE2010, HorowitzVaikuntanathanPRE2010, AbreuSeifertPRL2012, SagawaUedaPRL2012, SagawaUedaPRE2012} 
whereas for quantum treatments only the information stored in the memory 
was treated classically~\cite{SagawaUedaPRL2008, JacobsPRA2009}.
Indeed, the first property of a non-disturbing measurement implies that $\Delta F_S^\text{ms} = 0$ while 
the assumption of a classical memory implies that, after the measurement, the memory is diagonal in its 
computational basis $|o \rangle_U$. Therefore, the evolution caused by the Hamiltonian~(\ref{eq Hamiltonian fb}) 
will leave the entropy of the memory constant, i.e. $\Delta S_U^\text{fb} = 0$. 
Hence, we recover (\ref{eq 2nd law feedback control}), but having clearly identified the necessary 
additional assumptions. It is worth pointing out that even \emph{classical} measurements 
can also be disturbing~\cite{WisemanMilburnBook2010} as in fact any real measurement is. 
Then, $\Delta F_S^\text{ms} \neq 0$ and the amount of extractable work changes. 

We conclude this section by noting that (\ref{eq 2nd law fb interval final}) gives a bound 
on the extractable work \emph{during the feedback process} but neglects the work invested 
during the measurement step (which however can be zero). 
When adding (\ref{eq 2nd law fb interval final}) to (\ref{eq 2nd law ms interval final}), 
we find that the total extractable work is bounded by minus the entropy change in the unit
\begin{equation}
\beta W = \beta(W^\text{ms} + W^\text{fb}) \geq - \Delta S_U.
\end{equation}
This result is equivalent to Eq.~(\ref{eq 2nd law info reservoir}) and 
shows that our feedback control scheme (implemented by a stream of memories) 
is equivalent to the information reservoir described in Sec.~\ref{sec info reservoir}.
This connection between feedback control and information-driven engines was debated 
in Refs.~\cite{BaratoSeifertPRL2014, StrasbergEtAlPRE2014, HorowitzSandbergNJP2014, 
ShiraishiMatsumotoSagawaNJP2016} but is unambiguous here. 

A summary of the thermodynamics of feedback control within our framework is given in Fig.~\ref{fig meas and fb}. 
A model-system application will also be provided in Sec.~\ref{sec electronic Maxwell demon}.

\section{Effective master equations}
\label{sec effective master equation}

The thermodynamic framework introduced in Sec.~\ref{sec generalized thermodynamic framework} is very general
and allowed us to derive important exact identities. But in practice, it can only be used if one is able 
to solve the reduced dynamics of the joint system-unit complex. This is usually not an easy task. 
Our goal in this section will be to derive a closed reduced description for the system 
only, which includes its dynamics and thermodynamics. 
We will derive effective MEs for the system which do not rely on the weak coupling approximation, contrary 
to the results of Sec.~\ref{sec thermodynamics weak coupling limit}.
These MEs often have an apparent \emph{non-thermal} character. 
For instance they do not obey the condition~(\ref{eq local detailed balance}). 
Establishing a consistent thermodynamics for these MEs, when solely considered as mathematical objects, can thus be 
challenging and often requires the \emph{ad hoc} introduction of effective new quantities, which lack a solid physical 
interpretation~\cite{EspositoSchallerEPL2012, MunakataRosinbergJSM2012, StrasbergEtAlPRE2013, MunakataRosinbergJSM2013}. 
Progress has been achieved when the MEs result from the coarse graining of a larger network of states which 
originally obeys a thermodynamically consistent ME, especially if the network is bipartite~\cite{EspositoPRE2012, 
StrasbergEtAlPRL2013, HartichBaratoSeifertJSM2014, HorowitzEspositoPRX2014, ShiraishiSagawaPRE2015, HorowitzJSM2015}, 
or for particular information-driven engines~\cite{StrasbergEtAlPRE2014} and ``boundary-driven'' 
MEs~\cite{HorowitzPRE2012, HorowitzParrondoNJP2013, BarraSciRep2015, LorenzoEtAlPRA2015}. 
Our approach is similar in spirit since we will derive effective MEs starting from the framework 
of repeated interaction for which we established a consistent thermodynamics. 

In this section the energy and entropy of the system at the effective level will always be given by
\begin{align}
 E_S(t)	&=	\mbox{tr}_S\{H_S(t)\rho_S(t)\},	\\
 S_S(t)	&=	-\mbox{tr}_S\{\rho_S(t)\ln\rho_S(t)\}.
\end{align}

We will also allow the time interval $\tau$ between successive system-unit interactions to 
fluctuate according to the \emph{waiting time distribution} $w(\tau)$. 
The duration of the interaction itself $\tau' \leq \tau$ will be specified on a case by case basis. 
The time evolution of the system over a interval $\tau$ is given by some generic Kraus map $\Phi_S(\tau)$. 
We introduce the \emph{conditional density matrix} $\rho_S^{(n)}(t)$ which describes the system 
density matrix conditioned on the fact that $n$ interactions with the units happened so far. 
Then, $\rho_S^{(n)}(t)$ is related to $\rho_S^{(n-1)}(t-\tau)$ at an earlier time $\tau>0$ by  
\begin{equation}\label{eq rho n}
 \rho_S^{(n)}(t) = \int_{0}^{t-t_0} d\tau w(\tau) \Phi_S(\tau) \rho_S^{(n-1)}(t-\tau),
\end{equation}
where $t_0<t$ is an arbitrary initial time. 
The unconditional state of the system is recovered by summing over $n$: $\rho_S(t) = \sum_n \rho_S^{(n)}(t)$.

\subsection{Poisson-distributed interaction times}
\label{sec Poisson distributed interaction time}

\subsubsection{Setup}

In this subsection, we consider an exponential waiting time distribution $w(\tau) = \gamma e^{-\gamma\tau}$. 
This means that the number $N$ of units with which the system interacts during a fixed time window $T$ is 
Poisson-distributed, i.e. $P_N(T) = \frac{(\gamma T)^N}{N!}e^{-\gamma T}$. 
The average time between successive interactions is therefore $\int_0^\infty \tau w(\tau)d\tau = \gamma^{-1}$. 

\begin{figure}[t]
 \centering\includegraphics[width=0.44\textwidth,clip=true]{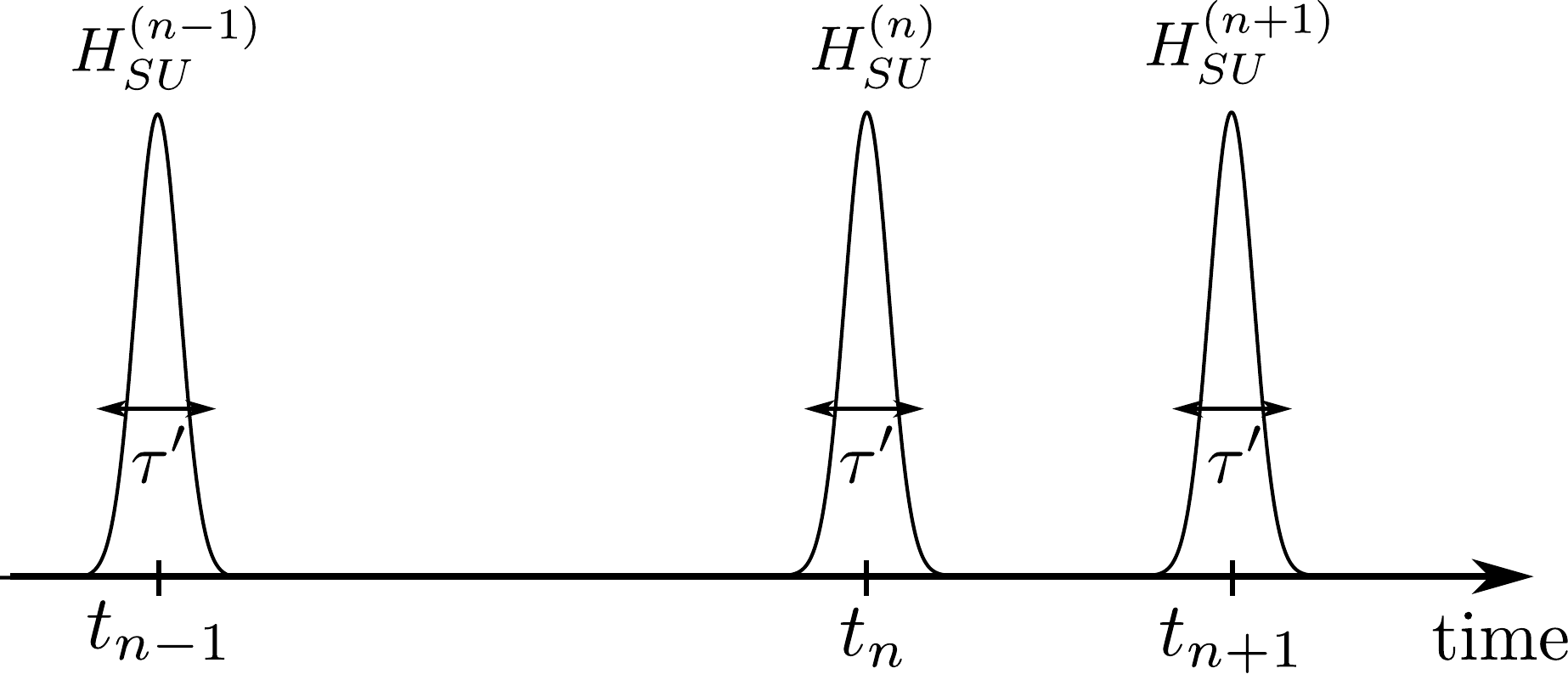}
 \label{fig Poisson times} 
 \caption{Sketch of the Poisson-distributed regime: 
 The system evolves most of the time freely but undergoes once in a while a short ($\tau'\rightarrow 0$) 
 and strong ($H_{SU} \sim \tau'^{-1}$) interaction with a unit where $\tau'$ denotes the duration of 
 the interaction as in Sec.~\ref{sec generalized thermodynamic framework}. 
 The duration $\tau$ between successive system-unit interactions fluctuates according to an exponential time 
 distribution with average duration $\gamma^{-1}$.}
\end{figure}

Furthermore, we assume that the system-unit interactions are very strong and short, but happen very rarely. 
In this way, we can assure that the units have a finite influence on the evolution of the density matrix. 
More specifically, if the random times at which a new unit interaction occur are denoted 
$t_n$ ($n=0,1,2,\dots$), the system-unit interaction Hamiltonian is a sum of delta kicks 
\begin{equation}\label{eq interaction Hamiltonian Poisson}
 H_{SU}(t) = \sum_n \delta(t-t_n) V_{SU}
\end{equation}
as sketched in Fig.~\ref{fig Poisson times}. 
This interaction creates an instantaneous unitary operation $U$ at times $t = t_n$ such that 
the system-unit state right after an interaction reads 
\begin{equation}\label{eq unitary part Poisson}
 \rho'_{SU}(t) = U\rho_S(t)\otimes\rho_U U^\dagger, ~~~ U = e^{-iV_{SU}}.
\end{equation}

Putting aside the brief system-unit interactions, most of the time the system will evolve in weak contact 
with a large thermal reservoir at inverse temperature $\beta$. Its dynamics will thus 
obey a ME of the form~(\ref{eq generic ME for X})
\begin{equation}\label{eq help 25}
d_t \rho_S(t) = \C L_0 \rho_S(t) \equiv -i[H_S(t),\rho_S(t)] + \C L_{\beta} \rho_S(t),
\end{equation}
where $\C L_{\beta}$ is the standard dissipator caused by the thermal reservoir. 
For notational simplicity we keep the time-dependence of superoperators implicit, $\C L = \C L(t)$. 

Similar interaction scenarios have been considered in the past but for different purposes.
See e.g. Refs.~\cite{FilipowiczJavanainenMeystrePRA1986, FilipowiczJavanainenMeystreJOSAB1986, 
MilburnPRA1987, AttalPautratAIHP2006, BaratoSeifertPRL2014, StrasbergEtAlPRE2014, GrimmerEtAlPRA2016}. 

Overall, the system evolution over an entire interval $\tau$ is given by the Kraus map 
\begin{equation}\label{eq Kraus map Poisson}
 \Phi_S(\tau)\rho_S = e^{\C L_0\tau}\C J_S\rho_S,
\end{equation}
where 
\begin{equation}
 \C J_S \rho_S(t) \equiv \mbox{tr}_U \{U \rho_S(t) \otimes \rho_U U^\dagger\}.	\label{eq def jump op poisson case}
\end{equation}
Thus, Eq.~(\ref{eq Kraus map Poisson}) describes the short kick felt by the system due to the interaction with the unit 
($\C J_S$) followed by the dissipative evolution of the system in contact with the reservoir ($e^{\C L_0\tau}$). 
If $\C L_0$ is time-dependent, we have to replace $e^{\C L_0\tau}$ by the corresponding time-ordered 
generator, see Eq.~(\ref{eq time evo ME general}). 
For later convenience, we also introduce the superoperator describing the effect of a system-unit interaction on the unit
\begin{equation}
 \C J_U \rho_U \equiv \mbox{tr}_S \{U \rho_S(t) \otimes \rho_U U^\dagger\}.	\label{eq def jump op poisson case U}
\end{equation}
Note that $\C J_U = \C J_U(t)$ might be time-dependent if $\rho_S(t)$ has not yet reached its steady 
state, but $\C J_S$ is not.

Using Eq. (\ref{eq Kraus map Poisson}) in Eq.~(\ref{eq rho n}) and substituting $\tau = t-t'$ to make the 
dependence on the actual time $t$ explicit, we obtain
\begin{equation}
 \rho_S^{(n)}(t) = \int_{t_0}^t dt' \gamma e^{-\gamma(t-t')} e^{\C L_0(t-t')} \C J_S \rho_S^{(n-1)}(t').
\end{equation}
By taking the time derivative, we find 
\begin{equation}
 d_t\rho_S^{(n)}(t) = -\gamma\rho_S^{(n)}(t) + \C L_0\rho_S^{(n)}(t) + \gamma\C J_S\rho_S^{(n-1)}(t),
\end{equation}
and by summing over $n$, we finally obtain the effective ME ruling the averaged time-evolution of the system 
\begin{equation}\label{eq help 2}
 d_t\rho_S(t) = \C L_0\rho_S(t) + \gamma(\C J_S - 1)\rho_S(t).
\end{equation}

We can make the new part $\gamma(\C J_S - 1)$ of the ME (\ref{eq help 2}) more explicit by writing the initial state of the unit as 
\begin{equation}\label{eq initial unit state Poisson ME}
 \rho_U = \sum_k p_k|k\rangle_U\langle k|
\end{equation}
where $\{|k\rangle_U\}$ is an arbitrary set of eigenstates of $\rho_U$ (not necessarily energy eigenstates of $H_U$). 
Then, we see that 
\begin{equation}
  \C J_S\rho_S(t) = \sum_{k,l} A_{kl}\rho_S(t) A^\dagger_{kl} \label{eq ME Poisson jump operator}
\end{equation}
has the form of a Kraus map~(\ref{eq Kraus map}) where the system operators defined as 
$A_{kl} \equiv \sqrt{p_k}\langle l|U|k\rangle_U$ fulfill the completeness relation 
$\sum_{k,l} A_{kl}^\dagger A_{kl} = \mathbf{1}_S$. Therefore, we can write 
\begin{equation}\label{eq help 21}
 \C L_\text{new} \rho_S(t) \equiv \gamma(\C J_S-1)\rho_S(t) = \gamma \sum_{k,l} \C D[A_{kl}]\rho_S(t)\,,
\end{equation}
where $\C D[A]\rho \equiv A\rho A^\dagger - \frac{1}{2}\{A^\dagger A,\rho\}$, thus explicitly showing that 
$\C L_\text{new}$ is of Lindblad form~\cite{BreuerPetruccioneBook2002, LindbladCMP1976, GoriniEtAlJMP1976}. 
{By choosing $\rho_U$ and $V_{SU}$ appropriately we can create arbitrary $A_{kl}$'s as long as they 
fulfill the completeness relation. Note that the class of generators $\C L_\text{new}$ created this way is not equivalent 
to the class of thermal generators~(\ref{eq thermal dissipator explicit}). In general, 
relations~(\ref{eq local detailed balance 2}) or~(\ref{eq local detailed balance}) will not be fulfilled for 
$\C L_\text{new}$.}
Furthermore, we remark that including multiple independent streams of units can be easily done within 
this scenario because the probability of a simultaneous interaction with more than one unit is negligible.

In summary, the new ME~(\ref{eq help 2}) can be written as 
\begin{equation}\label{eq ME case Poisson}
 d_t\rho_S(t) = -i[H_S(t),\rho_S(t)] + \C L_{\beta}\rho_S(t) + \C L_\text{new}\rho_S(t).
\end{equation}

\subsubsection{Thermodynamics}\label{subseconthermo}

We now turn to the thermodynamic description corresponding to the setup above. 

We start by considering energy and entropy changes in the units. 
During a short time interval $dt$, the probability that a unit interacts (resp. does not interact) 
is given by $\gamma dt \ll 1$ (resp. $1-\gamma dt$). In the former case the unit state 
changes from $\rho_U$ to $\C J_U\rho_U$ while it remains in $\rho_U$ in the latter case.
Since an energy and entropy change in the unit only occurs when an interaction 
takes place, the rates of unit energy and entropy change are given by
\begin{align}
 & d_t E_U(t) = \gamma (\mbox{tr}_U\{H_U\C J_U\rho_U\} - \mbox{tr}_U\{H_U\rho_U\}), \label{dEU}	\\
 & d_t S_U(t) =	\gamma (-\mbox{tr}_U\{(\C J_U\rho_U)\ln(\C J_U\rho_U)\} + \mbox{tr}_U\{\rho_U\ln\rho_U\}). 
\nonumber \\ \label{dSU} 
\end{align}

We now turn to the rate of work injected in the joint system-unit, which we again split into two parts 
\begin{equation}
 \dot W   = \dot W_S + \dot W_{SU}
\end{equation}
where
\begin{align}
 \dot W_S	=&~	\mbox{tr}_S\{\rho_S(t) d_t H_S(t)\}, \\
 \dot W_{SU}	=&~	\gamma \mbox{tr}_{SU}\{[H_S(t)+H_U][U\rho_S(t)\rho_U U^\dagger - \rho_S(t)\rho_U]\} \nonumber \\
		=&~	\gamma \mbox{tr}_S\{H_S(t)(\C J_S-1)\rho_S(t)\}	\nonumber	\\
		&+	\gamma \mbox{tr}_U\{H_U(\C J_U-1)\rho_U\}.
\end{align}
The first part is the work due to the time-dependence in the system Hamiltonian $H_S(t)$ 
while the second part is due to the system-unit interaction when it occurs. 
Since this latter gives rise to a unitary dynamics in the system-unit space which produces 
no heat, it is given by the energy change in the system and unit due to the system-unit interaction. 

The overall change in the energy of the system is naturally given by 
\begin{equation}
d_t E_S(t) = d_t \mbox{tr}_S \{H_S(t) \rho_S(t) \} \label{dES}\,
\end{equation}
and the heat entering the system from the reservoir is
\begin{equation}
 \begin{split}\label{HeatRatePoisMEq}
  \dot Q(t)	&=	\mbox{tr}_S\{H_S(t)\C L_{\beta}\rho_S(t)\}	\\
		&=	-\beta^{-1} \mbox{tr}_S\left\{[\C L_{\beta}\rho_S(t)] \ln \rho^S_{\beta}(t) \right\}.
 \end{split}
\end{equation}
Noting that $\dot W_{SU}(t) - d_t E_U(t) = \gamma \mbox{tr}_S\{H_S(t)\C L_\text{new}\rho_S(t)\}$,
we obtain the first law of thermodynamics  
\begin{equation}
d_t E_S(t) = \dot Q(t) + \dot W(t) - d_t E_U(t) , \label{eq 1st law Poisson ME}
\end{equation}
which constitutes the differential version of the general result Eq.~(\ref{eq 1st law modified}).

We now proceed to show that the differential version of the generalized second law~(\ref{eq system dissipation 2}), 
\begin{equation}
 \dot \Sigma_S(t) = d_t S_S(t) + d_t S_U(t) - \beta \dot Q \ge 0, \label{DiffGenSecLaw}
\end{equation}
also holds and that its non-negativity is ensured. Using the fact that 
$d_tS_S(t) = -\mbox{tr}\left\{[d_t \rho_S(t)] \ln\rho_S(t)\right\}$ 
together with (\ref{eq help 2}), we find that
\begin{align}
d_t S_S(t) =& - \mbox{tr}_S\left\{[\C L_{\beta}\rho_S(t)] \ln \rho_S(t) \right\} \label{dSS} \\
           &- \gamma\mbox{tr}_S\{[(\C J_S-1 )\rho_S(t)] \ln \rho_S(t)\}, \nonumber
\end{align}
Combining this with~(\ref{HeatRatePoisMEq}), we can rewrite (\ref{DiffGenSecLaw}) as
\begin{align}
  \dot\Sigma_S(t) =&-	\mbox{tr}_S\left\{\C L_{\beta}\rho_S(t)\left[\ln\rho_S(t) - \ln \rho^S_{\beta}(t) \right]\right\} \nonumber \\
		   &-	\gamma\mbox{tr}_S\{[(\C J_S-1)\rho_S(t)]\ln\rho_S(t)\} + d_t S_U(t)	\nonumber \\
		   \ge&	-	\gamma\mbox{tr}_S\{[(\C J_S-1)\rho_S(t)]\ln\rho_S(t)\} + d_t S_U(t),	\nonumber
\end{align}
where we used Spohn's inequality~(\ref{eq Spohns inequality}) at the end. Using~(\ref{dSU}) 
the remaining part can be expressed as
\begin{align}
 & \frac{\dot\Sigma_S(t)}{\gamma} \ge	\\
 & -\mbox{tr}_S\{[\C J_S\rho_S(t)]\ln[\C J_S\rho_S(t)]\} + D[\C J_S\rho_S(t)\|\rho_S(t)]	\nonumber	\\
 & -\mbox{tr}_S\{\rho_S(t)\ln\rho_S(t)\} - \mbox{tr}_U\{(\C J_U\rho_U)\ln(\C J_U\rho_U)\}	\nonumber	\\
 & +\mbox{tr}_U\{\rho_U\ln\rho_U\}.	\nonumber
\end{align}
Using now the fact that entropy does not change under unitary transformation so that 
\begin{align}
 & S[\rho_S(t)] + S[\rho_U]	\label{FunRel}	\\
 & = S[\rho_S(t)\otimes\rho_U] = S[U\rho_S(t)\otimes\rho_U U^\dagger]	\nonumber	\\
 & = S[\C J_S\rho_S(t)] + S[\C J_U\rho_U] - I_{S:U}[U\rho_S(t)\rho_U U^\dagger],	\nonumber
\end{align}
we can prove the non-negativity of $\dot\Sigma_S(t)$ since 
\begin{equation}
 \frac{\dot\Sigma_S(t)}{\gamma} \ge D[\C J_S\rho_S(t)\|\rho_S(t)] + I_{S:U}[U\rho_S(t)\rho_U U^\dagger] \ge 0.
\end{equation}

The present analysis underlines that our generalized thermodynamic framework of repeated interactions can be 
carried over to the limiting situation considered in this section of very short and Poisson-distributed system-unit 
interactions. The resulting description is closed in terms of the system density matrix obeying the dynamics 
(\ref{eq ME case Poisson}) and the non-negativity of the differential form of the second law (\ref{DiffGenSecLaw}) 
is a stronger result than the original integrated one (\ref{eq system dissipation}).     

A crucial point to emphasize is that knowing the physical mechanism underlying a ME dynamics 
such as (\ref{eq ME case Poisson}) is essential to establish its correct thermodynamics. 
Indeed, without the additional information about the units at hand, the thermodynamic analysis would be very different. 
Presupposing that we were able to disentangle the two dissipative mechanisms caused by $\C L_{\beta}$ and $\C L_\text{new}$, 
we would need to define an effective heat flow $\dot Q_\text{eff}(t) \equiv \mbox{tr}\{H_S\C L_\text{new} \rho_S(t)\}$ 
to explain the discrepancy in the first law. From our inclusive approach above, however, we know that 
$\dot Q_\text{eff}(t) = \dot W_{SU}(t) - d_t E_U(t)$. 
The effect on the second law would even be more drastic. 
Using Spohn's inequality~(\ref{eq Spohns inequality}), we know that the quantity 
\begin{equation}\label{eq help 14}
 \begin{split}
  \dot\Sigma^\text{eff}_S	\equiv&	-\mbox{tr}\{[\C L_0\rho_S(t)][\ln\rho_S(t)-\ln \rho^S_{\beta}(t)]\}	\\
				&-	\mbox{tr}\{[\C L_\text{new}\rho_S(t)][\ln\rho_S(t)-\ln\bar\rho_\text{new}]\}
 \end{split}
\end{equation}
would be always non-negative as a sum of two non-negative terms. Here, $\bar\rho_\text{new}$ denotes the steady state of 
$\C L_\text{new}$, i.e. ${\C L_\text{new}\bar\rho_\text{new} = 0}$. In fact, if $\C L_\text{new}$ would correspond to 
the dissipator caused by a standard thermal reservoir, Eq.~(\ref{eq help 14}) would correspond to the standard entropy 
production. However, this is not the case and thus, $\dot\Sigma^\text{eff}_S$ is not only numerically different from 
$\dot\Sigma_S$, but also lacks any a priori thermodynamic interpretation.

\subsubsection{Heat, work and information dominated interactions}
\label{sec Poisson heat work info dominated}

Following the line of Secs.~\ref{sec heat reservoir}-\ref{sec info reservoir}, we now consider 
specific scenarios which have a clear thermodynamic interpretation.

We start by studying the case of initially thermal units at inverse temperature $\beta'$ as in 
Sec.~\ref{sec heat reservoir}. Then, using~(\ref{dEU}) and~(\ref{dSU}), we deduce in accordance with 
Eq.~(\ref{eq help 15}) that 
\begin{equation}\label{eq help 26}
 d_t S_U(t) = \beta'd_t E_U(t) - \gamma D(\C J_U\rho_U\|\rho_U).
\end{equation}
Thus, in contrast to a weakly coupled macroscopic reservoir 
the units in general do not mimic an ideal heat reservoir unless additional 
assumptions are used, as we will see in Sec.~\ref{sec alternative setup ideal heat}. 

The difference between the effective ME (\ref{eq help 2}) compared to the weak coupling 
ME from Sec.~\ref{sec thermodynamics weak coupling limit} also becomes apparent by noting that 
the generator $\C L_\text{new}$ is not of the form of $\C L_{\beta}$ in Eq.~(\ref{eq thermal dissipator explicit}). 
Thus, initially thermal units will typically not imply that $\C L_\text{new} \rho_{\beta'}^S(t) = 0$. 
One very specific way to enforce it is to assume that the units and the system 
have identical Hamiltonians, $H_S = \sum_k E_k |E_k\rangle_S\langle E_k|$ and 
$H_U = \sum_k E_k |E_k\rangle_U\langle E_k|$, and that their delta kick interaction 
gives rise to a unitary evolution of the form 
$U = \sum_{k,l} |E_k\rangle_S\langle E_l|\otimes |E_l\rangle_U\langle E_k|$, 
which swaps energy between the system and the unit.

For the work reservoir, we require that $d_t S_U = 0$ while $d_t E_U \neq 0$. 
For initially thermal units this can be again ensured by choosing $\beta'\rightarrow0$ 
as in Sec.~\ref{sec work reservoir}. 
But in general, to ensure that the entropy of the unit remains constant while its 
energy can change, the effective unit dynamics should be given by a unitary 
operator $U_U$, $\C J_U \rho_U = U_U \rho_U U_U^\dagger$.
Beside the trivial choice $U = U_S \otimes U_U$, finding such conditions might not be easy.

In turn, the limit of an information reservoir where $d_t E_U = 0$ and $d_t S_U(t) \neq 0$ 
can be easily achieved for any interaction by considering a fully degenerate unit Hamiltonian, 
\begin{equation}
 H_U\sim \mathbf{1}_U ~~~ \text{(ideal information reservoir)},
\end{equation}
naturally always implying $d_t E_U = 0$. 

A last important class of interactions are those generated by the unitary operator 
\begin{equation}\label{eq help 20}
 U = \sum_k U_S^{(k)}\otimes |k\rangle_U\langle k|\,,
\end{equation}
where $U_S^{(k)}$ is an arbitrary unitary operator in the system Hilbert space whereas 
$|k\rangle_U$ denotes the eigenvectors of $\rho_U$~(\ref{eq initial unit state Poisson ME}). 
One easily verifies that the unit state does not change during the interaction, $\rho_U = \C J_U\rho_U$, 
and hence its energy and entropy also stays constant, $d_tE_U = 0, d_tS_U = 0$. 
In this case, the units are neither a work nor an information reservoir. 
Instead, the system state changes according to 
\begin{equation}
 \C J_S \rho_S = \sum_k p_k U_S^{(k)}\rho_S (U_S^{(k)})^\dagger,
\end{equation}
where $p_k=\langle k| \rho_U |k \rangle$.
This interaction will therefore in general inject energy as well as entropy into the system. 
Using (\ref{FunRel}), we see that the change in the system entropy caused by such a system-unit interaction 
is given by the mutual information established between the system and the unit after the interaction 
\begin{equation}
 S[\C J_S\rho_S(t)] - S[\rho_S(t)] = I_{S:U}[U\rho_S(t)\rho_U U^\dagger].
\end{equation}
Thus, the interaction~(\ref{eq help 20}) can be seen as a measurement of the unit by the system. Indeed, depending on the 
unit state $|k\rangle_U$, the system will in general change its state to $U_S^{(k)}\rho_S (U_S^{(k)})^\dagger$. 
Vice versa, by exchanging the labels $U$ and $S$ above, we can also implement a measurement 
of the system by the units. This will be used in Sec.~\ref{sec electronic Maxwell demon}.

\subsubsection{Ensemble of units and ideal heat reservoir}
\label{sec alternative setup ideal heat}

In Sec.~\ref{subseconthermo}, we considered the energy and entropy changes of 
those units which interacted with the system. 
An interesting alternative approach consists in evaluating energy and entropy changes with 
respect to a statistical ensemble composed of both, the units which did and did not interact.
One physically relevant scenario for this is the case where units are frequently sent to the system, but only a 
small Poisson-distributed fraction of them interacts whereas the rest remains unchanged. 

Mathematically, let us assume that every time-step $dt$ a unit passes the system with certainty, but only interacts with 
it with probability $\gamma dt\ll 1$, which is assumed to be infinitesimal such that the average evolution of the system 
is still differentiable and coincides with Eq.~(\ref{eq ME case Poisson}). If we do not record the precise interaction 
times, each outgoing unit $\rho'_U$ must be described by the state 
\begin{equation}
 \rho'_U = (1-\gamma dt)\rho_U + \gamma dt\C J_U\rho_U,
\end{equation}
where $\rho_U$ describes the initial state as usual. The change in unit entropy per time-step $dt$ then becomes 
\begin{equation}
 \begin{split}
  d_t\bar S_U(t)	&\equiv	\lim_{dt\rightarrow0} \frac{S(\rho'_U) - S(\rho_U)}{dt}	\\
			&=	-\gamma\mbox{tr}\{[(\C J_U-1)\rho_U]\ln\rho_U\}. 
 \end{split}
\end{equation}
Here, we used a bar to distinguish this definition from the previous case, Eq. (\ref{dSS}), 
in which every unit passing the system also interacts with the system. The difference between both is exactly
\begin{equation}
 d_t\bar S_U(t) - d_t S_U(t) = \gamma D(\C J_U\rho_U\|\rho_U) \ge 0,
\end{equation}
which can be interpreted as the entropy increase caused by mixing the units which interacted with the system 
with those which did not. In contrast, since the energy $E_U(t)$ of the units is a linear functional of $\rho_U(t)$, 
we easily deduce that $d_t\bar E_U(t) = d_t E_U(t)$. In other words, while the entropy 
balance differs between the two approaches, the energy balance remains the same. 
Since heat is also unaffected, the difference in entropy production between the 
two approaches reads
\begin{equation}
 \dot{\bar\Sigma}_S(t) = \dot\Sigma_S + \gamma D(\C J_U\rho_U\|\rho_U) \ge 0.
\end{equation}

An important implication of the present approach is that when considering units which 
are initially thermal $\rho_U = \rho^U_{\beta'}$, we find that 
\begin{equation}
 d_t\bar S_U(t) = \beta'd_t\bar E_U(t),
\end{equation}
in contrast to Eq.~(\ref{eq help 26}). 
Thus, the notion of an ideal heat reservoir requires to not only focus on those units which 
interacted but to consider the statistical mixture of units which did and did not interact.
This picture is also supported by an alternative approach.

Consider a reservoir made of an initial population of $N_0\gg 1$ identical and independent units. 
Let us assume that the particle content of the reservoir decays exponentially, $N_t = N_0 e^{-\gamma t}$,
and produces the sequence of units which all eventually interact with the system. 
In contrast to the case in Sec.~\ref{subseconthermo}, after having interacted, we sent back the units to the reservoir 
and let them mix with the remaining $N_t$ fresh units. The mixed state of this reservoir can be described as 
\begin{equation}
 \tilde\rho_U(t) = \frac{N_0 - N_t}{N_0}\C J_U\rho_U + \frac{N_t}{N_0}\rho_U,
\end{equation}
where we assumed that $\C J_U$ is time-independent for simplicity. For such a process we have $d_t N_t = -\gamma N_t$ 
and thus 
\begin{equation}
 d_t\tilde\rho_U(t) = \gamma\frac{N_t}{N_0}(\C J_U-1)\rho_U.
\end{equation}
Considering times over which $N_t \approx N_0$ to remain consistent with the assumption that only fresh units interact 
with the system, it is possible to recover the same mathematical results as above. 
Thus, physically separating the units from the system and dividing them into an incoming and outgoing 
stream is not essential in this picture. One could equally well consider a gas 
of noninteracting units surrounding the system and interacting with it at Poisson random times.

\subsection{Regular and frequent interaction intervals}
\label{sec perturbative expansion delta kicks}

\subsubsection{Setup}
\label{sec perturbative expansion delta kicks setup}

There is another class of the repeated interaction framework which has often been considered to derive 
MEs~\cite{AttalPautratAIHP2006, KarevskiPlatiniPRL2009, GiovannettiPalmaPRL2012, HorowitzPRE2012, 
HorowitzParrondoNJP2013, BarraSciRep2015, LorenzoEtAlPRA2015}. Its thermodynamics has been 
considered as well~\cite{HorowitzPRE2012, HorowitzParrondoNJP2013, BarraSciRep2015, LorenzoEtAlPRA2015}. 
In this scenario the duration between two consecutive system-unit interactions $\tau$ is taken constant 
as in Sec.~\ref{sec generalized thermodynamic framework} and equal to the duration of the system-unit 
interaction $\tau'$. Furthermore, the duration is short and $\tau' = \tau \equiv \delta t$ is used as 
a small expansion parameter where the interaction is assumed of the form 
\begin{equation}
 H_{SU}(t) = \sum_n \Theta(t-n\delta t)\Theta(n\delta t+\delta t-t)\frac{\tilde V}{\sqrt{\delta t}}.
\end{equation}
This Hamiltonian models very short, but permanent and strong interactions. The fact that every unit is 
replaced after a time $\delta t$ implies Markovianity. Furthermore, for a clean derivation of the ME,  
in this setting one has to assume no coupling to a thermal reservoir. 
Therefore, the system and unit evolve unitarily over one period via the operator 
\begin{equation}
 U(\delta t) = \exp\left\{-i\delta t\left[H_S(t) + H_U + \tilde V/\sqrt{\delta t}\right]\right\},
\end{equation}
where it is further assumed that $\delta t$ is much smaller than the rate of change of $H_S(t)$. 
Finally, one needs to expand the evolution of $\rho_S(t)$ up to first order in $\delta t$ under 
the assumption that $\mbox{tr}\{\tilde V\rho_U\}=0$.

Instead of following such a derivation, which is presented elsewhere~\cite{AttalPautratAIHP2006, 
KarevskiPlatiniPRL2009, GiovannettiPalmaPRL2012, HorowitzPRE2012, HorowitzParrondoNJP2013, 
BarraSciRep2015, LorenzoEtAlPRA2015}, we follow an alternative route by considering the setup 
of Sec.~\ref{sec Poisson distributed interaction time} in the limit of an infinitely fast Poisson 
process. This procedure yields identical mathematical results.
More specifically, we consider the limit where the Poisson rate scales as $\gamma = \epsilon^{-1}$ 
while at the same time assuming that the unitary interaction~(\ref{eq unitary part Poisson}) 
scales as $U = \exp(-i\sqrt{\epsilon}\tilde V)$, i.e.  $V_{SU} = \sqrt{\epsilon} \tilde V$ in 
Eq.~(\ref{eq interaction Hamiltonian Poisson}). Furthermore, we explicitly neglect the reservoir, 
i.e. we set $\C L_\beta = 0$ in the results of Sec.~\ref{sec Poisson distributed interaction time}. 
The Kraus map~(\ref{eq Kraus map Poisson}) then reads $\Phi_S(\tau)\rho_S = e^{-iH_S(t)\tau}[\C J_S\rho_S]e^{iH_S(t)\tau}$. 
Our goal will now be to derive an effective ME in the limit $\epsilon\rightarrow0$.

We start with Eq.~(\ref{eq def jump op poisson case}) by
expanding $U \rho_S(t) \rho_U U^\dagger$ in powers of $\sqrt{\epsilon}$. This yields 
\begin{equation}
 \begin{split}
  \C J_S \rho_S(t)	=&~	\rho_S(t) - i\sqrt{\epsilon}\mbox{tr}_U\left\{[\tilde V,\rho_S(t)\rho_U]\right\}	\\
			&-	\frac{\epsilon}{2}\mbox{tr}_U\left\{[\tilde V,[\tilde V,\rho_S(t)\rho_U]]\right\} + \dots
 \end{split}
\end{equation}
In order to derive a meaningful
differential equation we now also have to demand that the interaction $\tilde V$ or the initial unit state $\rho_U$ 
is chosen such that $\mbox{tr}_U\{\tilde V\rho_U\} = 0$, which removes the term proportional to $\sqrt{\epsilon}$. 
Then, we consider Eq.~(\ref{eq help 21}) which becomes 
\begin{equation}
 \begin{split}
  & \C L_\text{new}\rho_S	\\
  &~ = \frac{1}{\epsilon}\left(\rho_S-\frac{\epsilon}{2}\mbox{tr}_U\left\{[\tilde V,[\tilde V,\rho_S\rho_U]]\right\} + \dots -\rho_S\right)	\\
  &~ = \frac{1}{2}\mbox{tr}_U\left\{[\tilde V,[\tilde V,\rho_S(t)\rho_U]]\right\}\rho_S + \C O(\sqrt{\epsilon}).
 \end{split}
\end{equation}
We can make this effective ME more explicit by writing $\tilde V = \sum_k A_k\otimes B_k$ where 
$A_k$ and $B_k$ are arbitrary operators in the system and unit space such that $\tilde V$ is hermitian. Then, 
after taking into account the influence of the possibly time-dependent system Hamiltonian, 
we get from Eq.~(\ref{eq ME case Poisson}) with $\C L_\beta = 0$ 
our final ME 
\begin{align}
 d_t\rho_S(t)	&=	-i[H_S(t),\rho_S(t)]	\label{eq ME perturbative delta kicks 2nd order} \\
		&+	\sum_{k,l} \langle B_l B_k\rangle_U \left(A_k\rho_S(t) A_l - \frac{1}{2}\{A_lA_k,\rho_S(t)\}\right), \nonumber
\end{align}
where we defined $\langle B_l B_k\rangle_U \equiv \mbox{tr}_U\{B_lB_k\rho_U\}$. 
It agrees with Refs.~\cite{AttalPautratAIHP2006, KarevskiPlatiniPRL2009, HorowitzPRE2012, HorowitzParrondoNJP2013, 
BarraSciRep2015} and will be further used in Sec.~\ref{sec lasing without inversion}. 
Treating multiple streams of units can also be easily done within this setup.
We finally note that this ME is very similar (but not identical) to the singular coupling 
ME~\cite{BreuerPetruccioneBook2002}.

\subsubsection{Thermodynamics}
\label{sec Barra setup}

A thermodynamic analysis of such ``boundary driven MEs''~(\ref{eq ME perturbative delta kicks 2nd order}) was 
given in Ref.~\cite{BarraSciRep2015} for the case of initially thermal units $\rho_U = \rho_\beta^U$. We will now 
approach this problem from our perspective demonstrating that the thermodynamic framework in 
Ref.~\cite{BarraSciRep2015} is consistent, but overestimates the entropy production. 

For $\mbox{tr}_U\{\tilde V\rho_U\} = 0$ it follows immediately that (in the following we consider only the leading order 
terms) 
\begin{align}
 \C J_S\rho_S(t)	=&~	\rho_S(t) - \frac{\epsilon}{2}\mbox{tr}_U\left\{[\tilde V,[\tilde V,\rho_S(t)\rho_U]]\right\}, \\
 \C J_U\rho_U		=&~	\rho_U(t) - i\sqrt{\epsilon}\mbox{tr}_S\left\{[\tilde V,\rho_S(t)\rho_U]\right\}	\nonumber	\\
			&-	\frac{\epsilon}{2}\mbox{tr}_S\left\{[\tilde V,[\tilde V,\rho_S(t)\rho_U]]\right\}.	\label{eq change U Barra}
\end{align}
Thus, it is clear that all thermodynamic quantities defined in Sec.~\ref{sec Poisson distributed interaction time} for 
the system (i.e. $d_t E_S(t)$ and $d_t S_S(t)$) are well-behaved (i.e. do not diverge for $\epsilon\rightarrow0$), but 
for the unit this is less clear. 

We start by looking at unit-related quantities in the first law. 
For instance for Eq.~(\ref{dEU}) (the same terms appear in $\dot W_{SU}(t)$ too), 
\begin{equation}
 \begin{split}
  d_tE_U(t)	=&~	\gamma\mbox{tr}_U\{H_U(\C J_U-1)\rho_U\}	\\
		=&	- i\frac{1}{\sqrt{\epsilon}}\mbox{tr}_{SU}\left\{H_U[\tilde V,\rho_S(t)\rho_U]\right\}	\\
		&-	\frac{1}{2}\mbox{tr}_{SU}\left\{H_U[\tilde V,[\tilde V,\rho_S(t)\rho_U]]\right\},
 \end{split}
\end{equation}
where we replaced $\gamma = \epsilon^{-1}$ and used~(\ref{eq change U Barra}). Thus,
the first term will diverge as $\epsilon\rightarrow0$ unless $[H_U,\tilde V] = 0$ or $[H_U,\rho_U] = 0$. Note, 
however, that the divergences cancel out if we consider the first law~(\ref{eq 1st law Poisson ME}) for the system. 
Furthermore, within the framework of Ref.~\cite{BarraSciRep2015} we indeed have $[H_U,\rho_U] = 0$ since 
$\rho_U = \rho_\beta^U$ and hence, 
\begin{align}
 d_t E_U(t)	&=	-\frac{1}{2}\mbox{tr}_{SU}\left\{H_U[\tilde V,[\tilde V,\rho_S(t)\rho_U]]\right\},	\label{eq dEU Barra}	\\
 \dot W_{SU}(t)	&=	-\frac{1}{2}\mbox{tr}_{SU}\left\{(H_S + H_U)[\tilde V,[\tilde V,\rho_S(t)\rho_U]]\right\},
\end{align}
demonstrating that $d_t E_U(t)$ and $\dot W_{SU}(t)$ remain well-behaved. 
Furthermore, we can identify $\dot W_{SU}(t) = \dot W_\text{sw}(t)$ 
by using definition~(\ref{eq def work switch}) and some straightforward algebraic manipulations. 

Turning to the entropy change, we see from our result~(\ref{eq help 26}) that the only term which 
could cause a divergence is $\epsilon^{-1} D(\C J_U\rho_U\|\rho_U)$. With the use of Footnote~\ref{footnote 1} and 
Eq.~(\ref{eq change U Barra}) we find, however, that $D(\C J_U\rho_U\|\rho_U) = \C O(\epsilon)$ and thus, 
also the entropy change of the units is well-behaved. 

In total, the first and second law derived in Sec.~\ref{sec Poisson distributed interaction time} become for this setup
\begin{align}
 d_t E_S(t)		&=	\dot W_S(t) + \dot W_\text{sw}(t) + d_t E_U(t),	\\
 \dot\Sigma_S(t)	&=	d_t S_S(t) + \beta'd_t E_U(t) - \gamma D(\C J_U\rho_U\|\rho_U) \ge 0.	\label{eq 2nd law our Barra}
\end{align}
To compare our results with Ref.~\cite{BarraSciRep2015}, we set $\dot W_S(t) = 0$. 
In this reference, the change in the unit energy is identified with heat, $\dot Q_U(t) \equiv d_t E_U(t)$. 
The first law is the same as ours, but the second law reads $d_t S_S(t) + \beta'\dot Q_U(t) \ge 0$.
Interestingly this is the result obtained in Sec.~\ref{sec alternative setup ideal heat}, when 
the ensemble considered is not only that of the units which interacted but the entire set of units.
As we have seen, it overestimates our entropy production by a mixing term $\gamma D(\C J_U\rho_U\|\rho_U)$.

Finally, this example also illustrates that -- although the dynamics in the joint space of system and all 
units is unitary (and thus reversible) -- the dynamics of the system is irreversible precisely because we 
impose a unidirectional movement of the units. If we time-reverse the global evolution, 
we would recover the initial system state. 
This can be also seen in the system-specific entropy production, which can be rewritten as 
[see also Eq.~(\ref{eq entropy balance X Y})] $\dot\Sigma_S(t) = d_t I_{S:U}(t)$, i.e. for the entropy 
production rate of system and unit [compare with Eq.~(\ref{eq system dissipation})] we have $\dot\Sigma = 0$.

\subsection{Mimicking time-dependent Hamiltonians}
\label{sec driven systems}

\subsubsection{Setup}

In the last part of this section, we show that the stream of units can be engineered in 
a way that will effectively generate a time-dependent system Hamiltonian of the form 
\begin{equation}\label{eq help 16}
 H_S(t) = H_0 + f(t) A,
\end{equation}
where $f(t)$ is an arbitrary real-valued and differentiable function and $A$ an arbitrary hermitian system operator. 
We will further show that this stream of units acts as a work source thereby providing an alternative justification 
for treating time-dependent Hamiltonians as work sources as done in standard quantum 
thermodynamics~\cite{AlickiJPA1979, KosloffEntropy2013, GelbwaserKlimovskyNiedenzuKurizkiAdv2015}. 
{
We note that research in the direction of obtaining a time-dependent Hamiltonian out of a time-independent 
one has been carried out for different settings in Refs.~\cite{SalmilehtoSolinasMoettoenenPRE2014, 
AbergPRL2014, MalabarbaShortKammerlanderNJP2015}.}

The idea is that an arbitrary drive $f(t)$ can be effectively generated by a stream of units with system-unit 
interactions of the form $A\otimes F$, where $F$ is so far an unspecified hermitian unit operator. 
As in the previous subsection, we consider short and repeated interactions: $\tau' = \tau \equiv \delta t$.
We also consider no reservoir at the moment. However, since $f(t)$ can be arbitrary, one must relax the assumption 
that the units are prepared in the same initial state. Thus, only in this subsection, we allow that 
$\rho_{U_n}(n\delta t) \neq \rho_{U_m}(m\delta t)$ for $n\neq m$ (we set the initial time to be zero 
such that $\rho_{U_n}(n\delta t)$ denotes the initial state of the unit just before the interaction). 
The incoming units are however still assumed to be decorrelated. 
The time evolution of the system is given by
\begin{align}\label{eq rho n driving}
 & \rho_S(n\delta t+\delta t) =	\\
 & \mbox{tr}_U\left[e^{-i(H_0+AF+H_U)\delta t}\rho_S(n\delta t)\rho_{U_n}(n\delta t)e^{i(H_0+AF+H_U)\delta t}\right].	\nonumber
\end{align}
By expanding Eq.~(\ref{eq rho n driving}) to first order in $\delta t$ we arrive at 
\begin{align}\label{eq help 27}
 \rho_S(n\delta t+\delta t)	=&~	\rho_S(n\delta t)	\\
				&-	i\delta t[H_0 + \langle F\rangle_{U_n}(n\delta t)A,\rho_S(n\delta t)]\,.	\nonumber
\end{align}
We now choose the state of the unit such that 
\begin{equation}
 \langle F\rangle_{U_n}(n\delta t) \equiv \mbox{tr}_{U_n}\{F\rho_{U_n}(n\delta t)\} = f(n\delta t).
\end{equation}
Under these circumstances, we obtain from Eq.~(\ref{eq help 27}), after rearranging the terms 
in the limit $\delta t\rightarrow0$, 
\begin{equation}
 d_t\rho_S(t) = -i[H_0+f(t)A,\rho_S(t)],
\end{equation}
which is the desired evolution according to the Hamiltonian~(\ref{eq help 16}). 

In fact, if $\delta t$ is chosen small enough compared to any other time-scales, 
one could even include an additional reservoir in the description. 
In this case the Hamiltonian to be simulated becomes 
\begin{equation}\label{eq help 29}
H_{SR}(t) = H_0 + f(t) A + V_{SR} + H_R
\end{equation}
and the joint system-reservoir state evolves according to 
\begin{equation}
d_t\rho_{SR}(t) = -i[H_{SR}(t),\rho_{SR}(t)].
\end{equation}

\subsubsection{Thermodynamics}

In order to establish the thermodynamics of the present setup, we need to consider how the units change over time. 
Similarly as for Eqs.~(\ref{eq rho n driving}) and~(\ref{eq help 27}), we deduce that 
\begin{align}\label{eq help 28}
 \rho_{U_n}(n\delta t+\delta t)	=&~	\rho_{U_n}(n\delta t)	\\
				 &-	i\delta t[H_U + \langle A\rangle_S(n\delta t) F,\rho_{U_n}(n\delta t)].	\nonumber
\end{align}
Since the unit state changes unitarily, the entropy change of the units is zero, $d_tS_{U_n}(t) = 0$, and hence, according to 
the classification schemes from Sec.~\ref{sec applications part I}, the stream of units may behave as a work reservoir. 

To confirm it, we now consider energy balances. First, the change in unit energy is given by 
\begin{align}
 d_tE_{U_n}	&=	\delta t^{-1}\mbox{tr}_{U_n}\{H_{U}[\rho_{U_n}(n\delta t+\delta t) - \rho_{U_n}(n\delta t)]\} \nonumber \\
		&=	i\langle A\rangle_S^n \mbox{tr}_{U_n}\{[F,H_U]\rho_{U_n}^n\}.
\end{align}
For the last line, we used Eq.~(\ref{eq help 28}). Second, we need to consider the switching work steaming from the 
time-dependent coupling of the system and unit. However, we have to remember that the \emph{units change} for each interval. 
Naively applying the definition~(\ref{eq def work switch}) for $W_\text{sw}$, which is only valid for identical units, 
would yield a wrong result. Considering the interval starting in $n\delta t$ and lasting $\delta t$, we find that 
\begin{align}
 \delta t\dot W_\text{sw}	=&~	\mbox{tr}_{SU_{n+1}}\left\{A F \rho_S(n\delta t+\delta t)\rho_{U_{n+1}}(n\delta t+\delta t)\right\}	\nonumber	\\
				&-	\mbox{tr}_{SU_n}\left\{A F \rho_{SU_n}(n\delta t+\delta t)\right\}.
\end{align}
This quantity represents the work required to switch on the interaction for the next unit state  
$\rho_{U_{n+1}}(n\delta t+\delta t)$ minus the work required to switch off the interaction for the actual unit state 
$\rho_{U_{n}}(n\delta t+\delta t)$ at the end of the interval. To evaluate it, we use Eq.~(\ref{eq help 28}) and 
\begin{align}
 \rho_{SU_n}(n\delta t+\delta t)	&=	\rho_S(n\delta t)\rho_{U_n}(n\delta t)	\\
					&-	i\delta t[H_S + H_U + A F,\rho_S(n\delta t)\rho_{U_n}(n\delta t)]	\nonumber	\\
					&-	i\delta t\mbox{tr}_R\{[V_{SR},\rho_{SR}(n\delta t)]\}\rho_{U_n}(n\delta t),	\nonumber	\\
 \rho_{S}(n\delta t+\delta t)		&=	\rho_S(n\delta t)	\\
					&-	i\delta t[H_S + f(n\delta t)A,\rho_S(n\delta t)]	\nonumber	\\
					&-	i\delta t\mbox{tr}_R\{[V_{SR},\rho_{SR}(n\delta t)]\},	\nonumber
\end{align}
which follow from the Liouville-von Neumann equation for a time step $\delta t$. 
After a straightforward but tedious calculation, we arrive at (to first order in $\delta t$) 
\begin{equation}
 \begin{split}
  \dot W_\text{sw} = &~	f'(n\delta t) \langle A\rangle_S(n\delta t) \\
		     &+	i\langle A\rangle_S(n\delta t) \mbox{tr}_{U_n}\{[F,H_U]\rho_{U_n}(n\delta t)\}.
 \end{split}
\end{equation}
Here, we introduced the discretized derivative of $f(t)$ as 
\begin{equation}
 f'(n\delta t) \equiv \frac{f(n\delta t+\delta t) - f(n\delta t)}{\delta t}\,,
\end{equation}
which is well-behaved because we assumed $f(t)$ to be differentiable. 
Since the units act as a pure work reservoir, the total rate of work done on the system can be defined as 
\begin{equation}\label{eq work total driving}
 \dot W \equiv \dot W_\text{sw} - d_t E_{U_n} = f'(n\delta t) \langle A\rangle_S(n\delta t).
\end{equation}

Finally, let us compare this result with our general treatment from Sec.~\ref{sec two interacting systems} 
where we found out that for a time dependent Hamiltonian of the form~(\ref{eq help 29}), the rate of work 
done on the system is [compare with Eq.~(\ref{eq 1st law isolated})] 
\begin{equation}
 \dot W(t) = \mbox{tr}_{SR}\left\{\rho_{SR}(t)d_tH_\text{tot}(t)\right\} = \frac{df(t)}{dt}\mbox{tr}_S\{A\rho_S(t)\}.
\end{equation}
This expression exactly coincides for $t=n\delta t$ with Eq.~(\ref{eq work total driving}). 
This confirms that there is a clean (but somewhat artificial) way to simulate a driven system 
by a stream of units and to justify that the driving corresponds to a pure work source.

\section{Applications}
\label{sec applications part II}

In this section we demonstrate the use of the repeated interaction framework that we have developed in the 
previous sections by considering specific examples. We mostly emphasize the way in which the setups can be 
analyzed from our thermodynamic perspective and refer to the original literature for more details. 
Many more setups such as, e.g. the measurement and feedback scheme realized in 
Ref.~\cite{SayrinEtAlNature2011}, the squeezed reservoirs of Ref.~\cite{ManzanoEtAlPRE2016}, 
{the coherent states of Ref.~\cite{LiEtAlPRE2014}} or the entangled unit 
states of Ref.~\cite{DillenschneiderLutzEPL2009}\footnote{In this case each unit in our setup would actually correspond 
to two subunits which are both pairwise entangled with each-other, but interact with the system sequentially.} 
can be analyzed within our framework, but will not be considered here for brevity.

\subsection{Information reservoir: the Mandal-Jarzynski engine}
\label{sec appendix MJ engine}

We start by providing a microscopic model describing the information-driven engine first proposed by Mandal and 
Jarzynski~\cite{MandalJarzynskiPNAS2012} and show that it falls within the class of \emph{information reservoirs} 
considered in Sec.~\ref{sec info reservoir}. 

The system is modeled as a three-level system with Hamiltonian 
\begin{equation}
 H_S = \epsilon_S(|A\rl A| + |B\rl B| + |C\rl C|)
\end{equation}
and the unit by a two-level system (``bit'') with Hamiltonian 
\begin{equation}
 H_U = \epsilon_U(|0\rl0| + |1\rl1|).
\end{equation}
Thus, the bare system and unit Hamiltonians are completely degenerate. 
The system-unit interaction is switched on and off at the beginning and end of the full interval 
of length $\tau$, such that the time interval between the unit interactions is infinitely short (i.e. $\tau'\rightarrow\tau$).
During the interaction, the degeneracy of the system and unit states is lifted via 
\begin{equation}
 V_{SU} = \frac{\Delta w}{\epsilon_S}H_S|1\rl1|\,,
\end{equation}
such that the energy of the system-unit states $\ket{A1}$, $\ket{B1}$, $\ket{C1}$ becomes 
$\epsilon_S + \epsilon_U + \Delta w$, and the energy of the system-unit states $\ket{A0}$, $\ket{B0}$, $\ket{C0}$ is 
$\epsilon_S + \epsilon_U$.

The model is completed by adding a weakly coupled reservoir and by assuming that it induces thermally activated 
transitions between the following levels: $A0\leftrightarrow B0$, $B0\leftrightarrow C0$, $C0\leftrightarrow A1$, 
$A1\leftrightarrow B1$ and $B1\leftrightarrow C1$. For other possible physical setups see 
Refs.~\cite{LuMandalJarzynskiPT2014, StrasbergEtAlPRE2014}. 

Note that, although we have written down the model in a quantum mechanical way, the 
model in Ref.~\cite{MandalJarzynskiPNAS2012} is purely classical. In this spirit we neglect any subtleties arising 
from deriving a ME for degenerate quantum systems and use a classical rate equation where those levels are connected 
that interact with the reservoir as specified above. 
Then, the ME takes on the form $d_t\bb p(t) = \C R\bb p$ with (in suitable units)~\cite{MandalJarzynskiPNAS2012} 
\begin{align}
 \bb p(t) &= \left(\begin{array}{c}
                   p_{A0}(t) \\ p_{B0}(t) \\ p_{C0}(t) \\ p_{A1}(t) \\ p_{B1}(t) \\ p_{C1}(t) \\
                  \end{array}\right),	\\
 \C R &= \left(\begin{array}{cccccc}
               -1	&	1	&	0		&	0		&	0	&	0	\\
               1	&	-2	&	1		&	0		&	0	&	0	\\
               0	&	1	&	-2+\epsilon	&	1+\epsilon	&	0	&	0	\\
               0	&	0	&	1-\epsilon	&	-2-\epsilon	&	1	&	0	\\
               0	&	0	&	0		&	1		&	-2	&	1	\\
               0	&	0	&	0		&	0		&	1	&	-1	\\
              \end{array}\right).
\end{align}
The parameter $\epsilon\in(-1,1)$ is related to $\Delta w$ by local detailed balance 
[see Eq.~(\ref{eq local detailed balance 2})] via $-\beta\Delta w = \ln\frac{1-\epsilon}{1+\epsilon}$. 
The initial state of the incoming units is given by $\rho_U(0) = \frac{1+\delta}{2}|0\rl0| + \frac{1-\delta}{2}|1\rl1|$ with $\delta\in[-1,1]$.
The stationary solution for the system is obtained by solving this rate equation for a time $\tau$ with an initial 
condition $\rho_U(0)$ for the units and $\rho_S(0) = \rho_S(\tau)$ for the system. 
As demonstrated in Ref.~\cite{MandalJarzynskiPNAS2012}, it can be solved exactly. 

Our thermodynamic analysis from Sec.~\ref{sec generalized thermodynamic framework} now tells us immediately that 
$\Delta E_S = 0$ and $\Delta E_U = 0$ due to the degeneracy of the bare system and unit Hamiltonian. Since 
$\Delta E_U = 0$ it also follows from Eq.~(\ref{eq ideal info reservoir}) that the stream of bits constitutes an 
information reservoir unless $\Delta S_U = 0$, too. Furthermore, because we have no explicit driving, we also have 
$W_X = 0$ [see Eq.~(\ref{eq def work X})], and when the system has reached its steady state, we will also have 
$\Delta S_S = 0$. The first law~(\ref{eq 1st law modified}) then tells us that 
\begin{equation}
 Q = -W_\text{sw}.
\end{equation}
In fact, Mandal and Jarzynski imagined a little load attached to the system such that the heat absorbed from the 
system can be identified as work. Within our setup we indeed see that this ``attaching of the load'', i.e. 
the switching on and off of the system-unit interaction, is responsible for providing that energy and has a clear 
microscopic interpretation in our framework. From Eq.~(\ref{eq def work switch}) we can immediately compute 
\begin{equation}
 W_\text{sw} = -\Delta w(\langle 1|\rho_U(\tau)|1\rangle - \langle1|\rho_U(0)|1\rangle)\,.
\end{equation}
This term exactly equals (minus) the work identified in 
Ref.~\cite{MandalJarzynskiPNAS2012}. Finally, in accordance with our second 
law~(\ref{eq 2nd law info reservoir}) it was also shown in~\cite{MandalJarzynskiPNAS2012} that 
\begin{equation}
 \beta Q = -\beta W_\text{sw} \le \Delta S_U.
\end{equation}

\subsection{Work reservoir: the micromaser}
\label{sec micromaser}

The micromaser is historically, experimentally, and theoretically important and its operation is based on  
repeated interactions. Beyond quantum optics it has also been used, e.g.  as a model system for transport in 
superconducting single-electron transistors coupled to quantum resonators~\cite{RodriguesImbersArmourPRL2007}, where it 
displays an intriguing dynamics such as self-sustained oscillations and transitions to multistable behaviour. 
We now briefly elaborate on the fact that in its simplest version, the micromaser can be viewed as a system interacting 
with a stream of units which operate as a \emph{work reservoir}. Our approach will be qualitative since the detailed 
calculations can be found, e.g.  in Refs.~\cite{FilipowiczJavanainenMeystrePRA1986, MilburnPRA1987, ScullyZubairyBook1997}. 

\begin{figure}
 \centering\includegraphics[width=0.40\textwidth,clip=true]{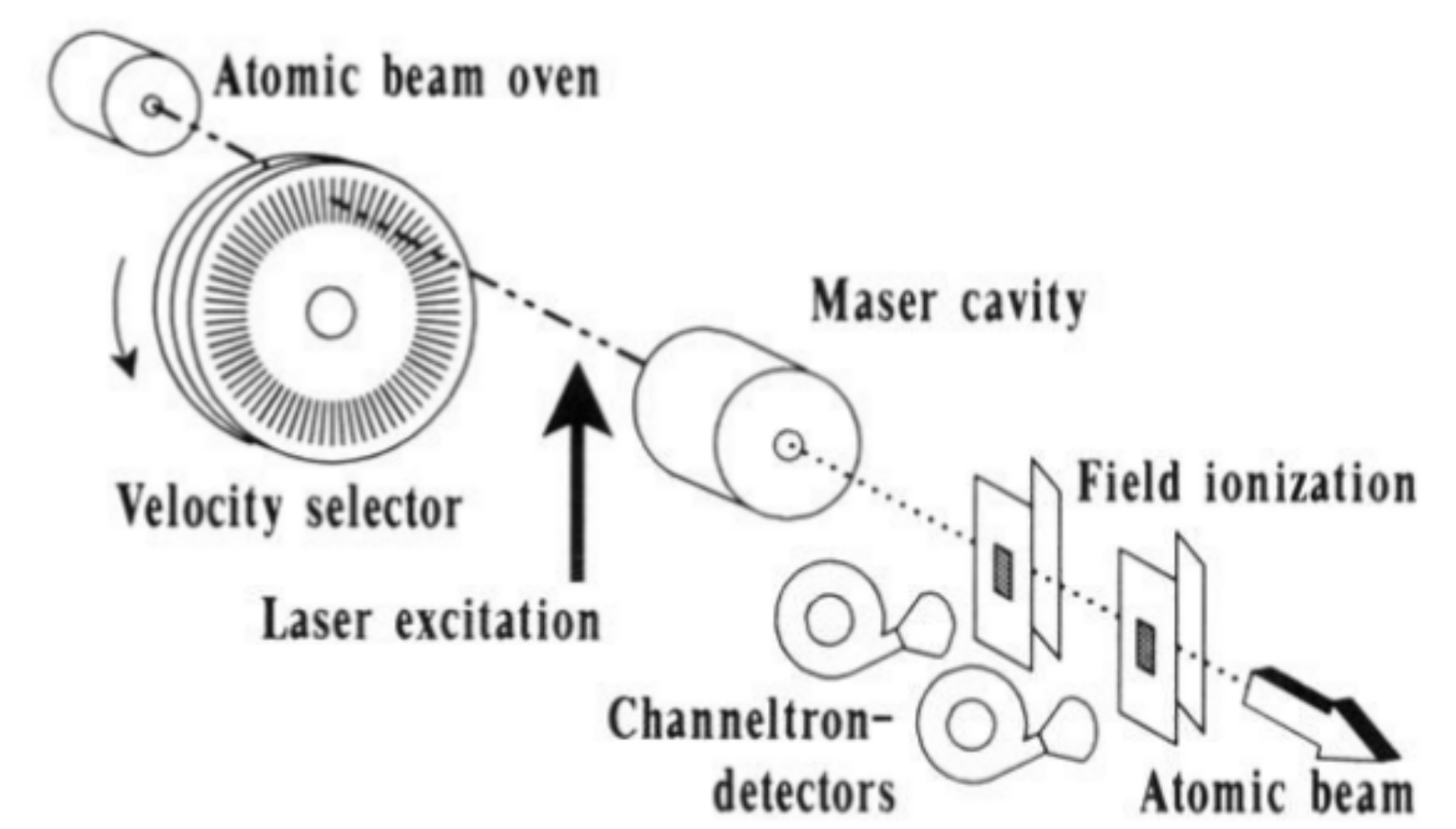}
 \label{fig micromaser} 
 \caption{Sketch of a micromaser setup. The units are produced in an atomic beam oven and are initialized with a 
 velocity selector and laser excitation. They then interact with a microwave (``maser'') cavity (the system) before they 
 are finally detected and their state is read out. Figure taken from Ref.~\cite{RempeSchmidtKalerWaltherPRL1990} 
 reporting on experimental observation of sub-Poissonian photon statistics in a micromaser. }
\end{figure}

The system $S$ in this case consists of a high quality cavity supporting a microwave field with Hamiltonian 
$H_S = \Omega a^\dagger a$ where $a^\dagger$ and $a$ are bosonic creation and annihilation operators and $\Omega$ denotes 
the frequency of the cavity. The microwave mode is coupled to an external reservoir of electromagnetic field modes 
at equilibrium with temperature $\beta^{-1}$ and ``high quality'' means that the coupling is very weak, 
especially such that it is negligible on timescales when the system interacts with an atom flying through the cavity. 
This atom corresponds to a unit $U$ and can be conveniently modeled as a two-level system (TLS) 
$H_U = \frac{\Delta}{2}(|1\rl1| - |0\rl0|)$ with energy gap $\Delta\approx\Omega$ (on resonance with the cavity). 
Depending on the experimental details the units can be prepared in different states and might arrive at the cavity at 
regular intervals or Poisson distributed, see Fig.~\ref{fig micromaser} for a sketch of the setup and 
Ref.~\cite{SayrinEtAlNature2011} for a recent experiment even involving measurement and feedback. 

In the standard setup the TLSs are prepared in a statistical mixture of excited and ground states and interact 
with the cavity during a short time (compared to the cavity lifetime) via a Jaynes-Cummings 
interaction Hamiltonian $V_{SU} = g(|0\rl1|a^\dagger + |1\rl0|a)$ causing the atom to emit or absorb 
photons~\cite{ScullyZubairyBook1997}. By tuning the interaction time (or the interaction strength $g$ respectively) 
correctly, one can make sure that an atom initially in the excited (ground) state will always emit (absorb) a cavity 
photon. In essence, the effect of the cavity is therefore to swap the population of the TLSs. 

This implies a difference in energy between incoming and outgoing TLSs ($\Delta E_U \neq0$), 
but not in entropy because entropy is invariant under exchange of state labels ($\Delta S_U=0$). 
Hence, the stream of atoms acts as a pure source of work which builds up a photon field inside the cavity. 
However, because the cavity is weakly coupled to an outside thermal reservoir, it constantly looses photons, too. 
To achieve a steady state occupation of the cavity \emph{above} the thermal average, the incoming TLSs must be in a 
population-inverted state, i.e. have a higher probability to be excited then in the ground state. 
More details, such as the exact threshold condition for a buildup of the cavity field, are given 
in Refs.~\cite{FilipowiczJavanainenMeystrePRA1986, MilburnPRA1987, ScullyZubairyBook1997}. 

\subsection{Quantum coherence as a resource: lasing without inversion}
\label{sec lasing without inversion}

As explained in the previous section, the thermodynamic working principle (but not the physical origin) 
of the micromaser can be understood within classical physics. 
However, it is also possible to populate a cavity above its thermal distribution by using a stream of atoms 
(units) \emph{without} population inversion. This phenomenon is known as lasing without 
inversion~\cite{ScullyZubairyBook1997} and results from a destructive interference of the photon absorption 
process due to a coherent superposition of the energy levels of the incoming unit. 
Thus, lasing without inversion is a pure quantum effect. 

The idea to use quantum coherence via lasing without inversion in order to extract more work from a heat 
engine than classically possible was proposed by Scully \emph{et al.} in Ref.~\cite{ScullyEtAlScience2003}, 
see also Refs.~\cite{ScullyPRL2010, ScullyEtAlPNAS2011} for similar models and Ref.~\cite{KorzekwaEtAlNJP2016} 
for a resource theory formulation of the problem. We now briefly sketch how to treat lasing without 
inversion~\cite{ScullyZubairyBook1997} and how to rederive the results from Ref.~\cite{ScullyEtAlScience2003} 
from our results of Sec.~\ref{sec perturbative expansion delta kicks}. 

The system we are considering is the same as for the micromaser (i.e. a single cavity with frequency $\Omega$). 
The units are three-level systems described by  
\begin{equation}
 H_U = E_a|a\rl a| + E_b |b\rl b| + E_c |c\rl c|.
\end{equation}
We assume a so-called ``$\Lambda$-configuration''~\cite{ScullyZubairyBook1997} where the two states $|b\rangle$ and 
$|c\rangle$ are nearly degenerate ($E_b\approx E_c$) and well separated from the excited state $|a\rangle$, which is 
nearly resonant with the cavity (i.e. $\Omega \approx E_a-E_b \approx E_a-E_c$). The initial state of the units is modeled as 
a statistical mixture between the energy eigenstates with an additional coherence allowed 
between the near-degenerate levels $|b\rangle$ and $|c\rangle$. 
Following the notation of Ref.~\cite{ScullyEtAlScience2003}, we use 
\begin{equation}
 \begin{split}\label{eq unit state LWI}
  \rho_U	=&~	P_a|a\rl a| + P_b |b\rl b| + P_c|c\rl c|	\\
		&+	\rho_{bc}|b\rl c| + \rho_{cb}|c\rl b|\,,
 \end{split}
\end{equation}
with $P_a+P_b+P_c = 1$, $P_{a,b,c}\ge0$ and $\rho_{bc}=\rho_{cb}^*\in\mathbb{C}$. 
Beyond that, the positive-definiteness of $\rho_U$ requires $\rho_{bc} \rho_{cb} \le P_b P_c$. 
Finally, the interaction between system and unit is modeled by a generalized Jaynes-Cummings Hamiltonian 
\begin{equation}
 \tilde V = -g\left[a(|a\rl b| + |a\rl c|) + a^\dagger(|b\rl a| + |c\rl a|)\right]\,,
\end{equation}
where we assumed that the direct transition between $|b\rangle$ and $|c\rangle$ is dipole-forbidden~\cite{ScullyZubairyBook1997} 
and assumed an interaction scenario as in Sec.~\ref{sec perturbative expansion delta kicks setup}. 
We note that in the interaction picture with respect to $H_S + H_U$ we have 
\begin{equation}
 \tilde V_\text{int}(t) \equiv e^{i(H_S+H_U)t} V_{SU} e^{-i(H_S+H_U)t} \approx \tilde V
\end{equation}
due to the resonance condition. 

We now assume a cavity of very high quality neglecting any dissipation due to the electromagnetic reservoir 
such that the system interacts with many atoms coherently. This corresponds to the stage $1\rightarrow2$ in Fig.~2 of 
Ref.~\cite{ScullyEtAlScience2003}. Using 
Eq.~(\ref{eq unit state LWI}), it is easy to confirm that $\mbox{tr}_U\{\tilde V\rho_U\} = 0$. Then, following 
Sec.~\ref{sec perturbative expansion delta kicks} we see that Eq.~(\ref{eq ME perturbative delta kicks 2nd order}) 
requires us to compute sixteen correlations functions $\langle B_lB_k\rangle_U$ where we identify $B_1 \equiv -g|a\rl b|$, 
$B_2 \equiv -g|a\rl c|$, $B_3 \equiv -g|b\rl a|$ and $B_4 \equiv -g|c\rl a|$. Only six are non-zero and 
the ME~(\ref{eq ME perturbative delta kicks 2nd order}) in the interaction picture becomes 
\begin{equation}
 \begin{split}
  & d_t\rho_S(t) = 	\\
  & \gamma_\text{eff}\left\{2P_a\C D[a^\dagger] + (P_b + P_c + \rho_{bc} + \rho_{cb})\C D[a]\right\}\rho_S(t)
 \end{split}
\end{equation}
with the dissipator $\C D$ defined below Eq.~(\ref{eq help 21}) and some effective and for our purposes unimportant rate 
$\gamma_\text{eff} > 0$. 
The Lindblad form is ensured by the non-negativity of the unit density matrix which implies 
$(P_b + P_c + \rho_{bc} + \rho_{cb}) \ge 0$.
Note that $\C D[a^\dagger]$ describes the absorption and $\C D[a]$ the emission of a cavity photon. 

If the unit is initially in a thermal state with occupations $P_a = e^{-\beta\Omega/2}/Z$ and $P_b = P_c = e^{\beta\Omega/2}/Z$ 
and without coherences, $\rho_{bc} = \rho_{cb} = 0$, the rates for emission and absorption satisfy local detailed balance and 
the equilibrium state $\rho_\beta^S$ is the steady state of the ME. 

In contrast, with coherences $\rho_{bc} + \rho_{cb} = 2\Re(\rho_{bc})$ (which can be positive as well as negative) the rate 
of absorption can be decreased or increased. If $2\Re(\rho_{bc}) < 0$, the rate of photon absorption by the units 
is lowered and lasing without inversion becomes possible. Note that, typically, such coherences in three-level lambda 
systems are created by two coherent light fields via stimulated Raman processes, i.e. time-dependent external lasers 
which thus ultimately provide a resource of energy.

To show the equivalence of these results with those of Ref.~\cite{ScullyEtAlScience2003}, we compute the 
evolution of the mean photon occupation of the cavity $N(t) \equiv \mbox{tr}_S\{a^\dagger a\rho_S(t)\}$. 
Using $[a,a^\dagger] = 1$ and the property $\mbox{tr}\{ABC\} = \mbox{tr}\{CAB\}$, we obtain  
\begin{align}
 & d_tN(t) =	\\
 & \gamma_\text{eff}\left\{2P_a[1+N(t)] - (P_b + P_c + \rho_{bc} + \rho_{cb})N(t)\right\}\,,	\nonumber
\end{align}
which is identical to Eq.~(5) in Ref.~\cite{ScullyEtAlScience2003} (after replacing $N(t)$ with $n_\phi$ and 
$\gamma_\text{eff}$ with $\alpha$) from which the results in Ref.~\cite{ScullyEtAlScience2003} are derived. 
The steady state occupation reads 
\begin{equation}
 \begin{split}\label{eq population inversion LWI}
  N_\text{eff}	&\equiv	\lim_{t\rightarrow\infty} N(t) = \frac{2P_a}{P_b+P_c-2P_a + 2\Re(\rho_{bc})}	\\
					&=	\left(e^{\beta\Omega}-1 + e^{\beta\Omega/2}Z\Re(\rho_{bc})\right)^{-1}\,.
 \end{split}
\end{equation}
For zero coherence it corresponds to the (equilibrium) Bose distribution. But for finite coherence, the cavity population 
can be lowered or raised. This means that the non-equilibrium free energy of the incoming atoms has been converted into 
a non-equilibrium free energy for the cavity. This feature alone does not yet yield a positive work output, 
but a thermodynamic cycle which does so is presented in Ref.~\cite{ScullyEtAlScience2003}. 
It basically relies on the fact that the population $N(t)$ of the cavity is related to a radiation pressure $P$ 
via $P \sim N(t)$, which can be used to drive a piston. Then, by putting the cavity first in contact with the stream 
of atoms populating it according to some effective temperature $T_\text{eff} > T$ [which can be inferred from 
Eq.~(\ref{eq population inversion LWI})] and afterwards with a standard heat reservoir at temperature $T$, we can extract 
work proportional to $N_{T_\text{eff}} - N_T$ where $N_T$ denotes the Bose distribution. 

The idea behind lasing without inversion thus provides an example of how our framework can account for coherences. 
Following Ref.~\cite{ScullyEtAlScience2003}, we have shown that these latter can be used as a thermodynamic resource. 
During the first part of the cycle as described above, the units do not correspond to any limiting classification scheme 
introduced in Sec.~\ref{sec applications part I} because the initial state of the units is not thermal and in general the 
energy as well as the entropy of the units will change, whereas during the rest of the cycle (where the systems expands 
back to an equilibrium distribution) there are no units interacting with the system.

\subsection{Measurement and feedback: electronic Maxwell demon}
\label{sec electronic Maxwell demon}

In the traditional thought-experiment of Maxwell, the demon shuffles gas particles from a cold to a hot 
reservoir with negligible consumption of energy~\cite{LeffRexBook2003, MaruyamaNoriVedralRMP2009}. 
In an isothermal setup, a similar violation of the traditional second law appears if a feedback mechanism 
shuffles particles from a reservoir with low chemical potential to a reservoir with high chemical potential. 
This is the central idea of the electronic Maxwell demon, which has been theoretically well studied for a number 
of different models~\cite{Datta2008, SchallerEtAlPRB2011, AverinMottonenPekolaPRB2011, KishGranqvistPloS12012, 
EspositoSchallerEPL2012, StrasbergEtAlPRL2013, BergliGalperinKopninPRE2013, StrasbergEtAlPRE2014, KutvonenEtAlSciRep2015}. 
The setup proposed in Ref.~\cite{StrasbergEtAlPRL2013} was recently experimentally realized in Ref.~\cite{KoskiEtAlPRL2015}. 
Below, we revisit one particular electronic Maxwell demon. 

\begin{figure}
 \centering\includegraphics[width=0.48\textwidth,clip=true]{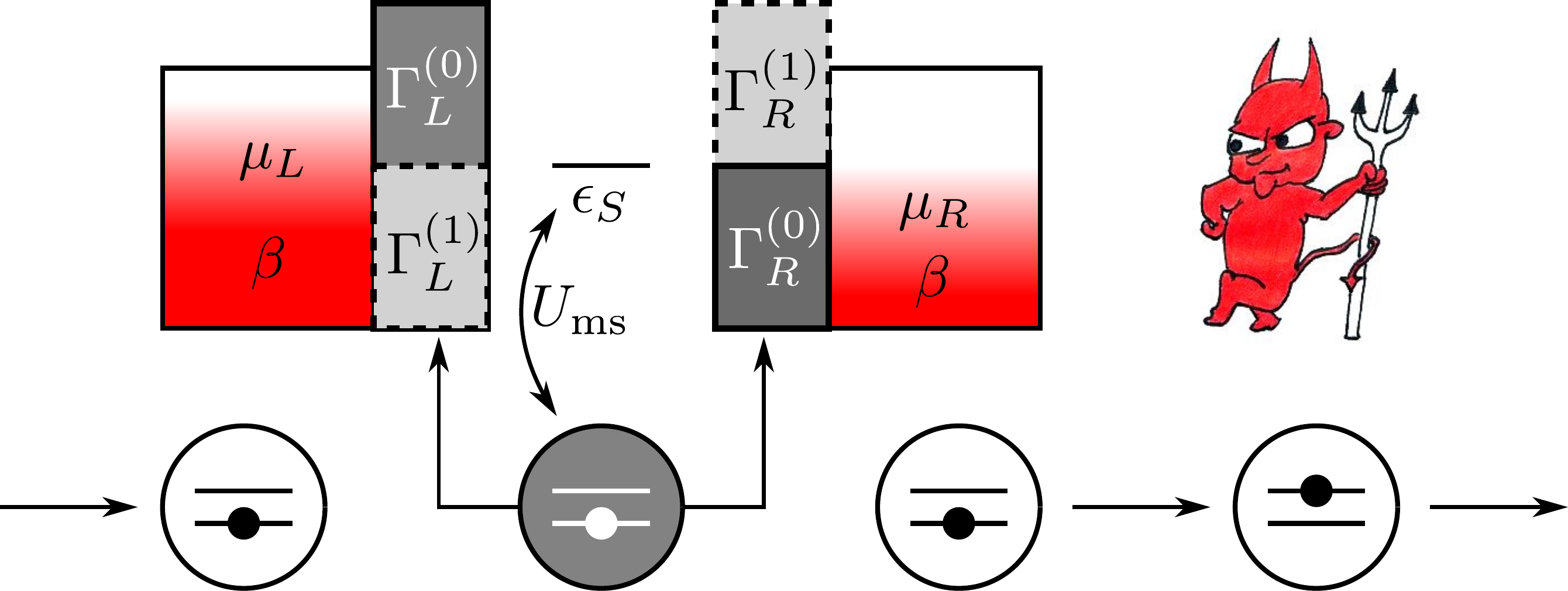}
 \label{fig electronic MD} 
 \caption{(Color Online) Sketch of the electronic Maxwell demon. The system consists of a single-level quantum dot 
 tunnel-coupled to two electronic reservoirs with chemical potentials $\mu_L \ge \mu_R$ and inverse temperature $\beta$. 
 The demon mechanism is implemented by a stream of units which monitors the state of the dot and depending on its state, 
 changes the tunneling barriers $\Gamma_\nu$ to make electronic transport against the bias possible. In absence of the demon 
 mechanism, the tunneling barriers would not depend on the state of the dot and electrons would flow from left to right. 
 The picture of the demon was provided by courtesy of Ania Brzezinska. }
\end{figure}

The system to be controlled is a conventional single-electron transistor (SET), 
which consists of a single-level quantum dot connected to two thermal reservoirs with 
chemical potential $\mu_\nu$ ($\nu\in\{L,R\}$) at the same inverse temperature $\beta$. 
The quantum dot can either be filled with an electron of energy $\epsilon_S$ or empty (corresponding to a zero energy state). 
A sketch of the setup (with the feedback mechanism described below) is shown in Fig.~\ref{fig electronic MD}. 
The ME governing the time evolution of the system in absence of feedback is 
\begin{equation}
 d_t\rho_S(t) = \sum_\nu \C L_\beta^{(\nu)}(\Gamma_\nu)\rho_S(t),
\end{equation}
where the thermal generators are defined as 
\begin{equation}
 \C L_\beta^{(\nu)}(\Gamma_\nu) \equiv \Gamma_\nu\left\{(1-f_\nu)\C D[|E\rangle_S\langle F|] + f_\nu\C D[|F\rangle_S\langle E|]\right\}.
\end{equation}
Here, $f_\nu = (e^{\beta(\epsilon_S-\mu_\nu)}+1)^{-1}$ is the Fermi function evaluated at the energy of the quantum 
dot, and $\Gamma_\nu \ge 0$ is a bare rate which depends on the details of the microscopic coupling Hamiltonian. 
Furthermore, $|E\rangle_S$ ($|F\rangle_S$) denotes the empty (filled) state of the dot and the dissipator 
$\C D$ is defined in the same way as below Eq.~(\ref{eq help 21}). 

We now would like to engineer a demon mechanism operating a feedback control on the system which modulates the energy 
barriers of the dot (i.e. the bare rates $\Gamma_\nu$) depending on the dot state as sketched in Fig.~\ref{fig electronic MD}. 
The phenomenological description of this mechanism was done in Ref.~\cite{SchallerEtAlPRB2011} 
and its thermodynamical analysis was performed in Ref.~\cite{EspositoSchallerEPL2012}.
A physical mechanism autonomously implementing this feedback was proposed Ref.~\cite{StrasbergEtAlPRL2013}.
It relies on a capacitive coupling to another single level quantum dot at a different temperature. 
This mechanism was further analyzed in Refs.~\cite{HorowitzEspositoPRX2014, KutvonenEtAlSciRep2015} and will be also 
used below for comparison.
We now propose a different mechanism implementing the same feedback on the system. 
As sketched in Fig.~\ref{fig electronic MD}, this one is based on repeated interactions 
with a stream of units consisting of two-level systems prepared in the state 
$\rho_U = (1-\epsilon)\Pi_0 + \epsilon\Pi_1$, where $\Pi_i = |i\rangle_U\langle i|$ is the projector 
on the state $i\in\{0,1\}$ and $\epsilon\in[0,1]$ is a free parameter quantifying the measurement error. 

At the beginning of the interaction interval, we assume that the interaction produces an instantaneous unitary 
operation $U_\text{ms} = \Pi_E\otimes \mathbf{1}_U + \Pi_F\otimes\sigma^x_U$ where $\Pi_E = |E\rangle_S\langle E|$ 
and $\Pi_F = |F\rangle_S\langle F|$ and $\sigma_U^x = |0\rangle_U\langle 1| + |1\rangle_U\langle 0|$, which 
can be interpreted as a measurement. 
Indeed, considering an initial system state of the form $\rho_S(t) = p_E(t)\Pi_E + p_F(t)\Pi_F$ where $p_E(t)$ [$p_F(t)$] 
is the probability to find the system in the empty [filled] state, the post measurement state of the system and unit reads 
\begin{equation}
 \begin{split}
  \rho_{SU}^\text{ms}(t)	&=	(1-\epsilon)p_E(t)\Pi_E\otimes\Pi_0 + \epsilon p_E(t)\Pi_E\otimes\Pi_1	\\
			&+	(1-\epsilon)p_F(t)\Pi_F\otimes\Pi_1 + \epsilon p_F(t)\Pi_F\otimes\Pi_0.
 \end{split}
\end{equation}
Note that $\epsilon = 0$ corresponds to a perfect measurement in which the state of the unit after the measurement is $0$ ($1$) 
if and only if the state of the system is $E$ ($F$). The reduced state of the system is always given by 
$\rho_S^\text{ms}(t) = \mbox{tr}_U\{\rho_{SU}^\text{ms}(t)\} = \rho_S(t)$, which means that the measurement 
\emph{does not disturb the system}. 
This feature circumvents the difficulty mentioned in Sec.~\ref{sec perturbative expansion delta kicks setup} and allows 
us to consider a continuously measured system in which $\delta t$ (the waiting time between two units) can be arbitrarily 
small despite the fact that the system is still interacting with its reservoir. 

Next, in spirit of Sec.~\ref{sec feedback control} and Eq.~(\ref{eq Hamiltonian fb}), we postulate a Hamiltonian of the 
form $H_\text{fb} = \epsilon_S|F\rangle_S\langle F| + \Pi_0 H_{SR}^{(0)} + \Pi_1 H_{SR}^{(1)} + H_R$ which acts during 
the remaining interaction time $\delta t$ and changes the system reservoir coupling $H_{SR}^{(i)}$ according to the 
state $i$ of the unit. Assuming that we can treat $H_{SR}^{(i)}$ as a weak perturbation, we are effectively changing the 
tunneling rates from $\Gamma_\nu$ to $\Gamma_\nu^{(i)}$ and the time evolution of the system and unit is then given by 
\begin{equation}
 \rho_{SU}(t+\delta t) = \sum_{i\in\{0,1\}} \exp\left\{\C L_\beta^{(\nu)}(\Gamma_\nu^{(i)})\delta t\right\}\Pi_i\rho_{SU}^\text{ms}(t)\Pi_i.
\end{equation}
Tracing over the unit space and expanding this equation to first order in $\delta t$ (assuming that $\delta t$ is 
sufficiently small) yields an effective evolution equation for the system of the form 
$d_t\rho_S(t) = \C L_\text{eff}\rho_S(t)$ with 
\begin{align}
 \C L_\text{eff}	\equiv&~		\sum_\nu\C L_\text{eff}^{(\nu)}	\label{eq ME effective electronic MD}	\\
		=&~		\sum_\nu[(1-\epsilon)\Gamma_\nu^{(1)} + \epsilon\Gamma_\nu^{(0)}](1-f_\nu)\C D[|E\rangle_S\langle F|]	\nonumber	\\
		&+		\sum_\nu[(1-\epsilon)\Gamma_\nu^{(0)} + \epsilon\Gamma_\nu^{(1)}]f_\nu\C D[|F\rangle_S\langle E|].	\nonumber
\end{align}
This equation is identical to the ME obtained for the system when it is subjected to the capacitive demon 
mechanism considered in Ref.~\cite{StrasbergEtAlPRL2013} which results from coarse-graining the demon dot and only 
retaining the SET degrees of freedom. 
In the error-free case ($\epsilon = 0$), it also reduces to the effective ME of Ref.~\cite{SchallerEtAlPRB2011}, 
but the above procedure constitutes an elegant way to generalize arbitrary piecewise-constant 
feedback schemes~\cite{SchallerPRA2012} to finite detection errors. 

We now turn to the thermodynamic analysis of our new demon mechanism. 
First of all, we can assume that the unit Hamiltonian $H_U \sim \mathbf{1}_U$ is fully degenerate. 
This implies $d_t E_U = 0$ at all times. 
Then, during the measurement step, the system and unit correlate such that [see Eq.~(\ref{eq 2nd law unitary meas})] 
$I_{S:U}^\text{ms} = d_t S^\text{ms}_U\delta t$, where we used the fact that $d_t S^\text{ms}_S = 0$ 
since the system density matrix is left unchanged by the measurement. 
This correlation can then be exploited during the feedback step. 
The second law for feedback (\ref{eq 2nd law fb interval}) in our situation reads in a differential form as 
\begin{equation}\label{eq 2nd law electronic MD previous}
 \dot\Sigma_S^\text{fb} = \beta(\mu_L-\mu_R) I_L + \frac{I_{S:U}^\text{ms}}{\delta t} \ge \frac{I^\text{fb}_{S:U}}{\delta t} \ge 0,
\end{equation}
where for simplicity we assumed that the system operates at steady state $d_t S_S^\text{fb} = 0$ and where 
we used $d_t S_U^\text{fb} = 0$ because the entropy of the unit does not change during the feedback step. 
Furthermore, the entropy flow reads $-\beta\dot Q^{(L)} - \beta\dot Q^{(R)} = \beta(\mu_L-\mu_R)I_L$, 
where $I_L$ is the matter current which entered the system from the left reservoir. 
Finally, $I^\text{fb}_{S:U}$ quantifies the remaining system-unit correlations after the feedback step. 

In spirit of Eq.~(\ref{eq system dissipation}) it will turn out to be useful to include the final correlation 
$I^\text{fb}_{S:U}$ in the second law and to define 
\begin{equation}\label{eq 2nd law electronic MD}
 \dot\Sigma = \beta(\mu_L-\mu_R)I_L + \C I \ge 0, ~~~ \C I \equiv \frac{I_{S:U}^\text{ms} - I^\text{fb}_{S:U}}{\delta t}
\end{equation}
Here, the newly defined quantity $\C I$ is the rate with which we use up the correlations established 
during the measurement. $\C I$ had not yet been considered in previous works on this system.  
In fact, the information current in Ref.~\cite{EspositoSchallerEPL2012} is purely phenomenological in nature, whereas 
the information flow in Ref.~\cite{HorowitzEspositoPRX2014} describes the same quantity but in a bipartite setting. 
Note that while both terms  $\frac{I_{S:U}^\text{ms}}{\delta t}$ and $\frac{I^\text{fb}_{S:U}}{\delta t}$ 
would diverge when $\delta t \to 0$, $\C I$ in general remains finite which motivates the use of 
Eq.~(\ref{eq 2nd law electronic MD}) instead of Eq.~(\ref{eq 2nd law electronic MD previous}). 
More specifically, we can compute 
 \begin{align}
  \C I	=&~	(1-\epsilon)\sum_\nu\left\{\Gamma_\nu^{(0)}f_\nu p_E + \Gamma_\nu^{(1)}(1-f_\nu) p_F\right\} \ln\frac{1-\epsilon}{\epsilon} \nonumber	\\
	&-	\epsilon\sum_\nu\left\{\Gamma_\nu^{(0)}(1-f_\nu)p_F + \Gamma_\nu^{(1)}f_\nu p_E\right\} \ln\frac{1-\epsilon}{\epsilon}	\nonumber \\
	=&~	2\Gamma(1-2\epsilon)\mbox{arctan}(1-2\epsilon)	\\
	&\times	\left(\cosh\delta - (1-2\epsilon)\sinh\delta\tanh\frac{\beta V}{4}\right).	\nonumber
 \end{align}
For the last equality we used the steady state solution of $p_E$ and $p_F$ and parametrized the rates as 
$\Gamma_L^{(0)} \equiv \Gamma_R^{(1)} \equiv \Gamma e^{-\delta}$ and $\Gamma_L^{(1)} \equiv \Gamma_R^{(0)} \equiv \Gamma e^{\delta}$,
such that $\delta \in \mathbb{R}$ characterizes a feedback strength.
Note that $\C I$ diverges for $\epsilon\rightarrow0$. This makes sense because if we monitor the quantum dot 
in an error-free way, we can also extract an infinite amount of work by letting $\delta\rightarrow\infty$. 

\begin{figure}
 \centering\includegraphics[width=0.48\textwidth,clip=true]{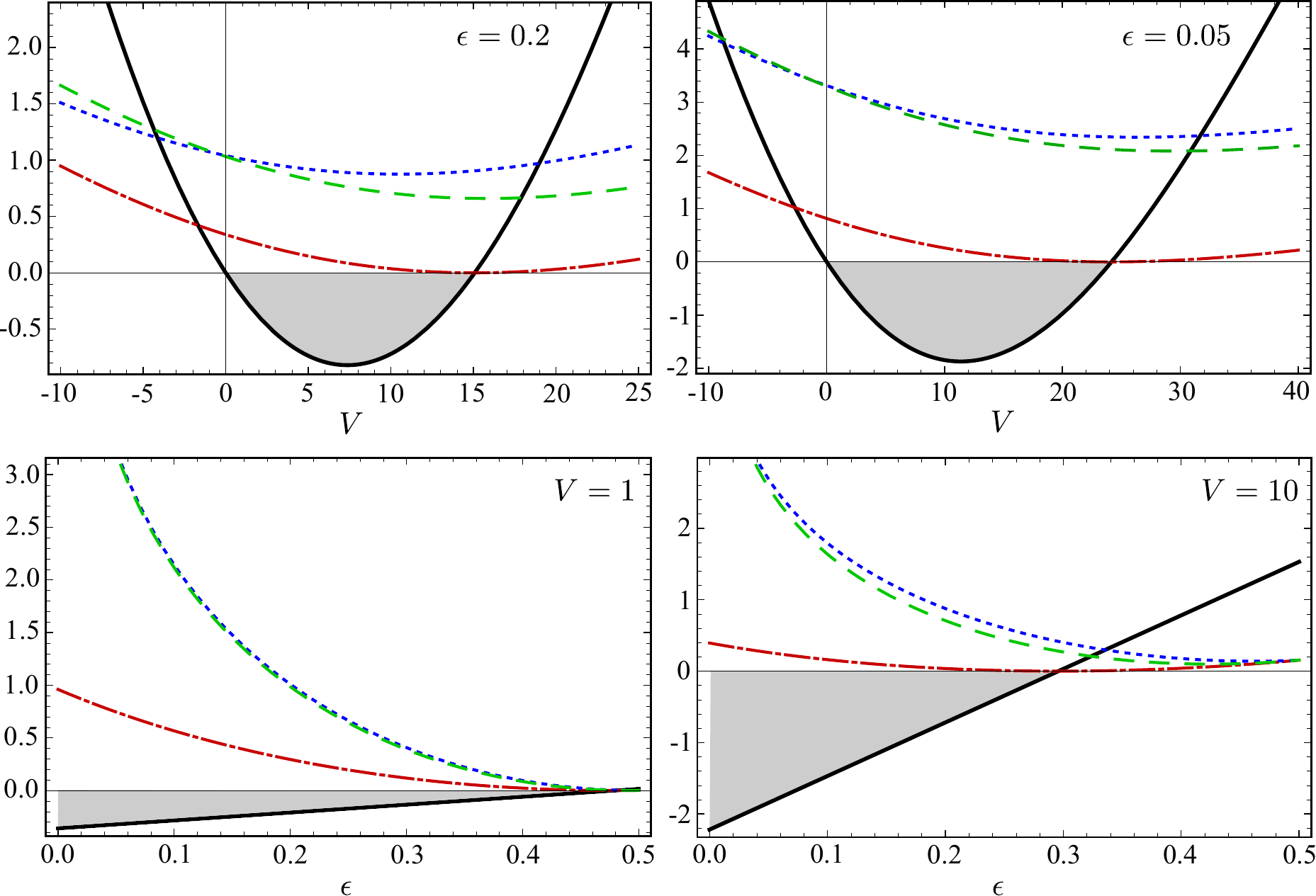}
 \label{fig comparison 2nd laws MD} 
 \caption{(Color Online) Plot of the rate of chemical work $\beta(\mu_L-\mu_R)I_L$ (thick and solid black line) 
 and of three different dissipations: the total one generated by the repeated interaction feedback mechanism 
 $\dot\Sigma$ (dotted blue line, Eq.~(\ref{eq 2nd law electronic MD})), the total one generated by the capacitive 
 feedback mechanism $\dot\Sigma^\text{cap}$ (dashed green line, 
 Refs.~\cite{StrasbergEtAlPRL2013, HorowitzEspositoPRX2014}), and the ``effective'' estimate obtained in these two cases 
 if the demon mechanisms were not known $\dot\Sigma_S^\text{eff}$ (dash-dotted red line, 
 Eq.~(\ref{eq 2nd law effective electronic MD}) and Ref.~\cite{EspositoSchallerEPL2012}). 
 The top row shows these quantities as a function of the bias voltage $V \equiv \mu_L - \mu_R$ for two different 
 measurement errors $\epsilon$. The bottom row shows them as a function of the measurement error $\epsilon \in [0,1/2)$ 
 for two different voltages $V$. The region in which we \emph{extract} work is shaded in gray. 
 Other parameters are $\Gamma = 1$, $\delta = \ln 2$, $\beta = 0.1$ and $U = 0.1$ and we choose a symmetric 
 configuration of the bias, $\mu_L = \epsilon_S + V/2$ and $\mu_R = \epsilon_S-V/2$, effectively eliminating 
 the dependence on $\epsilon_S$ in the equations.}
\end{figure}

It is now interesting to compare the total entropy production generated by the two different electronic demon 
mechanisms, i.e. the one due to capacitive coupling with another quantum dot~\cite{StrasbergEtAlPRL2013}, denoted here as 
$\dot\Sigma^\text{cap}$, and the one generated by repeated interactions considered above, $\dot\Sigma$. 
These two entropy productions can then be compared to the effective one $\dot\Sigma_S^\text{eff}$, obtained when 
the demon mechanism is not known and the only information at hand is that of the system's effective description. 
The best one can do in this case is to derive an ``effective'' second law in spirit of Eq.~(\ref{eq help 14}) 
or of Ref.~\cite{EspositoSchallerEPL2012} which at steady state can be written as 
\begin{equation}\label{eq 2nd law effective electronic MD}
 \dot\Sigma_S^\text{eff} \equiv \sum_\nu\mbox{tr}_S\{(\C L_\text{eff}^{(\nu)}\bar\rho)\ln\bar\rho_\nu\} \ge 0
\end{equation}
and is equivalent to Eq.~(14) of Ref.~\cite{StrasbergEtAlPRL2013}. 
Here $\bar\rho$ is the steady state fulfilling $\C L_\text{eff}\bar\rho = 0$ and $\bar\rho_\nu$ is 
the steady state with respect to reservoir $\nu$, i.e. $\C L_\text{eff}^{(\nu)}\bar\rho_\nu = 0$. 
Since this ``effective'' approach only quantifies the demon effect on the system and neglects the demon's 
dissipation, it will typically underestimate the true dissipation~\cite{EspositoPRE2012, StrasbergEtAlPRL2013}. 

To make the comparison between the various dissipations meaningful, we must compare them in the regime where they 
all give rise to the same effective dynamics on the system, i.e. to the same ME~(\ref{eq ME effective electronic MD}). 
For the repeated interaction mechanism, $\dot\Sigma$ is given by (\ref{eq 2nd law electronic MD}) 
while the effective dissipation $\dot\Sigma_S^\text{eff}$ is given by (\ref{eq 2nd law effective electronic MD}).
For the capacitive mechanism, $\dot\Sigma^\text{cap}$ is given by Eq.~(7) in Ref.~\cite{StrasbergEtAlPRL2013} 
in the fast demon limit $\Gamma_D \rightarrow \infty$, which is required to derive the 
ME~(\ref{eq ME effective electronic MD}). In this limit $\dot\Sigma^\text{cap}$ also coincides with Eq.~(24) in 
Ref.~\cite{HorowitzEspositoPRX2014} and therefore gives it an alternative interpretation in terms of the flow of mutual 
information. Note that we can link the measurement error $\epsilon$ to the parameters used 
in Ref.~\cite{HorowitzEspositoPRX2014} via the relation $\beta_D U = 2 \ln \frac{1-\epsilon}{\epsilon}$. 
The comparison is done in Fig.~\ref{fig comparison 2nd laws MD}. 
We observe that the effective second law $\dot\Sigma_S^\text{eff}$ greatly underestimates the 
true entropy production as expected, but also that the total entropy production generated by the two 
different demon mechanisms, $\dot\Sigma$ and $\dot\Sigma^\text{cap}$, are remarkably close.

\section{Discussions and Conclusions}
\label{sec final remarks}

\subsection{Connection to traditional thermodynamics}
\label{sec consistency standard second law}

We have seen in Sec.~\ref{sec heat reservoir} that if the units are initially thermal, our engine cannot surpass 
Carnot efficiency and deviations of the final unit state from the ideal thermal reservoir (i.e. nonequilibrium effects) 
always cause an even \emph{smaller} efficiency. 
However, if the units are prepared in an arbitrary state, we found that one can continuously extract work from a single 
heat reservoir by lowering the nonequilibrium free energy of the units according to Eq.~(\ref{eq 2nd law modified}). 
We explained that this does not violate the second law of thermodynamics because the overall entropy does not decrease. 
One may nevertheless wonder if this contradicts the classical formulation of the second law according to Kelvin 
and Planck stating that: \emph{There is no cyclic process in nature whose sole result is the conversion of heat from a 
single reservoir into work.} 

\begin{figure}
 \centering\includegraphics[width=0.32\textwidth,clip=true]{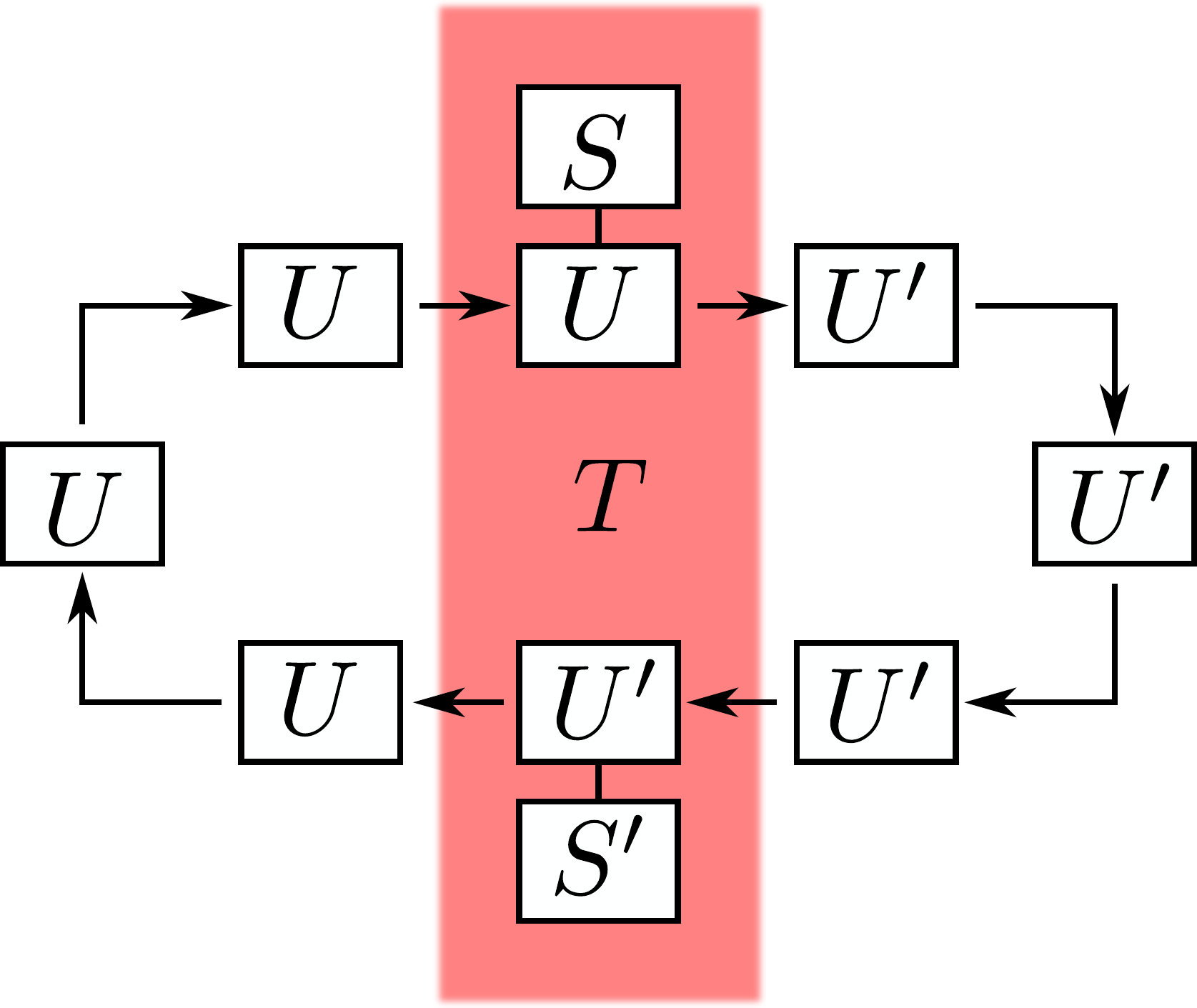}
 \label{fig cycle} 
 \caption{(Color Online) Sketch of the setup to demonstrate the Kelvin-Planck statement of the second law: the outgoing units $U'$ 
 after the interaction with system $S$ are reset to the state $U$ via interaction with a second system $S'$ coupled to 
 the same thermal reservoir as $S$. }
\end{figure}

To answer, let us close the ``cycle of units'' by feeding the outgoing units back into the system $S$ after they 
have interacted with an additional system $S'$ in contact with the same overall heat reservoir at temperature $T$, 
as illustrated on Fig.~\ref{fig cycle}. We assume to be at steady state and denote the outgoing units after the 
interaction with system $S$ by $U'$, which are in turn the incoming units for system $S'$. The additional system $S'$ is 
required to reset the units $U'$ to the state $U$, which again correspond to the incoming units of $S$. Then, for $S$ 
and $S'$ we find the two separate second laws~(\ref{eq 2nd law modified})
\begin{align}
 \beta(W - \Delta F_U)		&\ge I_{S:U}(\tau) \ge 0\,,	\\
 \beta(W' - \Delta F_{U'})	&\ge I_{S':U'}(\tau) \ge 0
\end{align}
which, when added together and using $\Delta F_{U'} = F_U - F_{U'} = - \Delta F_U$, lead to 
\begin{equation}
 \beta(W+W') \ge I_{S:U}(\tau) + I_{S':U'}(\tau) \ge 0.
\end{equation}
Hence, although $W$ might be negative, the sum $W+W'$ must be non-negative in perfect 
agreement with the Kelvin-Planck statement of the second law of thermodynamics.

\subsection{Quantum versus Classical Thermodynamics}
\label{sec quantum vs classical}

At this point it is worth revisiting the debated question of whether quantum thermodynamics offers advantages 
(e.g.  in terms of a higher power output or efficiency) in comparison to classical thermodynamics. 
There is ample evidence that states with quantum properties such as entanglement, coherence or squeezing can be used 
to extract more work than from thermal states, see e.g. Refs.~\cite{ScullyEtAlScience2003, DillenschneiderLutzEPL2009, 
RossnagelEtAlPRL2014, LiEtAlPRE2014, ManzanoEtAlPRE2016, ScullyPRL2010, ScullyEtAlPNAS2011, UzdinPRApp2016, 
KammerlanderAndersSciRep2016, KorzekwaEtAlNJP2016, NiedenzuEtAlNJP2016} (and also Sec.~\ref{sec lasing without inversion}). 
However, this by no means implies that quantum thermodynamics outperforms classical thermodynamics. 
Indeed, classical nonequilibrium properties are usually not considered, but if one only allows for nonequilibrium 
properties which are of a purely quantum origin, this amounts to an unfair competition. In the repeated interaction 
framework the nonequilibrium free energy~(\ref{eq def non eq free energy}) captures both quantum and classical effects 
and we will now use it to analyze the thermodynamics of work extraction. Note that, despite its limitations, the 
framework fully captures quantum effects in the unit state and in the system-unit interaction.

In the ideal scenario where no final system-unit correlations are present (since they always degrade the amount of 
extractable work)\footnote{We note that the maximum value of the quantum mutual information $I_{S:U}$ can be twice as 
large as its classical counterpart. Thus, quantum correlations in the outgoing state have the potential to degrade the 
amount of extractable work to a larger extend then classical ones.} 
the second law~(\ref{eq 2nd law modified}) bounds the extractable work as 
\begin{equation}
 -W \le -\Delta F_U = F_U(0) - F_U(\tau). 
\end{equation}
The extractable work $-W$ gets therefore maximized for a maximum initial and minimum final nonequilibrium free energy. 
The relevant question is therefore whether this procedure can be improved due to quantum effects? 

Let us consider an arbitrary unit which has $N$ levels and a Hamiltonian $H_U = \sum_n E_n|n\rl n|$ with 
$E_1 \le \dots \le E_N$. The state with maximum free energy corresponds to a state with maximum energy and 
minimum entropy as we can easily infer from Eq.~(\ref{eq def non eq free energy}). This state is given by 
$\rho_U(0) = |N\rangle\langle N|$ and thus, is also an allowed classical 
state.\footnote{If the highest energy belongs to a degenerate subspace, i.e. $E_N = E_{N-1} = \dots$, then there is an 
additional freedom in the choice of $\rho_U(0)$ which, however, does not change the nonequilibrium free energy.} 
Finding the state with minimum free energy is more tricky and in general context 
dependent.\footnote{{We remark that the argument presented here is even valid at finite times 
$\tau<\infty$ and for time-dependent Hamiltonians, i.e. out of equilibrium. Realizing that a Gibbs state is a state of 
minimum free energy with respect to a reservoir at inverse temperature $\beta$ is therefore of no help since the system might 
be driven and/or does not have the time to reach the Gibbs state. }}
We can nevertheless easily show that the state with minimum free energy must be ``classical''. 
For this purpose lets define classical states as states which are diagonal in the energy eigenbasis, 
i.e. states which can be written as $\rho_\text{cl} = \sum_n p_n |n \rl n|$. 
Let us denote by $\rho_\text{QM}$ quantum states (i.e. states for which there exists $n \neq m$ 
such that $\langle n|\rho_\text{QM}|m\rangle \neq 0$) which minimize the nonequilibrium free energy. 
Its corresponding classical state (obtained by neglecting all coherences) is given by 
$\rho_\text{cl} = \sum_n \langle n|\rho_\text{QM}|n\rangle |n\rl n|$. 
If we now assume that the nonequilibrium free energy corresponding to $\rho_\text{QM}$ 
is strictly smaller than that of $\rho_\text{cl}$, we get
\begin{equation}\label{eq help 23}
 E(\rho_\text{QM}) - E(\rho_\text{cl}) < T[S(\rho_\text{QM}) - S(\rho_\text{cl})].
\end{equation}
By construction we know that the left hand side is zero since 
\begin{equation}
 E(\rho_\text{QM}) = \mbox{tr}\{H_U\rho_{QM}\} = \sum_n E_n \langle n|\rho_\text{QM}|n\rangle = E(\rho_\text{cl}). 
\end{equation}
Since furthermore we have that $S(\rho_\text{QM}) < S(\rho_\text{cl})$ 
(Theorem 11.9 in~\cite{NielsenChuangBook2000}), Eq.~(\ref{eq help 23}) leads to a contradiction 
and the state of minimum free energy must necessarily be classical. 

We thus proved that within our framework there is no benefit in using quantum over classical ``resources'' 
{in terms of the \emph{bound} dictated by the generalized second law}. 
That is to say, for any scenario where quantum effects are used to extract work, an equivalent classical scenario 
can be conceived (using classical units with the \emph{same} number $N$ of basis states but a different 
interaction) which extracts at least the same amount of work. {It should be noted, however, that we 
did not investigate the question how this bound can be reached and there is evidence that quanutum systems can offer 
advantages in terms of speed, i.e. if we want to extract work at finite 
power~\cite{HovhannisyanEtAlPRL2013, BrunnerEtAlPRE2014, BinderEtAlNJP2015, UzdinLevyKosloffPRX2015, 
NiedenyuGelbwaserKlimoskyKurizkiPRE2015, GelbwaserKlimovskyEtAlSciRep2015}. }

\subsection{Summary and outlook}
\label{sec summary}

We start by summarizing the paper together with its key results. 
In Sec.~\ref{sec standard thermodynamic description}, we reviewed the exact identities describing the correlated 
dynamics of two interacting systems, one of which could be considered as a reservoir that is initially thermal. 
We also considered the 
weak-coupling limit which implicitly assumes macroscopic reservoirs.
In Sec.~\ref{sec generalized thermodynamic framework}, we extended these concepts to describe a system which in 
addition of being continuously interacting with an initially thermal reservoir is subjected to repeated 
interactions with identical units prepared in arbitrary states. 
By establishing exact energy and entropy balances, we showed that the stream of units can be seen as a 
nonequilibrium reservoir or a resource of free energy. 
In Sec.~\ref{sec applications part I}, we identified the limits where these units operate as a pure work, 
heat or information reservoir and also formulated Landauer's principle. 
Most importantly, we showed that our setup can be used to formulate quantum feedback control, derived a new generalized 
bound for the extractable work and provided a clean connection to the theory of  information reservoirs. 

Up to that point, the discussion was based on exact identities which are conceptually powerful but of limited practical use.  
In Sec.~\ref{sec effective master equation}, we started focusing on limits where the system obeys a closed effective dynamics.
We derived effective MEs for the system and established their corresponding thermodynamics: 
This has been done in Subsec.~\ref{sec Poisson distributed interaction time} when the system weakly and continuously interacts with a 
thermal reservoir while rarely 
interacting with units arriving at Poisson distributed times,
and in Subsec.~\ref{sec perturbative expansion delta kicks} when the system frequently interacts solely with the units.  
We also discussed in Subsec.~\ref{sec driven systems} the limit where a time dependent system Hamiltonian 
can be effectively mimicked by a stream of units behaving as a pure work source.

Finally, in Sec.~\ref{sec applications part II} we used our framework to analyze important 
models which were previously considered for their non-conventional thermodynamic features.
In Subsec.~\ref{sec appendix MJ engine} we proposed a microscopic model effectively implementing the Mandal-Jarzynski engine. We showed that 
the work extracted from the entropy of the tape originates from the switching on and off of the system-unit interaction. 
In Subsec.~\ref{sec micromaser} we used our framework to study the thermodynamics of the micromaser which is 
probably the most popular setup making use of repeated interaction. Building on an extension of this model, we showed 
in Subsec.~\ref{sec lasing without inversion} that work can be extracted from purely quantum features based on the idea of 
``lasing without inversion". Finally, in Subsec.~\ref{sec electronic Maxwell demon} we considered a Maxwell demon effect 
on an electronic current crossing a single level quantum dot which was theoretically studied in the past and also 
experimentally realized. We analyzed this effect thermodynamically when the demon mechanism is operated by repeated 
interactions with a stream of units and showed how it differs from the previously considered mechanisms. 

The framework of repeated interactions presented in this paper is quite general and provides a unifying 
picture for many problems currently encountered in the literature. It nevertheless has limitations. 
For instance, our results crucially rely on the fact that the individual units in the incoming stream are decorrelated. 
This assumption is often justified, but recent works started to investigate the role of correlated units, 
classically~\cite{MerhavJSM2015, BoydMandalCrutchfieldNJP2016, GarnerThompsonVedralGuArXiv2015, StrasbergPhD2015, 
BoydMandalCrutchfieldPRE2017} as well as quantum mechanically~\cite{ChapmanMiyakePRE2015}. 
This leads in general to a refined second law with tighter bounds on the amount of extractable 
work~\cite{MerhavJSM2015, BoydMandalCrutchfieldNJP2016, BoydMandalCrutchfieldPRE2017}.
The cost for creating these correlated units (``patterns'') was considered in Ref.~\cite{GarnerThompsonVedralGuArXiv2015},
and Ref.~\cite{StrasbergPhD2015} proposed a simple device to exploit these based on techniques of predictive coding 
(which is a special coding technique, see, e.g.  Sec.~4.7.2 in Ref.~\cite{FeynmanBook1985}). 
In turn, Ref.~\cite{ChapmanMiyakePRE2015} investigated a model where work extraction and information 
erasure is simultaneously possible (and is even enhanced by quantum correlations). It would also be interesting to 
investigate the role of temporal correlations in the unit string due to non-exponential waiting time distribution.

Another limitation of our results when deriving effective master equations is that, although system-unit interactions 
can be arbitrary strong, the system-reservoir interaction must be weak and the resulting dynamics Markovian.  
Uzdin \emph{et al.}~\cite{UzdinLevyKosloffEntropy2016} recently tried to tackle this issue via ``heat exchangers'' 
which could be strongly coupled to the system and which are equivalent to our units. Therefore, their work faces similar 
limitations as ours. Besides few exact identities, the correct thermodynamic description of a system in strong contact 
with a continuous (perhaps non-Markovian) reservoir is still an open and active field of 
research~\cite{JarzynskiJSM2004, CampisiTalknerHaenggiPRL2009, EspositoLindenbergVandenBroeckNJP2010, SchallerEtAlNJP2013, 
PucciEspositoPelitiJSM2013, LudovicoEtAlPRB2014, EspositoOchoaGalperinPRL2015, EspositoOchoaGalperinPRB2015, 
BruchEtAlPRB2016, SeifertPRL2016, StrasbergEtAlNJP2016, CarregaEtAlPRL2016, TalknerHaenggiPRE2016, KatoTanimuraJCP2016, 
JarzynskiPRX2017}. 

As a final remark, let us note that connecting our present work to the recently developed quantum 
resource theories~\cite{GourMuellerEtAlPR2015, GooldEtAlJPA2016} is an interesting perspective, 
as already indicated in Sec.~\ref{sec discussion}. 
The goal of these theories is to establish an axiomatic mathematical framework to study quantum 
thermodynamics based on the study of the interconvertibility of states under certain constraints. 
In fact, if we switch off the permanent coupling to the heat reservoir, assume initially thermal units 
and demand that $\Delta E_S + \Delta E_U = 0$, our framework becomes identical to the one used in the 
resource theory of thermal operations~\cite{GooldEtAlJPA2016}. 
However, we have demonstrated that a consistent thermodynamic framework can be also established for a much 
larger class of situations which are also of experimental relevance. Applying the tools and results from  
resource theory to such problems could prove useful.

\subsection*{Acknowledgements}

Tobias Brandes passed away on February $2$, $2017$ during the finalizing stage of this paper. 
We mourn the loss of a great friend and colleague. 
We thank Juan M. R. Parrondo for stimulating discussions concerning the statistical 
interpretation of the state of the units and its relation to entropy changes.
This work benefited from the COST Action MP1209.
Financial support by the DFG (SCHA 1646/3-1, SFB 910, and GRK 1558), 
by the National Research Fund Luxembourg (project FNR/A11/02) 
and by the European Research Council (project 681456) is gratefully acknowledged.


\bibliography{books,open_systems,thermo,info_thermo,general_refs,further}

\end{document}